\documentclass{aa}
\usepackage[varg]{txfonts}
\usepackage{natbib}
\usepackage{amsmath, amssymb}
\usepackage{graphicx}
\usepackage[breaklinks=true]{hyperref}
\usepackage{xfrac}

\def\st{{\rm \hspace{0.1ex}\circ\hspace{-0.9ex}-}}
\def\dG{\Delta_{\rm f}^{\st}\hspace{-0.2ex}G}

     \def\aap{A\&A}
     \def\apjl{ApJL}
     \def\apjs{ApJS}

\hypersetup{
	pdftoolbar=false,
	pdfmenubar=true,
	pdfstartview={FitP},
	breaklinks=true,
	frenchlinks=false,
        colorlinks=true,
        linkcolor=red,
        citecolor=blue,
        filecolor=cyan,
        urlcolor=magenta,
        pdftitle={Dust in brown dwarfs and extra-solar planets VI},
	pdfsubject={planets and satellites: atmospheres  -- methods: numerical},
	pdfauthor={Graham Lee, Jasmina Blecic, Christiane Helling}
        }

\begin{document}

\title{Dust in brown dwarfs and extra-solar planets}
\subtitle{VI. Assessing seed formation across the brown dwarf and exoplanet regimes}
\author{G. K. H. Lee\inst{1,4}\thanks{E-mail:
graham.lee@physics.ox.ac.uk}, J. Blecic\inst{2,3} \and Ch. Helling\inst{4}}
\institute{$^{1}$Atmospheric, Oceanic \& Planetary Physics, Department of Physics, University of Oxford, Oxford OX1 3PU, UK \\
$^{2}$NYU Abu Dhabi, PO Box 129188, Abu Dhabi, UAE \\
$^{3}$Max-Plank-Institut f\"{u}r Astronomie, K\"{o}nigstuhl 17, D-69117 Heidelberg, Germany \\
$^{4}$Centre for Exoplanet Science, SUPA, School of Physics and Astronomy, University of St Andrews, North Haugh, St Andrews KY16 9SS, UK}

\date{Received: 21 Sep 2017 / Accepted: 25 Jan 2018}

\abstract{
The cloud formation process starts with the formation of seed particles, after which, surface chemical reactions grow or erode the cloud particles.
If seed particles do not form, or are not available by another means, an atmosphere is unable to form a cloud complex and will remain cloud free.
}
{
We investigate which materials may form cloud condensation seeds in the gas temperature and pressure regimes (T$_{\rm gas}$ = 100-2000 K, p$_{\rm gas}$ = 10$^{-8}$-100 bar) expected to occur in planetary and brown dwarf atmospheres.
}
{
We apply modified classical nucleation theory which requires surface tensions and vapour pressure data for each solid species, which are taken from the literature.
Input gas phase number densities are calculated assuming chemical equilibrium at solar metallicity.
}
{
We calculate the seed formation rates of TiO$_{2}$[s] and SiO[s] and find that they efficiently nucleate at high temperatures of T$_{\rm gas}$ = 1000-1750 K.
Cr[s], KCl[s] and NaCl[s] are found to efficiently nucleate across an intermediate temperature range of T$_{\rm gas}$ = 500-1000 K.
We find CsCl[s] may serve as the seed particle for the water cloud layers in cool sub-stellar atmospheres.
Four low temperature ice species, H$_{2}$O[s/l], NH$_{3}$[s], H$_{2}$S[s/l] and CH$_{4}$[s], nucleation rates (T$_{\rm gas}$ = 100-250 K) are also investigated for the coolest sub-stellar/planetary atmospheres.
}
{
Our results suggest a possibly, (T$_{\rm gas}$, p$_{\rm gas}$) distributed hierarchy of seed particle formation regimes throughout the sub-stellar and planetary atmospheric temperature-pressure space.
With TiO$_{2}$[s] providing seed particles for the most refractory cloud formation species (e.g. Al$_{2}$O$_{3}$[s], Fe[s], MgSiO$_{3}$[s], Mg$_{2}$SiO$_{4}$[s]), Cr[s] providing the seed particles for MnS[s], Na$_{2}$S[s] and ZnS[s] sulfides, and K/Na/Rb/Cs/NH$_{4}$-Cl binding solid species providing the seed particles for H$_{2}$O[s/l] and NH$_{4}$-H$_{2}$PO$_{4}$/SH[s] clouds.
A detached, high-altitude aerosol layer may form in some sub-stellar atmospheres from the nucleation process, dependent on the upper atmosphere temperature, pressure and availability of volatile elements.
In order to improve the accuracy of the nucleation rate calculation, further research into the small cluster thermochemical data for each cloud species is warranted.
The validity of these seed particle scenarios will be tested by applying it to more complete cloud models in the future.
}


\maketitle

\section{Introduction}
\label{sec:Intro}

The formation of cloud particles is a natural result of the thermochemical environments of both sub-stellar and planetary atmospheres, and it is a major component in defining their local and global atmospheric properties.
The total energy budget of a sub-stellar atmosphere is largely determined by the extent, composition and size of cloud particles that are present.
Cloud particles can be composed of a mix of material and reach a diversity of sizes ranging from haze-like ($\mu$m) to halestone-like (mm).
Clouds contribute to greenhouse and anti-greenhouse effects, blocking photons from escaping or passing through the atmosphere and also scattering photons back into space.

The presence of clouds has large observational consequences on sub-stellar and planetary objects.
For example, in exoplanet transmission spectroscopy, the observation of a Rayleigh-like slope or flat, featureless spectrum at optical wavelengths \citep[e.g.][]{Pont2013,Kreidberg2014,Chen2017,Gibson2017,Kirk2017} and muted infrared water absorption features \citep[e.g.][]{Deming2013, Sing2016, Wakeford2017b}.
The non-negligible optical wavelength geometric albedo (A$_{g}$ $>$ 0.1) of some hot Jupiter exoplanets observed using the \textit{Hubble Space Telescope} (HST) (e.g. HD 189733b; \citealt{Evans2013}) and \textit{Kepler} planets (e.g. \citealt{Demory2013,Heng2013,Angerhausen2015,Esteves2015, Parmentier2016}) suggest the presence of an optical wavelength scattering cloud component on the dayside of some planets.
Forward scattering of photons by large, high altitude aerosols has been observed to make Titan appear brighter during twilight than daylight phases in \textit{Cassini} wavelength bands \citep{Munoz2017}.
\citet{Helling2014} and \citet{Marley2015} provide a review on the observational and modelling efforts for cloudy sub-stellar atmospheres.

Several chemical equilibrium (CE) studies have investigated the composition and locations of cloud layers in sub-stellar atmospheres.
The \citet{Sharp1990} and \citet{Burrows1999} family of models carried out number density calculations for sub-stellar objects and included the effect of `rain out' of condensable species.
The \citet{Fegley1994, Lodders2002, Visscher2006} series of papers have also studied the temperature and pressure dependencies on gas phase number densities including the depletion of gas phase elements by solid species, with each study in the series focusing on specific sets of molecules in phase equilibrium for sub-stellar atmospheres.
\citet{Allard2001}, and subsequent papers, model brown dwarf cloud species in CE and their effect on spectral signatures.
\citet{Mbarek2016} applied a CE approach to model the cloud composition of Super-Earth/Warm-Neptune atmospheres of varying chondritic elemental number density, which resulted in different cloud compositions occurring depending on the H/O and C/O ratios.
Recently, \citet{Mahapatra2017} modelled the gas phase CE number densities and cloud formation properties of hot Super-Earths such as 55 Cnc e and CoRot-7b.
Other elemental ratios, for example Fe/H or Mg/O will change due to the cloud formation process, affecting the gas phase molecular number densities \citep[e.g.][]{Helling2006}.
A carbon rich cloud formation scenario has also been investigated by \citet{Helling2017}.

Inspired by chemical equilibrium cloud species studies across the brown dwarf and exoplanet effective/equilibrium temperature range, we investigate the capacity of the commonly identified cloud species to form seed particles under a variety of thermodynamic conditions in sub-stellar atmospheres.
In this paper, we focus on and assess the nucleation rates and properties of these cloud species across a range of atmospheric temperatures (100-2000 K) and pressures (10$^{-8}$-100 bar).
The completeness of our study is largely determined by the availability of thermochemical data of each material in the literature.

In Section \ref{sec:nuctheory}, we summarise the background of nucleation theory.
Section \ref{sec:nucsuit} discusses each species considered for homogenous nucleation in this study.
In Section \ref{sec:results}, we present the nucleation rates of our 11 chosen species, and compare them to temperature-pressure profiles of sub-stellar objects.
Section \ref{sec:discussion} contains the discussion and Section \ref{sec:conclusion} our summary and conclusions.
Throughout this paper we consider the nucleation of solid species, denoted by the suffix `[s]', except for H$_{2}$O and H$_{2}$S where a solid or liquid phase may be the preferred seed forming phase, denoted by the suffix `[s/l]'.
Gas phase molecules are referenced without a suffix.

\section{Nucleation theory}
\label{sec:nuctheory}

We define nucleation as the chemical transformation of a gas phase into the solid phase, a phase transition, similar to earlier papers in this series \citep{Woitke2004,Helling2006,Helling2008,Witte2009,Helling2013,Lee2015a,Helling2017}.
This follows the concepts introduced by \citet{Gail1984} to model dust formation in Asymptotic Giant Branch (AGB) star winds \citep{Gail1986}.
Reactions amongst molecules, atoms or ions can lead the formation of small clusters (molecules with sizes of N $>$ 1 monomers) through complex networks which eventually lead to the formation of a solid surface onto which other supersaturated gases can subsequently condense.
Clusters containing N > 1 monomers have different molecular geometries \citep[e.g.][]{Goeres1993,John1997,Lee2015a,Decin2017} each with their own thermodynamic properties.

It may be possible to identify reaction paths where each reaction step for the same molecule, atom or ion leads to the formation of the next larger cluster.
This process requires that the monomers are present in the gas phase.
Such homogeneous nucleation occurs, for example, for TiO$_{2}$[s] in which TiO$_{2}$ monomers are present in the gas phase.

To calculate the rate of seed formation, the effective flux $J_{i}^{C}$ [cm$^{-3}$ s$^{-1}$] through this cluster space or along a reaction path is described by the equation \citep{Helling2013}
\begin{equation}
\label{eq:JC}
J_{i}^{C}(N,t) = \sum_{r_{i}-1}^{R_{i}} \left(\frac{f(N - i,t)}{\tau_{gr}(r_{i},N - i,t)} - \frac{f(N,t)}{\tau_{ev}(r_{i},N,t)}\right),
\end{equation}
where $f(N)$ [cm$^{-3}$] is the number density of a cluster of size N and $r_{i}$ is the chemical reaction involving the i-mer.
The nucleation process is then described by the sum over all $r_{i}$ of these reaction networks.
$\tau_{gr}(r_{i},N - i, t)$ [s] is the growth timescale of the growth reaction $r_{i}$ leading from the cluster size N - i to cluster size N.
$\tau_{ev}(r_{i},N, t)$ [s] is the evaporation timescale leading from size N to size N - i.
$\tau_{gr}(r_{i},N - i, t)$ is given by
\begin{equation}
\label{eq:tau_gr}
\frac{1}{\tau_{gr}(r_{i},N - i, t)} = A(N - i)\alpha(r_{i},N - i) v_{rel}(n_{f}(r_{i}),N - i) n_{f}(r_{i}),
\end{equation}
and $\tau_{ev}(r_{i},N, t)$ given by
\begin{equation}
\frac{1}{\tau_{ev}(r_{i},N, t)} = A(N)\beta(r_{i},N) v_{rel}(n_{r}(r_{i}),N) n_{r}(r_{i}),
\end{equation}
where $A(N)$ is the surface area of cluster of size $N$, $\alpha$ and $\beta$ the reaction efficiency factors for the forward and backward reactions respectively,
$n_{f}(r_{i})$ and $n_{r}(r_{i})$ the number density of the molecule of the growth (forward) process and of the evaporation (reverse) process for reaction $r_{i}$, respectively and  v$_{\rm rel}$ = $\sqrt{k_{b}T / 2\pi m_{\rm f}}$ the average relative velocity between the growing/evaporating molecule and the cluster.

The chemical reactions that form larger and larger cluster sizes proceed at a rate that is determined by the local thermodynamic properties of the gas, but also by the thermochemical properties (Gibbs formation energies) of the clusters themselves.
The slowest reaction rate in the whole chemical reaction network, or along one particular reaction path, is defined as the bottleneck reaction \citep[][Section 3]{Goeres1993}.
Associated with the bottleneck reaction is a critical cluster size, N$_{*}$, which is defined as the cluster size at which the energy barrier to proceed with the next reaction step is largest.
The growth of clusters of size N$_{*}$ therefore represent the limiting step of the whole reaction network.
The value of N$_{*}$ is dependent on the local thermochemical conditions \citep{Gail2014}.
If this cluster size is surpassed (i.e. N = N$_{*}$ + 1), it is energetically favourable for all larger clusters to grow to become seed particles.

A high supersaturation ratio, S $\gg$ 1, is needed before nucleation can take place, requiring that the temperature of the gas phase to be cooler than the temperature threshold for thermal stability (T = T(S=1)) at a given gas pressure.
The required temperature difference from the S = 1 phase equilibrium to the S $\gg$ 1 regime for efficient nucleation is described as `supercooling', hence T(S$\gg$1) $\ll$ T(S=1).
The efficient nucleation zones of a species will therefore occur at greater heights in an atmosphere than the the S = 1 zone of that species, which produces differences in the simulation results between thermal stability models and models that consider nucleation \citep{Helling2008c}.
Such large supersaturation ratios are required to overcome the bond-energy defect of the curved surfaces of the clusters \citep{Goeres1996, Gail2014}.

The \citet{Gail1984, Gauger1990, Patzer1998} paper series details much of the physical and mathematical basis of modern nucleation theory in an astrophysical context, which directly underpins the current study.
The majority of the small cluster thermochemical data required for nucleation theories have also been calculated by these groups.
\citet{John1997} investigated the energetics of small Fe clusters, finding that the low binding energies of the dimer, Fe$_{2}$, make nucleation inefficient.
\citet{Chang1998} and \citet{Patzer1999} used density functional theory (DFT) to model the cluster energetics of small Al$_{2}$O$_{3}$ clusters.
They showed that the (Al$_{2}$O$_{2}$)$_{N}$ clusters are more energetically favoured than (Al$_{2}$O$_{3}$)$_{N}$ clusters, suggesting that homogenous nucleation of Al$_{2}$O$_{3}$[s] seed particles is energetically unfavourable.
The formation of Al$_{2}$O$_{3}$[s] seed particles would therefore likely occur via a heteromolecular nucleation path, the product of surface reactions from other gas phase constituents, for example, in the reaction: Al$_{2}$O$_{2}$ + H$_{2}$O $\rightarrow$ Al$_{2}$O$_{3}$[s] + H$_{2}$.
\citet{Decin2017} recently investigated and discussed the small cluster properties of (Al$_{2}$O$_{3}$)$_{N}$.
Following on from ALMA observations of the AGB stars R Dor and IK Tau, they present emission spectra of (Al$_{2}$O$_{3}$)$_{N}$ N = 1-4 using DFT and suggest that (Al$_{2}$O$_{3}$)$_{N}$ clusters of N $>$ 34 may explain their observed 11 $\mu$m feature.
\citet{Jeong2000} and \citet{Jeong2003} performed DFT calculations on polymer chain-like (TiO$_{2}$)$_{N}$ small cluster geometries, finding that TiO$_{2}$[s] is an excellent candidate for seed particle formation in AGB outflows.
(TiO$_{2}$)$_{N}$ clusters were recalculated in \citet{Lee2015a} using updated lower energy state (TiO$_{2}$)$_{N}$ cluster geometries from \citet{Calatayud2008} and \citet{Syzgantseva2011}.

In this section, we summarise three approaches to modelling homogenous nucleation, kinetic nucleation theory (Section \ref{sec:KNT}), classical nucleation theory (Section \ref{sec:CNT}) and modified nucleation theory (Section \ref{sec:MCNT}), which apply different levels of approximation to overcome, where required, the challenge of missing material data.
In depth descriptions and reviews of the various nucleation approaches can be found in \citet{Patzer1998}, \citet{Helling2013} and \citet{Gail2014}.

\subsection{Kinetic nucleation theory}
\label{sec:KNT}

Should small cluster thermochemical data be available, either from experiment or from computational chemistry modelling \citep[e.g.][]{Jeong2000,Lee2015a,Mauney2015,Bromley2016}, it is possible to apply the detailed kinetic approach in order to derive the rate at which seed particles form, without any additional assumptions.
Kinetic nucleation theory (KNT) considers the detailed growth and evaporation rates of each cluster size in the chemical network.
However, rather than considering every possible reaction of different cluster isomers, a more practical method is considering a single reaction path where the formation of a cluster size depends on its chemical kinetic formation and destruction rates.

Applying the Becker-D\"{o}ring method \citep{Becker1935}, the principles of detailed balance and local thermochemical equilibrium \citep[e.g. see Section 3][]{Helling2013} to the set of growth and evaporation rate equations for each cluster in the chemical network (Eq. \ref{eq:JC}), the stationary nucleation rate is derived.
The stationary nucleation rate, $J$ [cm$^{-3}$ s$^{-1}$], is given by \citep{Gail2014}
\begin{equation}
J^{-1}(t) = \sum_{N=1}^{N_{max}} \left( \frac{\tau_{gr}(r_{i}, N, t)}{\stackrel{\circ}{f}(N)} \right),
\label{eq:J*KNT}
\end{equation}
where $\tau_{gr}(r_{i}, N, t)$ [s] is the growth timescale for clusters of size $N$ given by Eq. \eqref{eq:tau_gr}, and $\stackrel{\circ}{f}(N)$ [cm$^{-3}$] the number density of clusters of size $N$ in chemical equilibrium.

The chemical equilibrium number density of clusters of size $N$, $\stackrel{\circ}{f}(N)$ [cm$^{-3}$], is calculated from the law of mass action with the partial pressure of clusters of size $N$ given by
\begin{equation}
\stackrel{\circ}{p}(N) = p^{\st}\left(\frac{\stackrel{\circ}{p}(1)}{p^{\st}}\right)^{N}\exp\left({-\frac{\Delta G(N)}{RT}}\right),
\label{eq:p(n)}
\end{equation}
where $\stackrel{\circ}{p}(1)$ = $\stackrel{\circ}{f}(1)$k$_{b}$T [dyn cm$^{-2}$] is the partial pressure of the monomer in thermal equilibrium, $\Delta G(N)$ [erg mol$^{-1}$] the Gibbs free energy, R [erg mol$^{-1}$ K$^{-1}$] the molar gas constant and T [K] the local gas temperature.
Using the identity $\stackrel{\circ}{f}(N)$ = $\stackrel{\circ}{p}(N)$ / k$_{b}$T, the number density of clusters of size $N$ in local thermochemical equilibrium is found.
The Gibbs free energy, $\Delta G(N)$, is related to the Gibbs free energy of formation, $\dG(N)$ [erg mol$^{-1}$], by
\begin{equation}
\Delta G(N) = \dG(N) - N\dG(1),
\label{eq:DeltaG}
\end{equation}
where $\dG(1)$ is the Gibbs free energy of formation of the monomer, with $^\st$ denoting the reference state, here 298.15 K at 1 bar.

Considering Eq. \ref{eq:J*KNT}, the largest term on the RHS sum for cluster size N will be the rate limiting step of the chemical network.
The growth reaction of this cluster size N is the bottleneck of the chemical network, and this cluster size N will be the critical cluster size, N$_{*}$.
Since it is energetically favourable for clusters of size N =  N$_{*}$ + 1 to continue on to form larger clusters, the calculation of this limiting step largely determines the nucleation rate of seed particles.

\subsection{Classical nucleation theory}
\label{sec:CNT}

Classical nucleation theory (CNT) was developed primarily in the early 20th century, pioneered by, e.g., \citet{Volmer1926,Becker1935,Zeldovich1943}.
CNT is an effective theoretical tool to calculate nucleation rates for species for which no thermochemical cluster data are available to date, for example Cr[s].

In CNT, the stationary nucleation rate, J$_{*}$ [cm$^{-3}$ s$^{-1}$], for the critical cluster of size N$_{*}$, is given by \citep{Helling2013}
\begin{equation}
J_{*}^{c}(t) = \frac{\stackrel{\circ}{f}(1,t)}{\tau_{gr}(1,N_{*},t)} Z(N_{*}) \exp\left((N_{*} - 1) \ln S(T) - \frac{\Delta G(N_{*})}{RT}\right),
\label{eq:J*final}
\end{equation}
where $Z(N_{*})$ is the kinetic Zeldovich factor, $S(T)$ = $\stackrel{\circ}{f}(1)$ k$_{b}$T / p$_{\rm vap}$ the supersaturation ratio, $\Delta G(N_{*})$ [erg mol$^{-1}$] the Gibbs free energy and $\tau_{gr}$ [s] the growth rate of a cluster of size N$_{*}$.
The Zeldovich efficiency factor (the loss of nuclei during their Brownian motion) is given by
\begin{equation}
Z(N_{*}) =  \left(\frac{1}{2\pi RT}\left|\frac{\partial^{2}\Delta G}{\partial N^{2}}\right|_{N_{*}}\right)^{1/2} .
\label{eq:ZelCNT}
\end{equation}

The characteristic of classical nucleation theory is the utilisation of the surface tension of the species' bulk solid, $\sigma_{\infty}$ [erg cm$^{-2}$], as representative of difference in bond energies between the cluster composed of N monomers and the surface of the condensed phase \citep{Gail2014}.
The assumption in CNT is the use of this surface tension to represent the potentially unknown $\Delta$G(N), given by the relation \citep{Helling2013, Gail2014}
\begin{equation}
\frac{\Delta G(N)}{RT} = -N \ln(S) + \theta_{\infty} N^{2/3} \quad \mbox{with} \quad  \theta_{\infty} = \frac{4\pi a_{0}^{2} \sigma_{\infty}}{k_{b}T}.
\label{eq:dGclass}
\end{equation}
The value of the surface tension is frequently taken from experimental results \citep[e.g.][]{Janz2013}.

\subsubsection{Modified classical nucleation approach}
\label{sec:MCNT}

\citet{Draine1977} and \citet{Gail1984} developed the modified classical nucleation theory (MCNT) which links the macroscopic bulk and microscopic cluster properties of the nucleating species.
This theory takes into account the curvature on the surface energy for small clusters.
A fitting factor, N$_{\rm f}$, is defined as the integer monomer cluster size where the surface energy is reduced to half the bulk value \citep{Gail1984}, or left as a free parameter \citep{Gail2014} if small cluster data is not available.
The Gibbs free energy is then given by
\begin{equation}
 \frac{\Delta G(N)}{RT} = \theta_{\infty} \frac{N-1}{ (N-1)^{1/3} + N_{f}^{1/3}},
 \label{eq:DGtheta}
\end{equation}
and critical cluster size, N$_{*}$, as
\begin{equation}
 N_{*} - 1 = \frac{N_{*,\infty}}{8}\left( 1 + \sqrt{1 + 2\left(\frac{N_{f}}{N_{*,\infty}}\right)^{1/3}} - 2\left(\frac{N_{f}}{N_{*,\infty}}\right)^{1/3}\right)^{3} ,
 \label{eq:N*}
\end{equation}
with
\begin{equation}
 N_{*,\infty} = \left(\frac{\frac{2}{3}\theta_{\infty}}{\ln S(T)}\right)^{3}.
\end{equation}

The Zeldovich factor in Eq. \eqref{eq:ZelCNT} is then
\begin{equation}
 Z(N_{*}) = \left(\frac{\theta_{\infty}}{9 \pi(N_{*} - 1)^{4/3}}\frac{(1+2(\frac{N_{f}}{N_{*} - 1})^{1/3})}{(1 + (\frac{N_{\rm f}}{N_{*} - 1})^{1/3})^{3}} \right)^{1/2}.
\end{equation}

The input quantities required for modified classical nucleation theory are therefore the bulk surface tension, $\sigma_{\infty}$; the gas phase monomer number density, $\stackrel{\circ}{f}(1,t)$ (assumed to be in chemical equilibrium); the monomer radius, $a_{0}$; the vapour pressure; $p_{\rm vap}$, and the fitting factor, $N_{\rm f}$.

It is also possible to fit a pseudo bulk surface tension to small cluster data using Eq. \eqref{eq:DGtheta}, in order to provide a more accurate estimate of the nucleation rate.
This exercise has been carried out in \citet{Jeong2003} and \citet{Lee2015a} who found good agreement to KNT for TiO$_{2}$ nucleation in the S $\gg$ 1 regime (also see discussion in Section \ref{sec:smallcluster}).

More details on kinetic and classical nucleation theory can be found in \citet{Helling2013} and \cite{Gail2014}.

\section{Possible seed particle species in the brown dwarf and planetary regimes}
\label{sec:nucsuit}

In this section, we present a summary of the present state of knowledge about potential seed forming species across the sub-stellar and planetary atmospheric temperature and pressure regimes.
Additionally, we describe our chemical equilibrium scheme, required to calculate the number density of seed forming gas species as input to the nucleation theory.

\subsection{Titanium Dioxide and Silicon Oxide}

The nucleation properties of TiO$_{2}$[s] and SiO[s] were previously examined and discussed in \citet{Helling2013, Lee2015a}.
We repeat the methodology presented in those studies for a wider range of temperatures and pressures in order to asses the nucleation properties of TiO$_{2}$[s] and SiO[s] across a greater cloud formation parameter space and to compare to other seed forming species.

TiO$_{2}$[s] is a refractory material, thermally stable at high temperatures ($>$ 1500 K), which has been investigated as a condensation seed for AGB star wind models \citep{Gail1998,Jeong2003} and sub-stellar clouds \citep{Woitke2004}.
In chemical equilibrium, gas phase TiO$_{2}$ is typically 1-2 orders of magnitude less abundant than, for example, SiO \citep{Helling2013}.
TiO$_{2}$ and other Ti binding gas species, such as TiC, TiO, Ti and TiH are present in the gas phase.
At temperatures greater than $\sim$ 1300 K, TiO is the most abundant Ti species \citep{Lee2015a}.
At temperatures $<$ 1300 K TiO$_{2}$ is generally more abundant than TiO by $\sim$ 2 orders of magnitude.
Ti bearing species are some of the first species to condense and form clouds in hot Jupiter exoplanets \citep[e.g.][]{Woitke2004,Wakeford2017a}.

SiO is an abundant gas phase species in chemical equilibrium \citep{Burrows1999}.
Its homogenous nucleation rates have been investigated experimentally and computationally \citep[e.g.][]{Gail2013, Bromley2016}, with an emphasis on AGB star outflows.
Recently, \citet{Bromley2016} performed detailed kinetic nucleation calculations of SiO clusters up to N = 20.
Their updated nucleation rates suggest that SiO is an unfavourable seed particle provider for AGB stellar outflow thermochemical conditions.
SiO is generally the most abundant Si bearing gas species in sub-stellar atmospheres, with SiO$_{2}$ 2 orders of magnitude less abundant in typical L brown dwarf temperature-pressure conditions \citep{Lee2015a}.

\subsection{Chromium}

Chromium bearing species have been modelled as a cloud species for sub-stellar atmospheres in many CE studies such as \citet{Burrows1999,Lodders2006,Morley2012}.
Cr molecules found in the gas phase in chemical equilibrium include CrO, CrO$_{2}$, CrO$_{3}$ and CrH.
Atomic Chromium is the most refractory Cr bearing species, and is thermochemically more favourable to condense at higher temperatures compared to Cr$_{2}$O$_{3}$ \citep{Lodders2006}, suggesting that the atomic form may be the key Cr[s] cloud forming species in sub-stellar atmospheres.
The gas-solid S = 1 phase equilibrium of Cr[s] occurs $\sim$ 100 K lower than S = 1 equilibrium threshold, for a given gas pressure, of the magnesium silicate cloud layer \citep{Morley2012}.

\subsection{K/Na Chlorides}

Sea salt spray is known to be a source of seed particles for Earth water clouds near the planet surface \citep{ODowd1997}.

The condensation of Potassium Chloride, KCl[s], has been suggested to alter the atomic gas phase K number densities for some hot Jupiter atmospheres, altering their observed K absorption feature in transmission spectroscopy \citep[c.f.][]{Sing2015a,Gibson2017}.
KCl occurs with K, KOH and KH gas phase species, and has the largest number density of these species in the gas phase at temperatures $<$ 700 K \citep{Lodders1999}.
The number density of NaCl follows similar trends in temperature and pressure dependencies to KCl.

One aspect when considering NaCl[s] nucleation or growth is the fact that Na$_{2}$S[s] has been shown to be a potential cloud particle species for T brown dwarf atmospheres \citep{Lodders1999,Morley2012}.
Should Na$_{2}$S[s] condense and rain-out, it would remove a large amount of Na from the gas phase.
It can therefore be expected that NaCl[s] cloud formation, which occurs at a lower temperature compared to Na$_{2}$S[s], would be suppressed due to a lack of Na-bearing species in the gas phase above the Na$_{2}$S[s] cloud layer.
The bulk thermodynamic properties of KCl[s] and NaCl[s] have been extensively experimentally investigated, such as molten salt surface tension experiments \citep{Janz2013}.

\subsection{H$_{2}$O, NH$_{3}$, H$_{2}$S and CH$_{4}$ ices/liquids}

We consider four ice/liquid species in this study, H$_{2}$O[s/l], NH$_{3}$[s], H$_{2}$S[s/l] and CH$_{4}$[s].
Gas phase H$_{2}$O, NH$_{3}$, H$_{2}$S and CH$_{4}$ have higher number densities than other species, e.g. TiO$_{2}$, since their constituent atoms are far more abundant at solar metallicity \citep[e.g.][]{Burrows1999, Lodders2006}.

Should an ice species be highly supersaturated and supercooled, it may perform the role of seed particles, providing surfaces for subsequent growth of that ice cloud particle.

\subsection{Gas phase chemical equilibrium number densities}
\label{sec:TEA}

\begin{figure*}[ht] 
   \centering
   \includegraphics[width=0.49\textwidth]{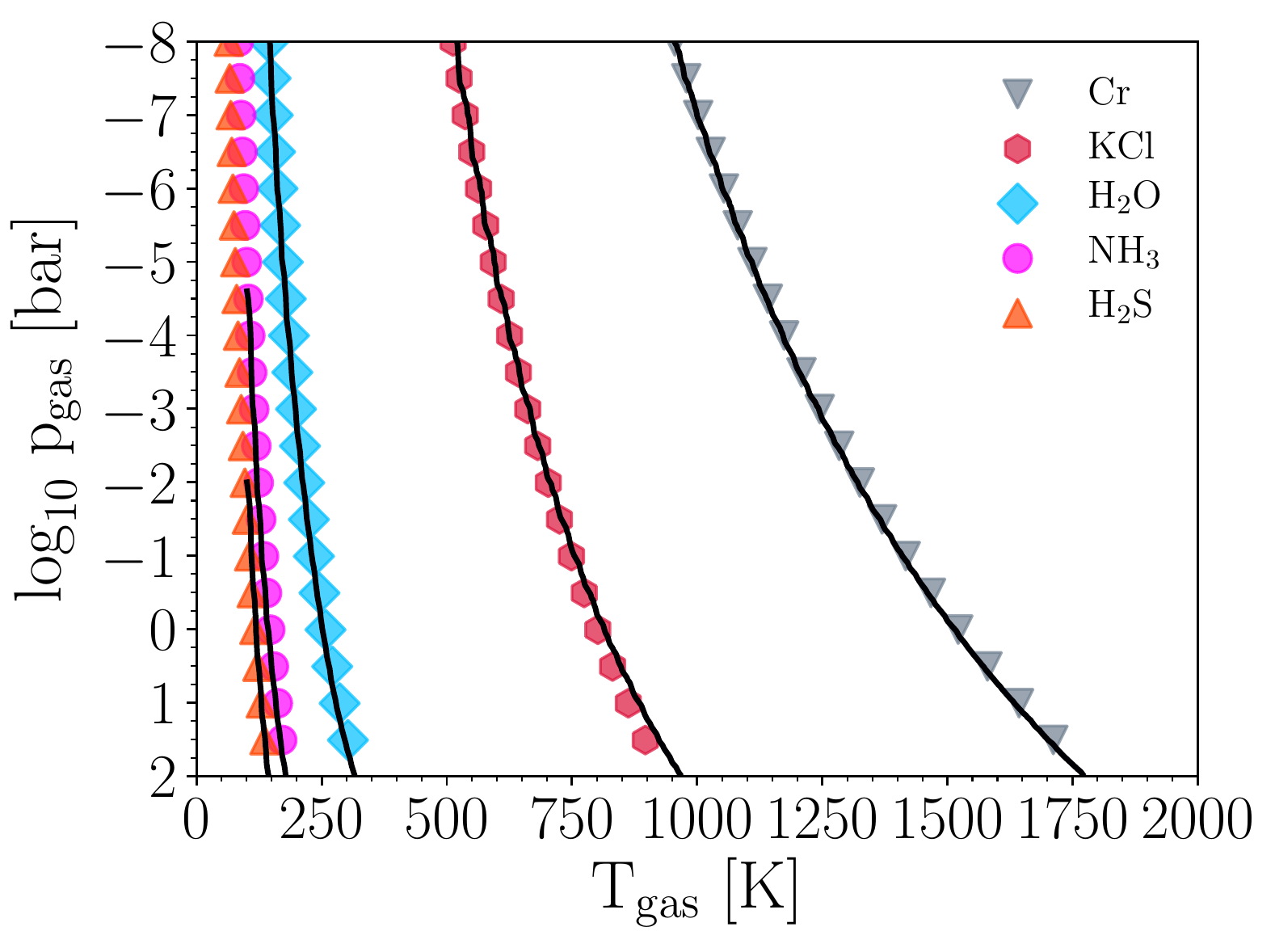}
   \includegraphics[width=0.49\textwidth]{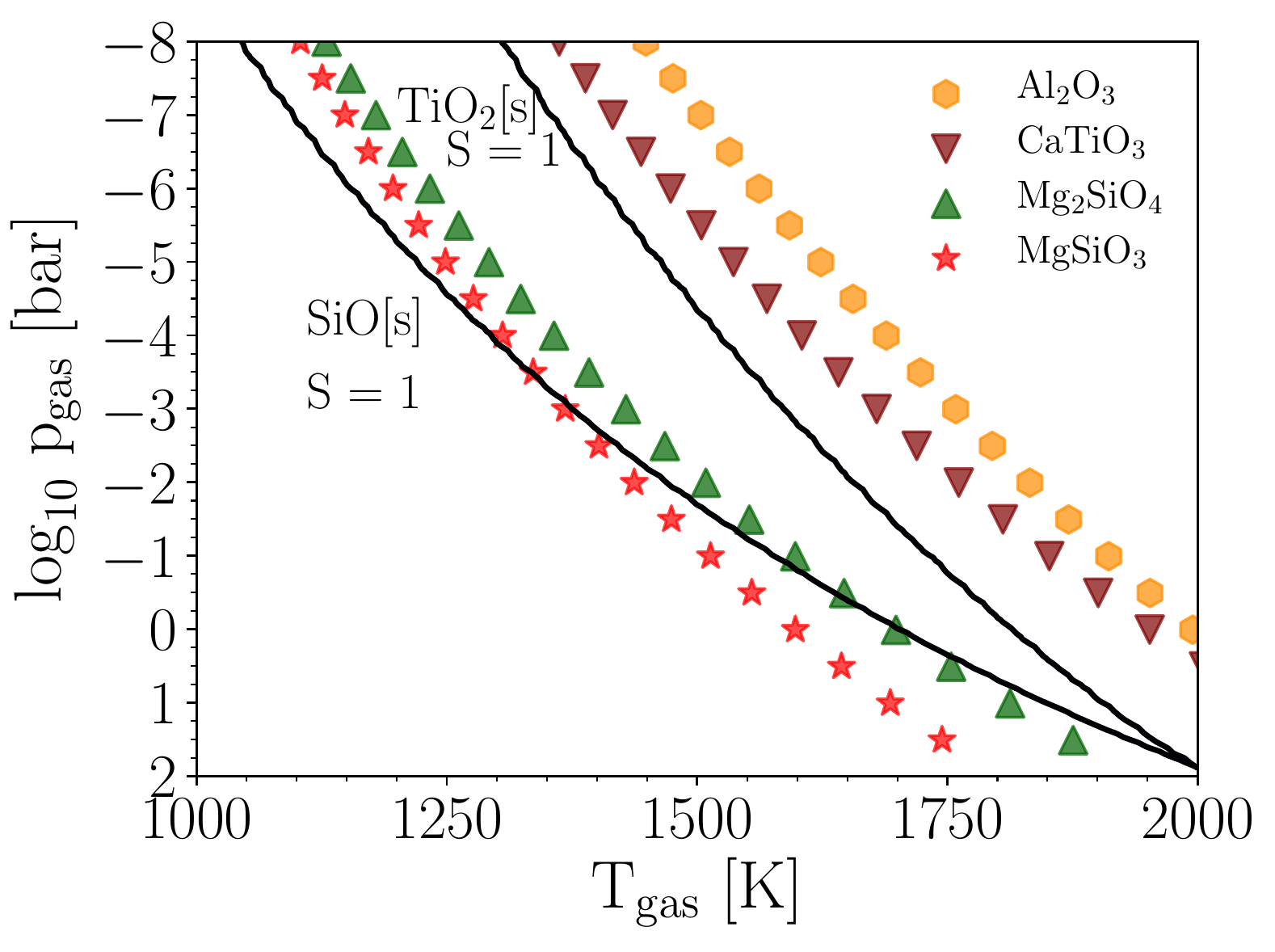}
   \caption{\textbf{Left:} S = 1 phase equilibrium zones of Cr[s], KCl[s], H$_{2}$O[s/l] and H$_{2}$S[s/l] cloud species (black lines) as a result the TEA species number density calculations.
   \textbf{Right:} S = 1 phase equilibrium zones of TiO$_{2}$[s] and SiO[s]  cloud species (black lines) as a result of our \citet{Helling2016} species number density calculations.
   Symbols denote values from the S = 1 phase equilibrium fitting equations from \citet{Lodders2002}; H$_{2}$O[s/l], H$_{2}$S[s/l], \citet{Visscher2010}; MgSiO$_{3}$[s], Mg$_{2}$SiO$_{4}$[s], \citet{Morley2012}; Cr[s], KCl[s] and \citet{Wakeford2017a}; Al$_{2}$O$_{3}$[s], CaTiO$_{3}$[s] at solar metallicity.
   No direct comparisons with TiO$_{2}$[s] and SiO[s] between the studies are available.}
   \label{fig:TEA_results}
\end{figure*}

We use two chemical equilibrium schemes to calculate the gas phase species number densities required.
For TiO$_{2}$ and SiO number densities we apply the chemical equilibrium scheme used in \citet{Helling2016}.
For Cr, KCl, NaCl, CsCl, H$_{2}$O, NH$_{3}$, H$_{2}$S and CH$_{4}$ gas phase number density calculations, we apply the open source Thermochemical Equilibrium Abundances (TEA) code of \citet{Blecic2016}.
Both codes use the \citet{Asplund2009} solar metallicity atomic abundances as input.
However, Solar System gas/ice giants are super-solar in metallicity, which affects the number density of gas phase species important to the nucleation process (Section \ref{sec:discussion}).

The \citet{Helling2016} scheme contains a total of 167 gas phase species, with the following Ti species:

\begin{itemize}
\item Ti, TiH, TiO, TiO$_{2}$, TiC, TiC$_{2}$, TiS, Ti+ ,
\end{itemize}
and Si species:
\begin{itemize}
\item Si, SiH, SiH$_{4}$, SiO, SiO$_{2}$, SiN, SiS, Si+ .
\end{itemize}

Currently TEA has 889 gas phase species available in its database.
We performed tests comparing the full species list to reduced sub-set of our molecules of interest, in order to enable reasonable computational times, while retaining negligible differences, $<$ 1 \%, in number density results of the gas phase molecules involved in the seed formation to the full species list.
For the TEA computations for temperatures $>$ 500 K the following species list is applied

\begin{itemize}
\item H, He,  C,  O, N, Ti, Si, Cr, K, Na, Cl, H$_{2}$, CO, CO$_{2}$, CH$_{4}$, N$_{2}$, H$_{2}$O, NH$_{3}$, TiO$_{2}$, SiO, KCl, NaCl, HCl ,
\end{itemize}
while for temperatures $<$ 500 K the following species list is used
\begin{itemize}
\item H, He, C, O, N, S, Cr, K, Na, Cl, P, H$_{2}$, CO, CO$_{2}$, CH$_{4}$, N$_{2}$, H$_{2}$O, NH$_{3}$, HCN, C$_{2}$H$_{2}$, C$_{2}$H$_{4}$, KCl, NaCl, H$_{2}$S, HS, HCl, PH$_{3}$ .
\end{itemize}

One needs to be careful when optimizing the speed of the thermochemical calculation by excluding species which are not of their imminent interest, as a non-realistic and incomplete list of species can affect the equilibrium number density of the species of interest (and other minor species) by several orders of magnitude, and produce an incorrect result.

This two-code approach is used since the TEA species list, taken from the JANAF-NIST database, does not currently include some of the important Ti and Si species compared to the \citet{Helling2016} species list.
This results in differences between the two codes for the gas phase number densities of TiO$_{2}$ and SiO gas phase species.

Figure \ref{fig:TEA_results} presents the S = 1 phase equilibrium results of TiO$_{2}$[s], SiO[s], Cr[s], KCl[s], H$_{2}$O[s/l], NH$_{3}$[s] and H$_{2}$S[s/l] compared to expressions found in \citet{Lodders2002, Visscher2010, Morley2012} and \citet{Wakeford2017a}.
The contour line results show good agreement, suggesting that our cloud species gas phase number densities are consistent with previous studies.
We also note that \citet{Helling2008} found good agreement between the gas phase equilibrium codes of \citet{Helling2016} and \citet{Lodders2002}.
The slight deviations are possibly due to the different input solar elemental abundance used in each study and potentially different data sources for the vapour pressure of each species.
Although no direct comparison is available for TiO$_{2}$[s] and SiO[s], our results and trends are consistent with their related species (Figure \ref{fig:TEA_results}, right).

\section{Assessing seed particle formation rates for exoplanet and brown dwarf atmospheres}
\label{sec:results}

In this section, we present an assessment of the suitability of materials as seed forming species for cloud formation in brown dwarf and planet atmospheres.
This assessment is based on the present state material properties available for the gas phase, for clusters and for solid and liquid phases.
We investigate the nucleation rates of each of the eight candidate species across a 2-dimensional grid of gas temperature and gas pressure.
We compare the S = 1 phase equilibrium limit and the (T$_{\rm gas}$,p$_{\rm gas}$) efficient nucleation regions to the temperature pressure profiles of sub-stellar objects.
Throughout, we assume solar metallicity at the element abundance values given in \citet{Asplund2009}.

In Section \ref{sec:hierarchy}, we compare our nucleation rate results to the S = 1 phase equilibrium limits of several mineral species.
When invoking MCNT for Cr[s], KCl[s], NaCl[s], H$_{2}$O[s/l], NH$_{3}$[s], H$_{2}$S[s/l] and CH$_{4}$[s], we apply a fitting factor of $N_{f}$ = 1 for all species, following the advice in \citet{Gail2014}.
The sensitivity of our nucleation rates to a higher value of the free parameter, N$_{f}$ = 10, is given in Appendix \ref{app:paramaters}, which does not alter the conclusions discussed in the main text.
Table \ref{tab:vapp}, \ref{tab:s_tensions} and \ref{tab:radii} provide the required input vapour pressures, surface tensions and monomer radii for each species used in the present paper.

We use our two CE codes to produce gas phase number densities across a 2-dimensional grid of temperature and pressure points, from T$_{\rm gas}$ = 500-2000 K in steps of 25 K and p$_{\rm gas}$ = 10$^{-8}$-100 bar for 100 grid point equally spaced in $\log_{10}$ pressure.
For the temperature range T$_{\rm gas}$ = 100-500 K, the temperature step is reduced to 10 K to increase the temperature resolution of the results.
Using these gas phase number densities, we then calculate nucleation rates, J$_{*}$, for each seed particle nucleation species.
Throughout this paper, we define an `efficient' nucleation rate to be J$_{*}$ $>$ 1 cm$^{-3}$ s$^{-1}$.
Table \ref{tab:res_sum} presents a tabular overview of our main results for the following sections.

\begin{table*}
\caption{Summary of all solid, ice and liquid species nucleation methodology and results in this study.
T(S = 1) is the temperature range of the S = 1 phase equilibrium zone across the gas pressure range (10$^{-8}$-100 bar).
 $\Delta$T$_{\rm cool}$ is the difference in temperature required (supercooling) before an efficient nucleation rate (J$_{*}$ $>$ 1 cm$^{-3}$ s$^{-1}$) is reached from the S = 1 phase equilibrium zone.}
\begin{center}
\begin{tabular}{l c c c c c}  \hline \hline
Species & Method & $N_{f}$ &  T(S = 1) [K] & $\Delta$T$_{\rm cool}$ [K] & Figure \\ \hline
TiO$_{2}$[s] & MCNT/KNT & 0  & 1350-2000 & 100-250 & \ref{fig:TiO2_res}/\ref{fig:TiO2_KNT} \\
SiO[s] & \citep{Gail2013} & -  & 1100-2000 & 200-250 &  \ref{fig:SiO_res}\\
Cr[s] & MCNT &  1 & 1050-1750 & 500-700 & \ref{fig:Cr_res}\\
KCl[s]  & MCNT & 1 & 600-900 & 50-100 & \ref{fig:KCl_res} \\
NaCl[s]  & MCNT & 1  & 600-900 & 50-100 & \ref{fig:NaCl_res}  \\
CsCl[s]  & MCNT & 1  & 450-600 & 50-100 & \ref{fig:Cs_seeds} \\ \hline
H$_{2}$O[s/l] & MCNT & 1  & 160-310 & 10-30 & \ref{fig:H2O_res}  \\
NH$_{3}$[s]  & MCNT & 1  & $<$100-180 & 10-20 & \ref{fig:NH3_res} \\
H$_{2}$S[s/l]  & MCNT & 1  & $<$100-140 & 25-35 & \ref{fig:H2S_res}  \\
CH$_{4}$[s]  & MCNT & 1  & - & - & - \\  \hline
\end{tabular}
\end{center}
\label{tab:res_sum}
\end{table*}%

\subsection{TiO$_{2}$[s]}
\label{sec:TiO2_res}

\begin{figure*}[ht] 
   \centering
   \includegraphics[width=0.49\textwidth]{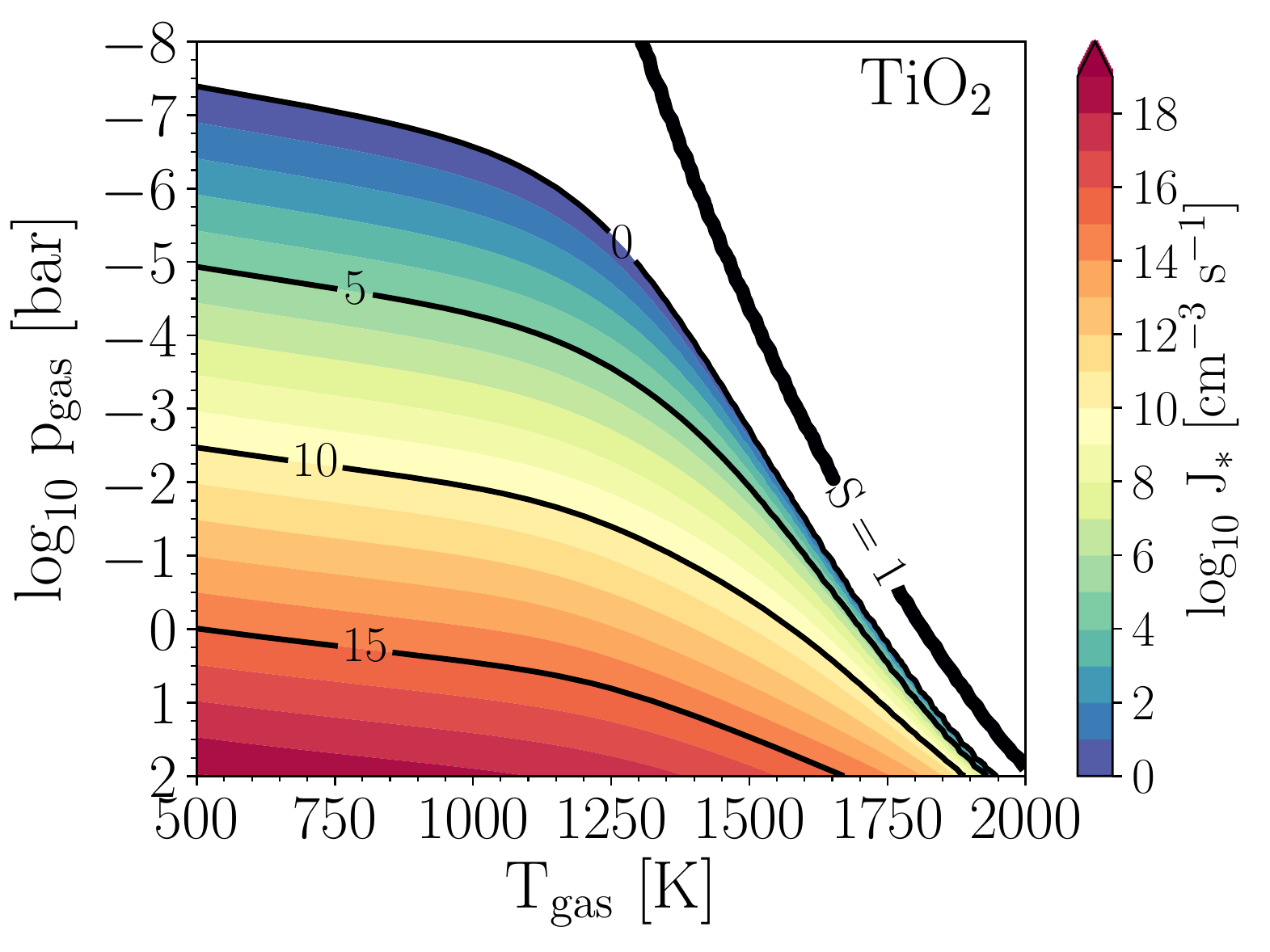}
   \includegraphics[width=0.49\textwidth]{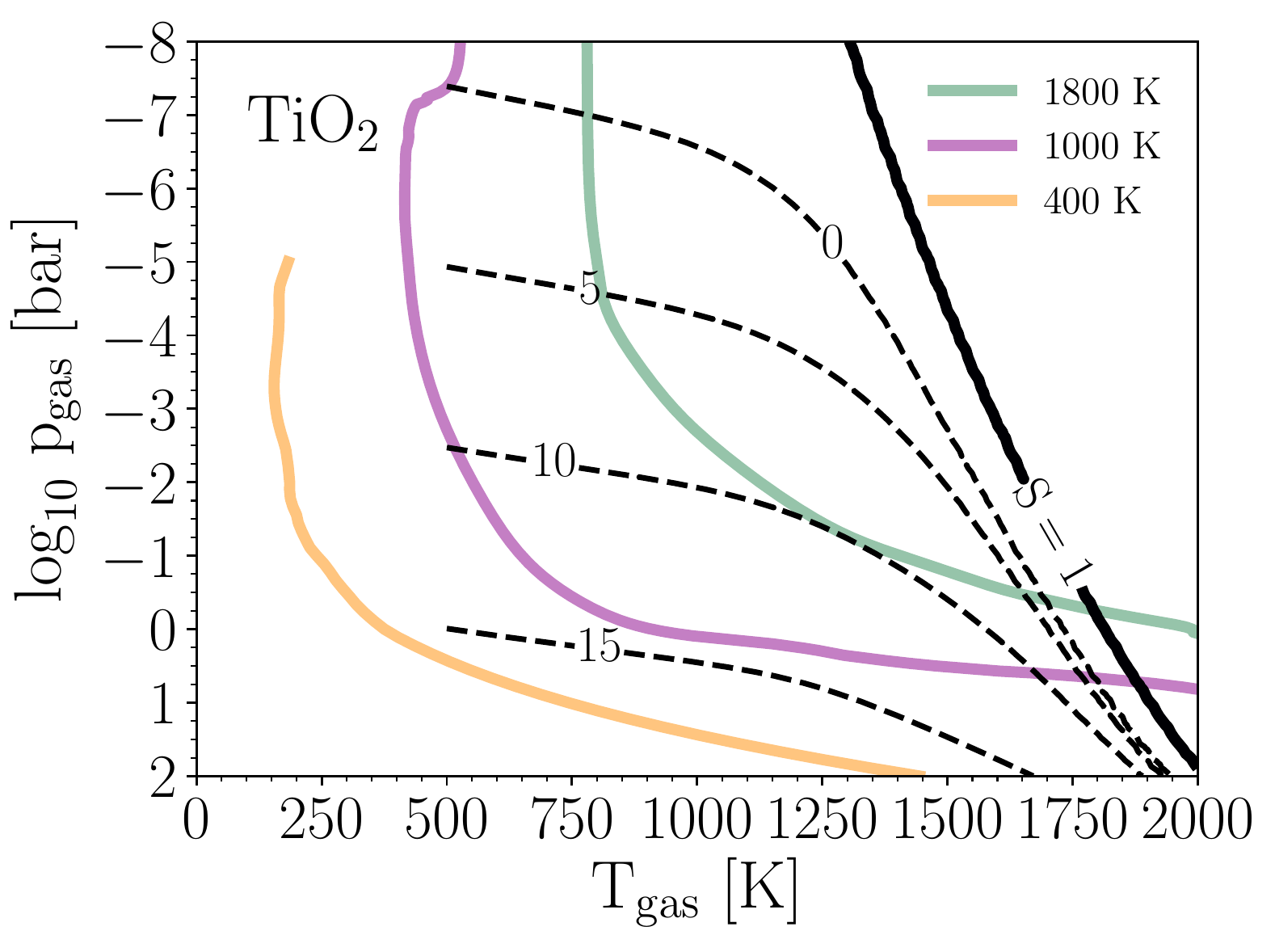}
   \caption{Nucleation rates of TiO$_{2}$[s] using MCNT (Eq. \ref{eq:J*final}).
   \textbf{Left:} Contours of $\log_{10}$ J$_{*}$ [cm$^{-3}$ s$^{-1}$] of TiO$_{2}$[s] (colour bar and thin black lines) across the temperature-pressure grid, and the gas-solid phase equilibrium S(T$_{\rm gas}$,p$_{\rm gas}$) = 1 zone (thick black, solid).
   \textbf{Right:} S = 1 phase equilibrium result (black, solid) and selected $\log_{10}$ J$_{*}$ contour lines (black, dashed) compared to T$_{\rm eff}$ = 1800 K, log g = 5 (green, solid), T$_{\rm eff}$ = 1000 K, log g = 5 (purple, solid), \textsc{Drift-Pheonix} \citep{Witte2009} and T$_{\rm eff}$ = 400 K, log g = 5 (dark orange, solid), \textsc{Exo-REM} \citep{Baudino2015,Baudino2017} brown dwarf (T$_{\rm gas}$,p$_{\rm gas}$) profiles.}
   \label{fig:TiO2_res}
\end{figure*}

Figure \ref{fig:TiO2_res} (left) presents our temperature and pressure grid results for TiO$_{2}$[s] nucleation.
TiO$_{2}$[s] requires a supercooling of 100-250 K before efficient nucleation can occur.
Figure \ref{fig:TiO2_res} (right) shows the TiO$_{2}$[s] J$_{*}$ contours and S = 1 phase equilibrium results compared to [M/H] = 0.0, log g = 5, T$_{\rm eff}$ = 1000K and 1800 K brown dwarf atmospheres from the \textsc{Drift-Pheonix} \citep{Witte2009} models.
This comparison suggest that TiO$_{2}$[s] is one of the first refractory species to potentially nucleate, since it is thermochemically favourable for it to nucleate in L and T brown dwarf atmospheres.

\subsection{SiO[s]}

\begin{figure*}[ht] 
   \centering
   \includegraphics[width=0.49\textwidth]{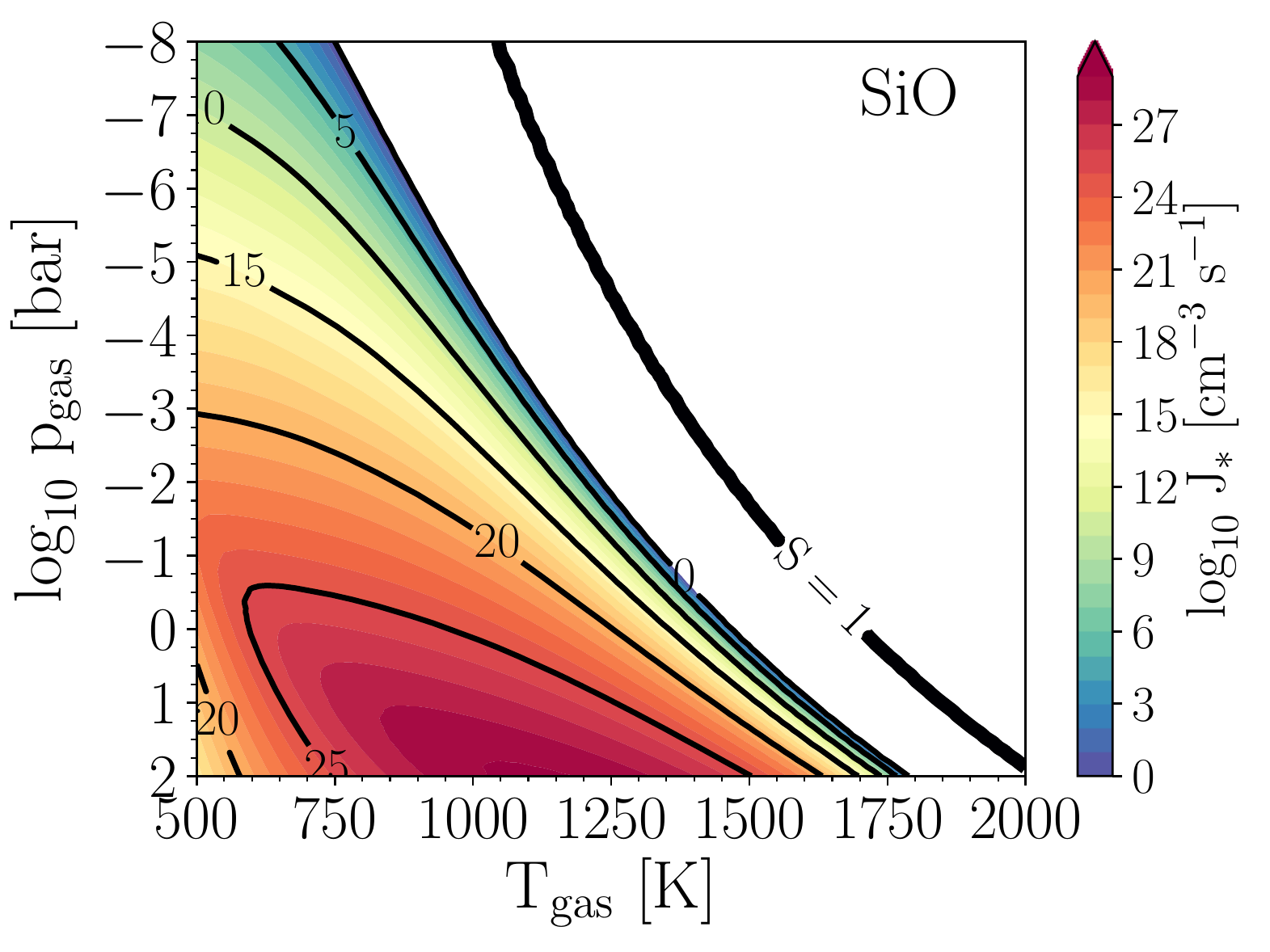}
   \includegraphics[width=0.49\textwidth]{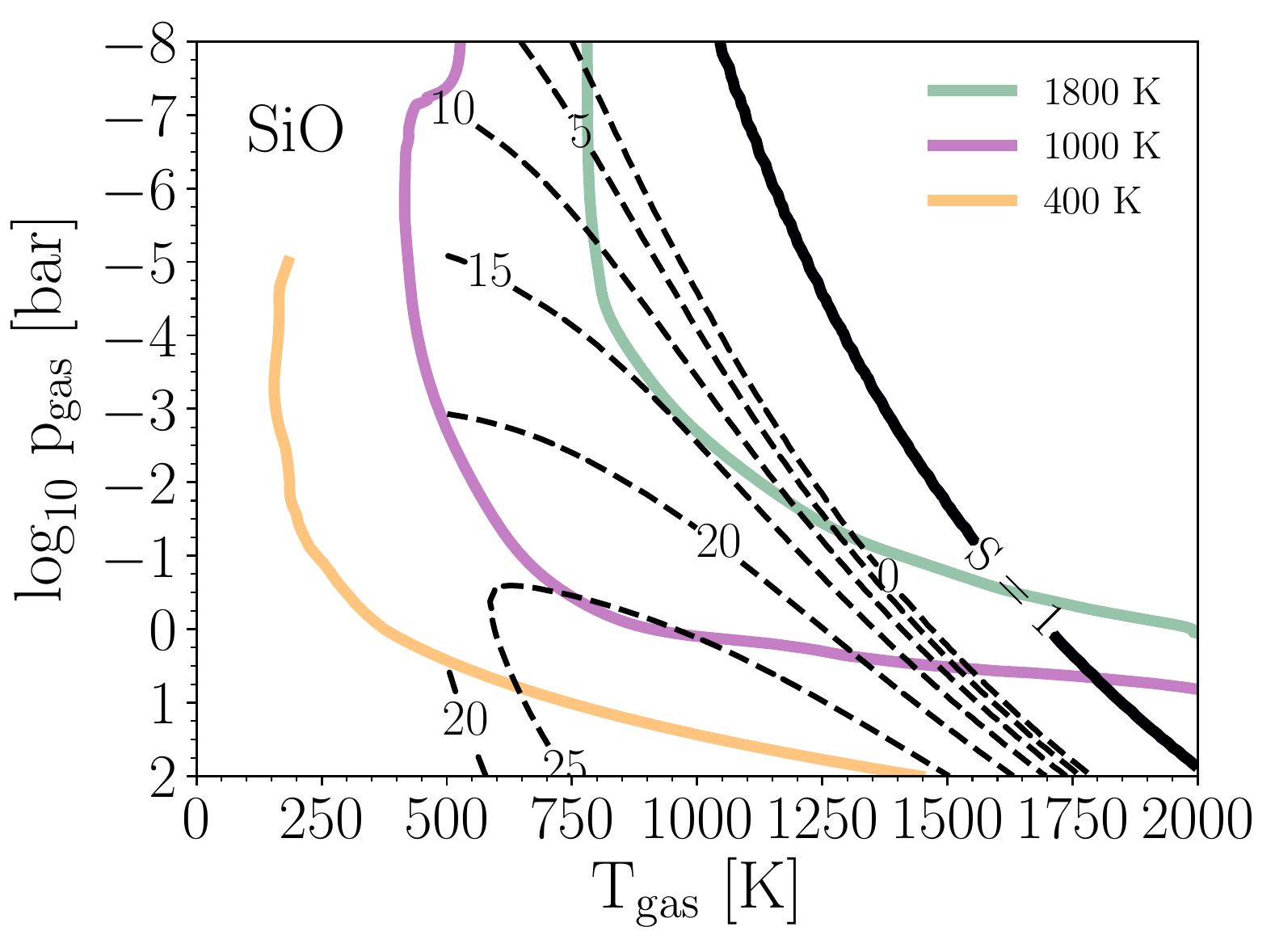}
   \caption{Nucleation rates of SiO[s] using the \citet{Wetzel2013,Gail2013} expressions.
   \textbf{Left:} Contours of $\log_{10}$ J$_{*}$ [cm$^{-3}$ s$^{-1}$] of SiO[s] (colour bar and thin black lines) across the temperature-pressure grid, and the gas-solid phase equilibrium S(T$_{\rm gas}$,p$_{\rm gas}$) = 1 zone (thick black, solid).
   \textbf{Right:} S = 1 phase equilibrium result (black, solid) and selected $\log_{10}$ J$_{*}$ contour lines (black, dashed) compared to T$_{\rm eff}$ = 1800 K, log g = 5 (green, solid), T$_{\rm eff}$ = 1000 K, log g = 5 (purple, solid), \textsc{Drift-Pheonix} \citep{Witte2009} and T$_{\rm eff}$ = 400 K, log g = 5 (dark orange, solid), \textsc{Exo-REM} \citep{Baudino2015,Baudino2017} brown dwarf (T$_{\rm gas}$,p$_{\rm gas}$) profiles.}
   \label{fig:SiO_res}
\end{figure*}

Figure \ref{fig:SiO_res} (left) presents our J$_{*}$ temperature and pressure grid results for SiO[s] nucleation.
SiO[s] requires a supercooling of 200-250 K before efficient nucleation can occur.
In Fig. \ref{fig:SiO_res} (right), we compare the nucleation rate for SiO[s], J$_{*}$(SiO), with two example brown dwarf atmospheric (T$_{\rm gas}$, p$_{\rm gas}$) profiles.

\subsection{Cr[s]}

\begin{figure*}[ht] 
   \centering
   \includegraphics[width=0.49\textwidth]{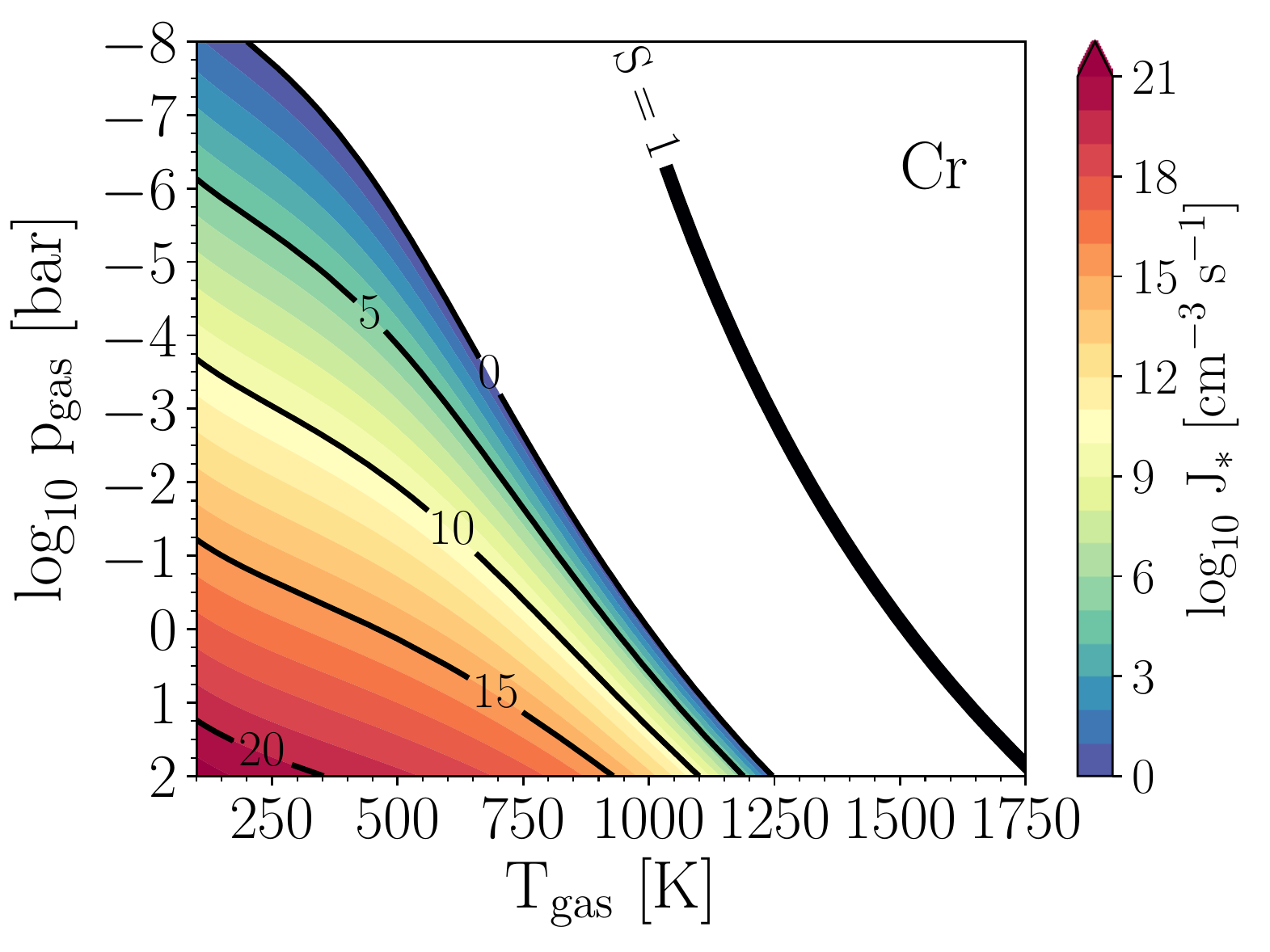}
   \includegraphics[width=0.49\textwidth]{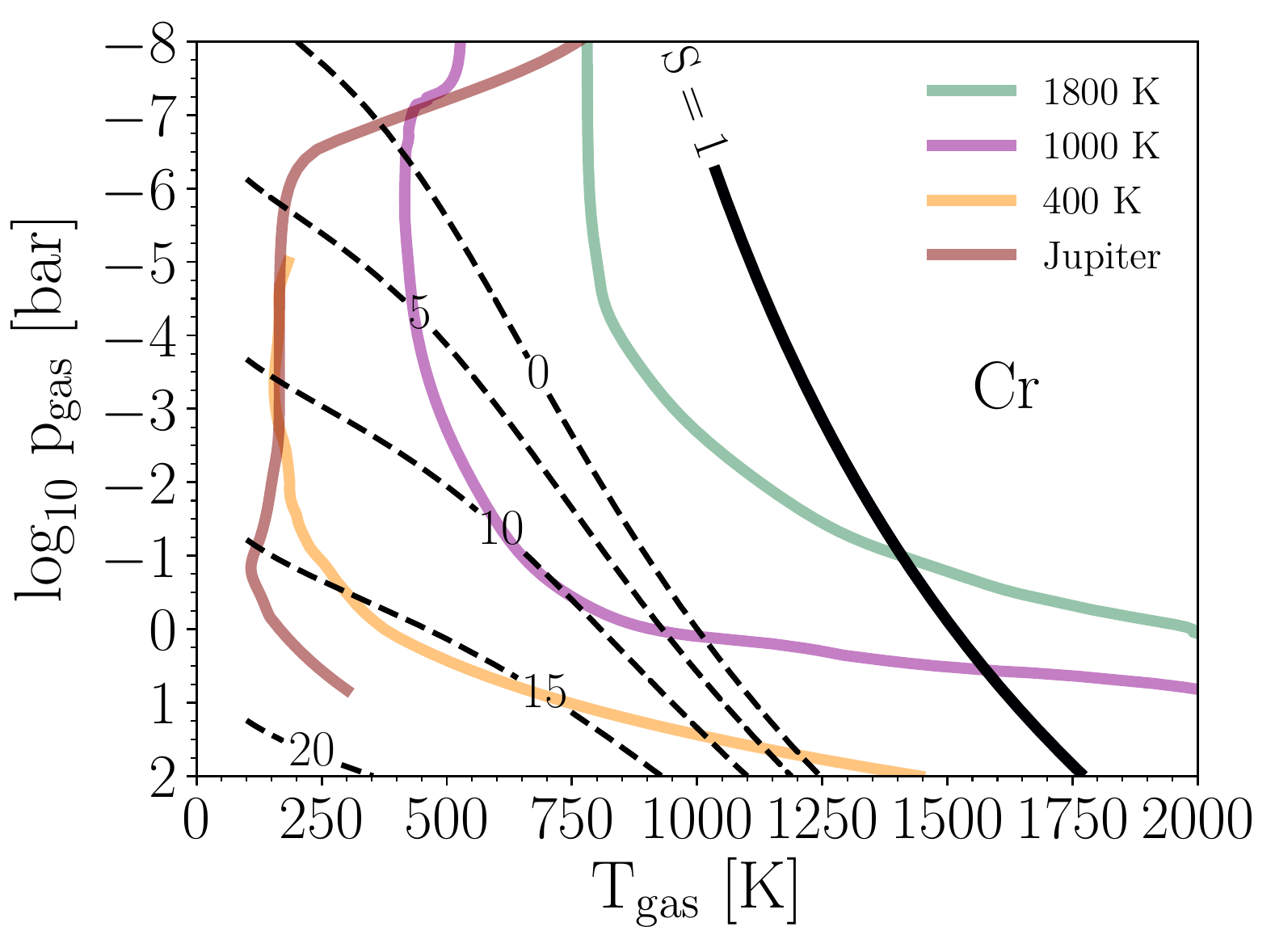}
   \caption{Nucleation rates of Cr[s] using MCNT (Eq. \ref{eq:J*final}.).
   \textbf{Left:} Contours of $\log_{10}$ J$_{*}$ [cm$^{-3}$ s$^{-1}$] of Cr[s] (colour bar and thin black lines) across the temperature-pressure grid, and the gas-solid phase equilibrium S(T$_{\rm gas}$,p$_{\rm gas}$) = 1 zone (thick black, solid).
   \textbf{Right:} S = 1 phase equilibrium result (black, dashed) and selected $\log_{10}$ J$_{*}$ contour lines (black, dotted) compared to log g = 5, T$_{\rm eff}$ = 1800 K (green, solid), 1000 K (purple, solid), \textsc{Drift-Pheonix} \citep{Witte2009} and T$_{\rm eff}$ = 400 K, log g = 5 (dark orange, solid), \textsc{Exo-REM} \citep{Baudino2015,Baudino2017} brown dwarf (T$_{\rm gas}$, p$_{\rm gas}$) profiles and Jupiter (brown, solid) \citep{Rimmer2016}.}
   \label{fig:Cr_res}
\end{figure*}

Our results presented in Figure \ref{fig:Cr_res} (left) show that Cr[s] requires significantly more supercooling before efficient nucleation compared to the other species considered in this study.
The difference in temperature between the S = 1 phase equilibrium limit and efficient nucleation (J $>$ 1 cm$^{-3}$ s$^{-1}$) is 500-700 K.
Figure \ref{fig:Cr_res} (right) shows the comparison to the brown dwarf (T$_{\rm gas}$,p$_{\rm gas}$) profiles, suggesting that Cr[s] is unlikely to nucleate efficiently in the T$_{\rm eff}$ = 1800 K profile, as the temperature is still too high for nucleation at pressures $<$ 10$^{-4}$ bar.
However, in present conditions Cr[s] is able to nucleate efficiently for the cooler T$_{\rm eff}$ = 1000 K (T$_{\rm gas}$,p$_{\rm gas}$) profile.

\subsection{KCl[s]}

\begin{figure*}[ht] 
   \centering
   \includegraphics[width=0.49\textwidth]{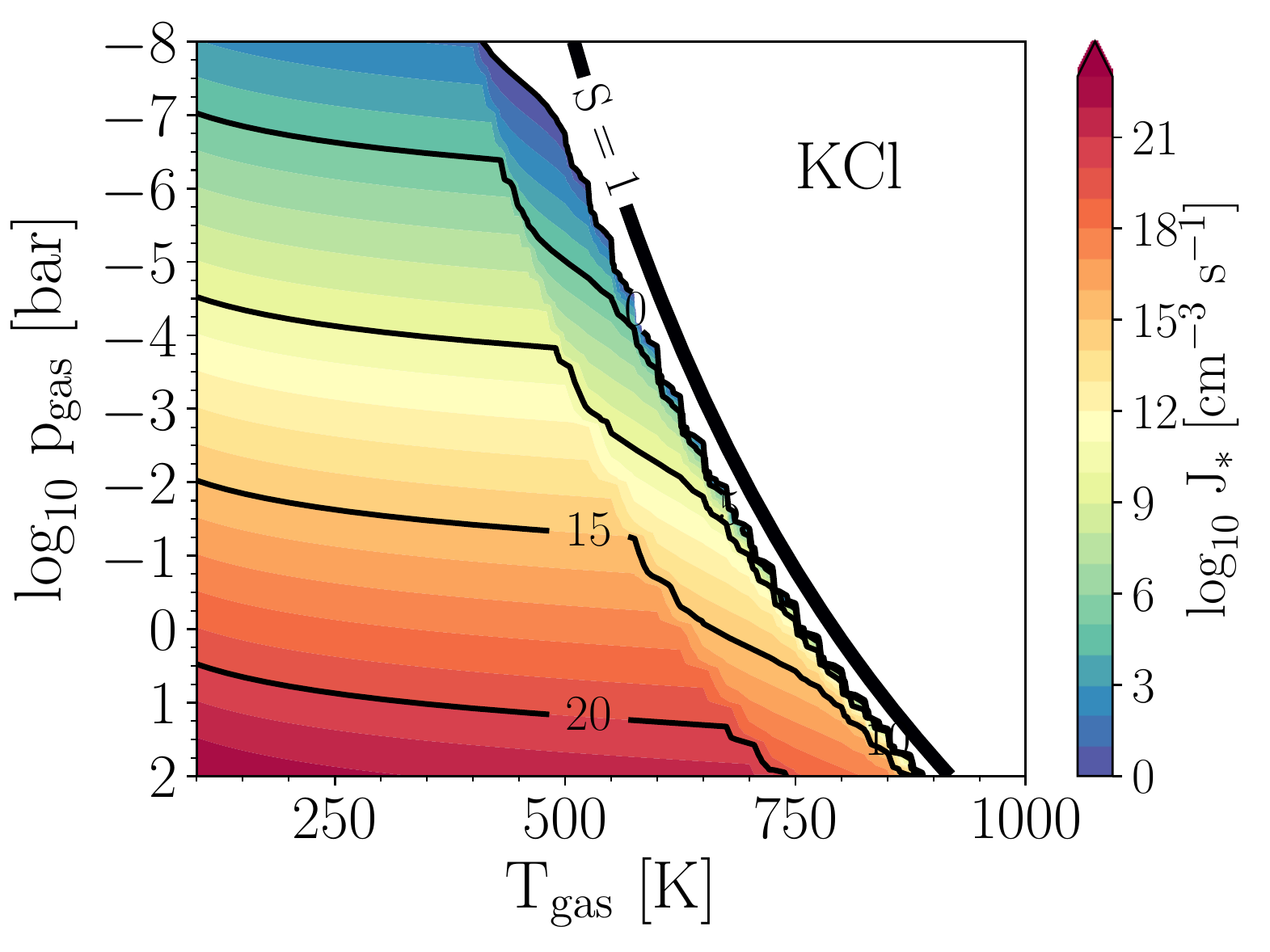}
   \includegraphics[width=0.49\textwidth]{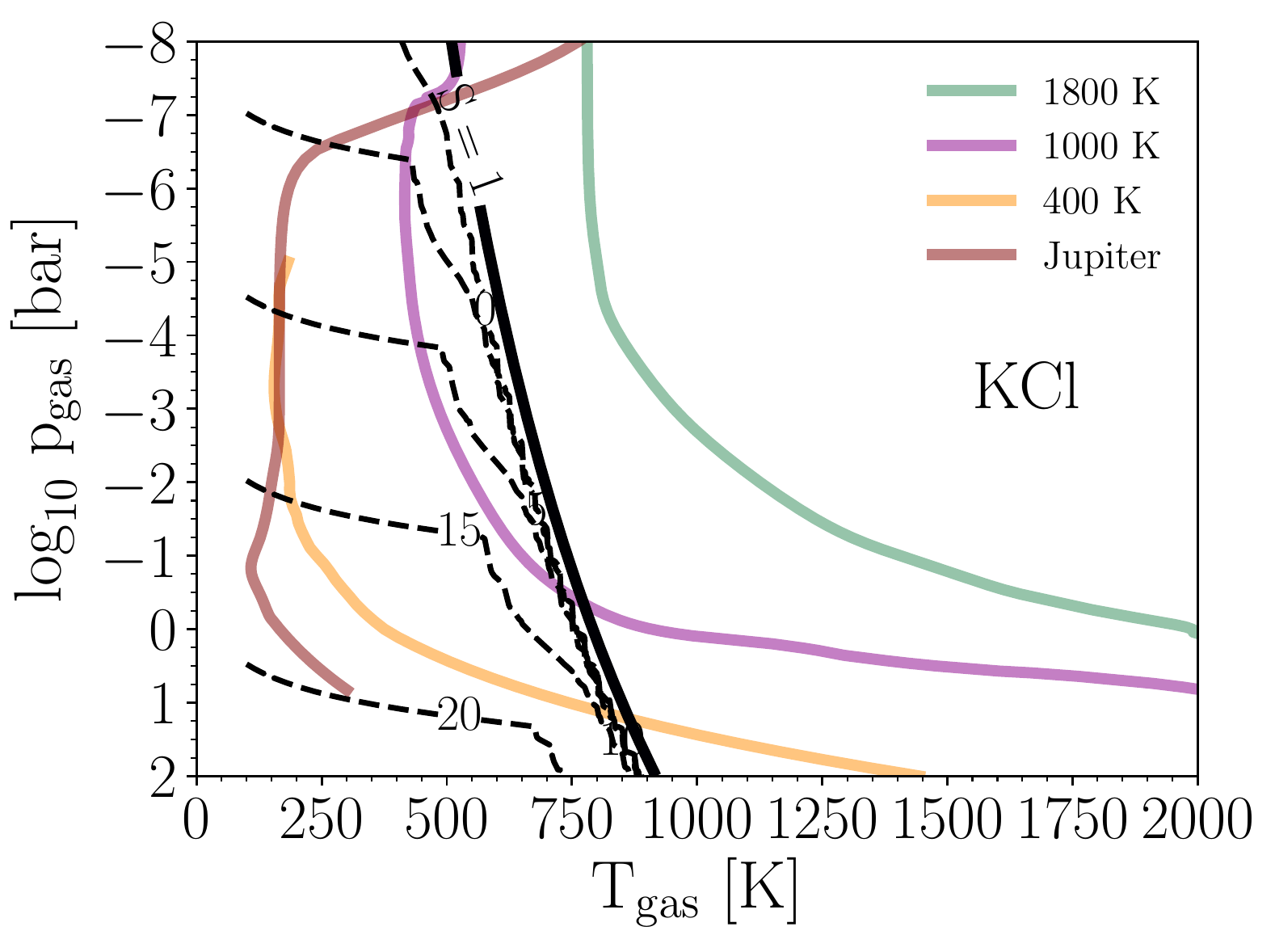}
   \caption{Nucleation rates of KCl[s] using MCNT (Eq. \ref{eq:J*final}.).
   \textbf{Left:} Contours of $\log_{10}$ J$_{*}$ [cm$^{-3}$ s$^{-1}$] of KCl[s] (colour bar and thin black lines) across the temperature-pressure grid, and the gas-solid phase equilibrium S(T$_{\rm gas}$,p$_{\rm gas}$) = 1 zone (thick black, solid).
   \textbf{Right:} S = 1 phase equilibrium result (black, solid) and selected $\log_{10}$ J$_{*}$ contour lines (black, dashed) compared to log g = 5, T$_{\rm eff}$ = 1800 K (green, solid), 1000 K (purple, solid), \textsc{Drift-Pheonix} \citep{Witte2009} and T$_{\rm eff}$ = 400 K, log g = 5 (dark orange, solid), \textsc{Exo-REM} \citep{Baudino2015,Baudino2017} brown dwarf (T$_{\rm gas}$, p$_{\rm gas}$) profiles and Jupiter (brown, solid) \citep{Rimmer2016}.}
   \label{fig:KCl_res}
\end{figure*}

Figure \ref{fig:KCl_res} (left) shows that efficient KCl[s] nucleation occurs very close to the S = 1 phase equilibrium region, suggesting that KCl[s] does not require too much supercooling (50-100 K) before KCl[s] seed particles form in the atmosphere.
Figure \ref{fig:KCl_res} (right) suggests that KCl[s] nucleation can potentially occur in the upper atmospheres of T brown dwarf atmospheres and in the deep regions of Jupiter-like planets at $\gtrsim$ 100 bar.

\subsection{NaCl[s]}

\begin{figure*}[ht] 
   \centering
   \includegraphics[width=0.49\textwidth]{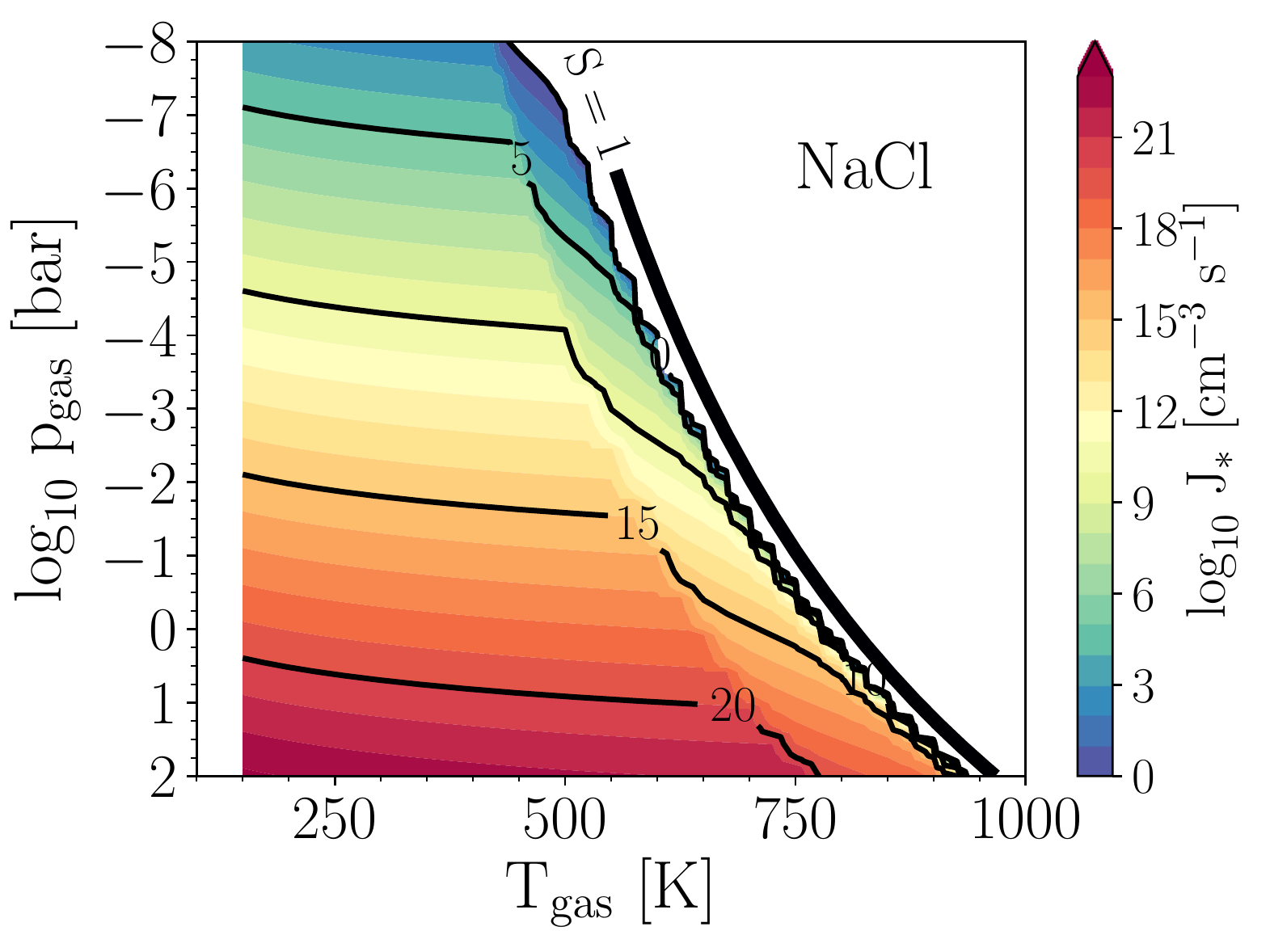}
   \includegraphics[width=0.49\textwidth]{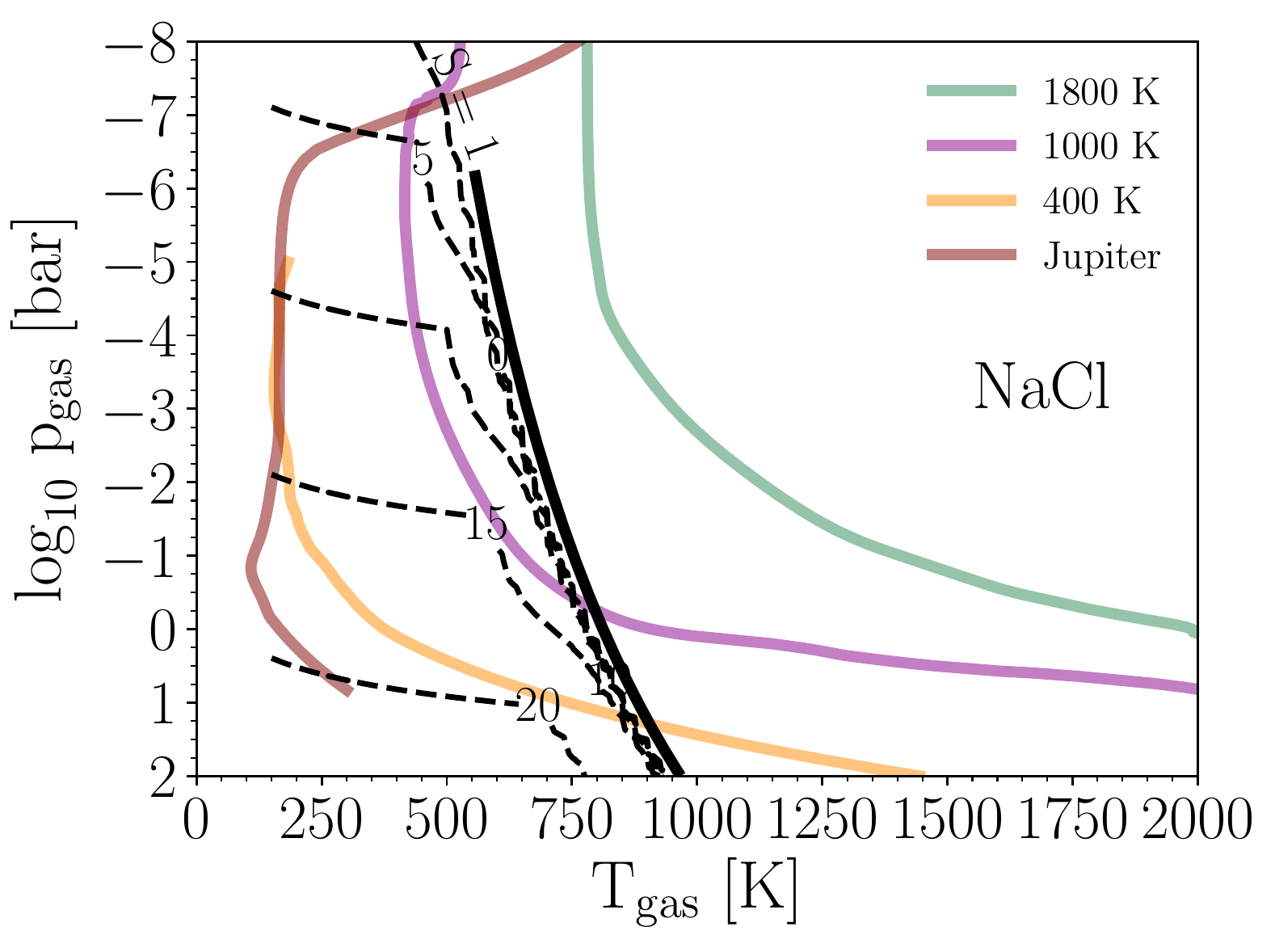}
   \caption{Nucleation rates of NaCl[s] using MCNT (Eq. \ref{eq:J*final}.).
   \textbf{Left:} Contours of $\log_{10}$ J$_{*}$ [cm$^{-3}$ s$^{-1}$] of NaCl[s] (colour bar and thin black lines) across the temperature-pressure grid, and the gas-solid phase equilibrium S(T$_{\rm gas}$,p$_{\rm gas}$) = 1 zone (thick black, solid).
   \textbf{Right:} S = 1 phase equilibrium result (black, solid) and selected $\log_{10}$ J$_{*}$ contour lines (black, dashed) compared to log g = 5, T$_{\rm eff}$ = 1800 K (green, solid), 1000 K (purple, solid), \textsc{Drift-Pheonix} \citep{Witte2009} and T$_{\rm eff}$ = 400 K, log g = 5 (dark orange, solid), \textsc{Exo-REM} \citep{Baudino2015,Baudino2017} brown dwarf (T$_{\rm gas}$, p$_{\rm gas}$) profiles and Jupiter (brown, solid) \citep{Rimmer2016}.}
   \label{fig:NaCl_res}
\end{figure*}

NaCl[s] shows very similar nucleation trends to KCl[s], with only a slight difference in the S = 1 phase equilibrium location.
Figure \ref{fig:NaCl_res} (left) shows that efficient NaCl[s] nucleation occurs very close to the S = 1 phase equilibrium limit, suggesting that NaCl[s] does not require too much super-cooling (50-100 K) before NaCl[s] seed particles form in the atmosphere.
Figure \ref{fig:NaCl_res} (right) suggests that NaCl[s] nucleation can potentially occur in T brown dwarf-like atmospheres, and possibly in the deep regions of Jupiter-like conditions at $\gtrsim$ 100 bar.

\subsection{H$_{2}$O[s/l]}

\begin{figure*}[ht] 
   \centering
   \includegraphics[width=0.49\textwidth]{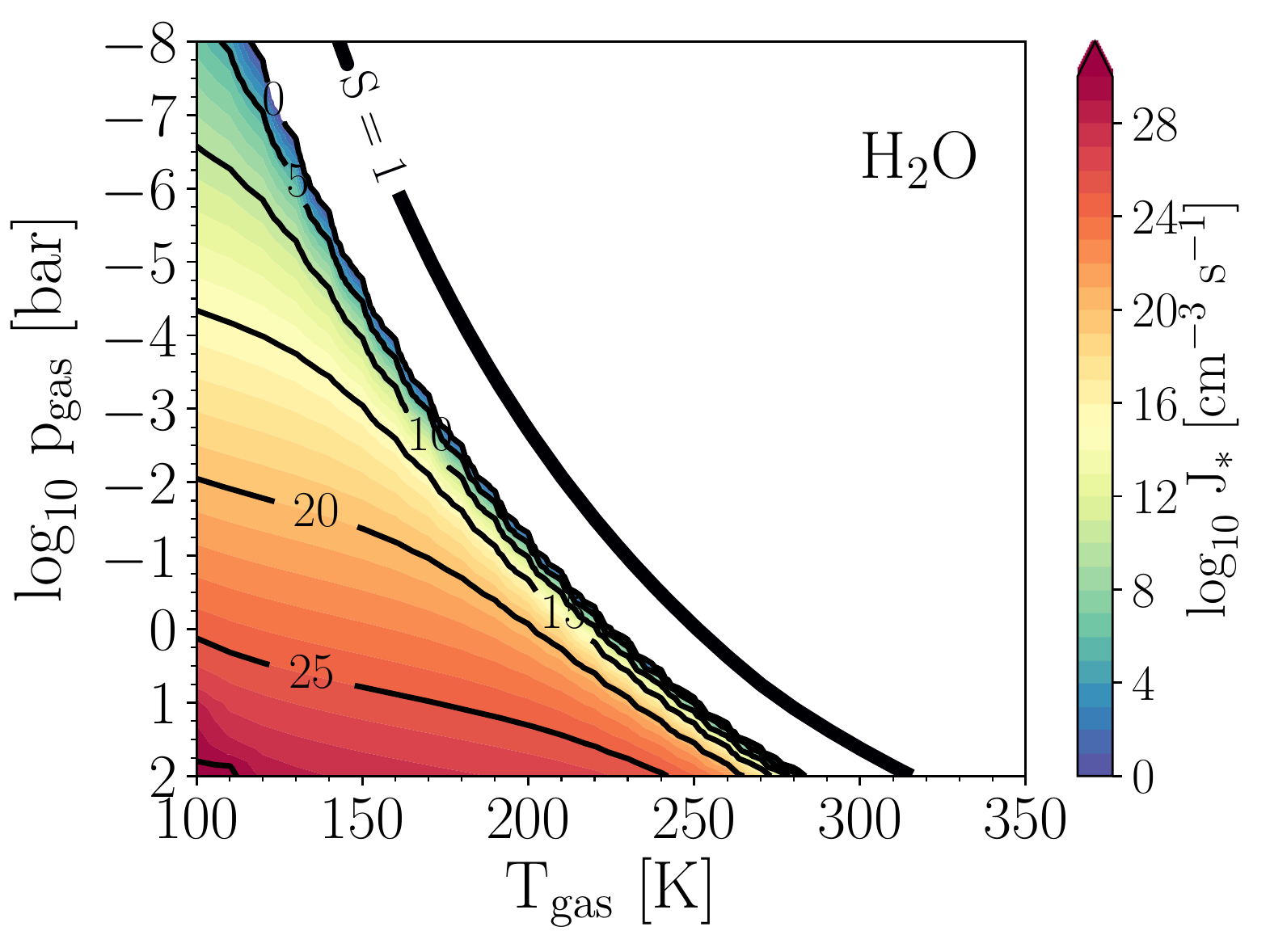}
   \includegraphics[width=0.49\textwidth]{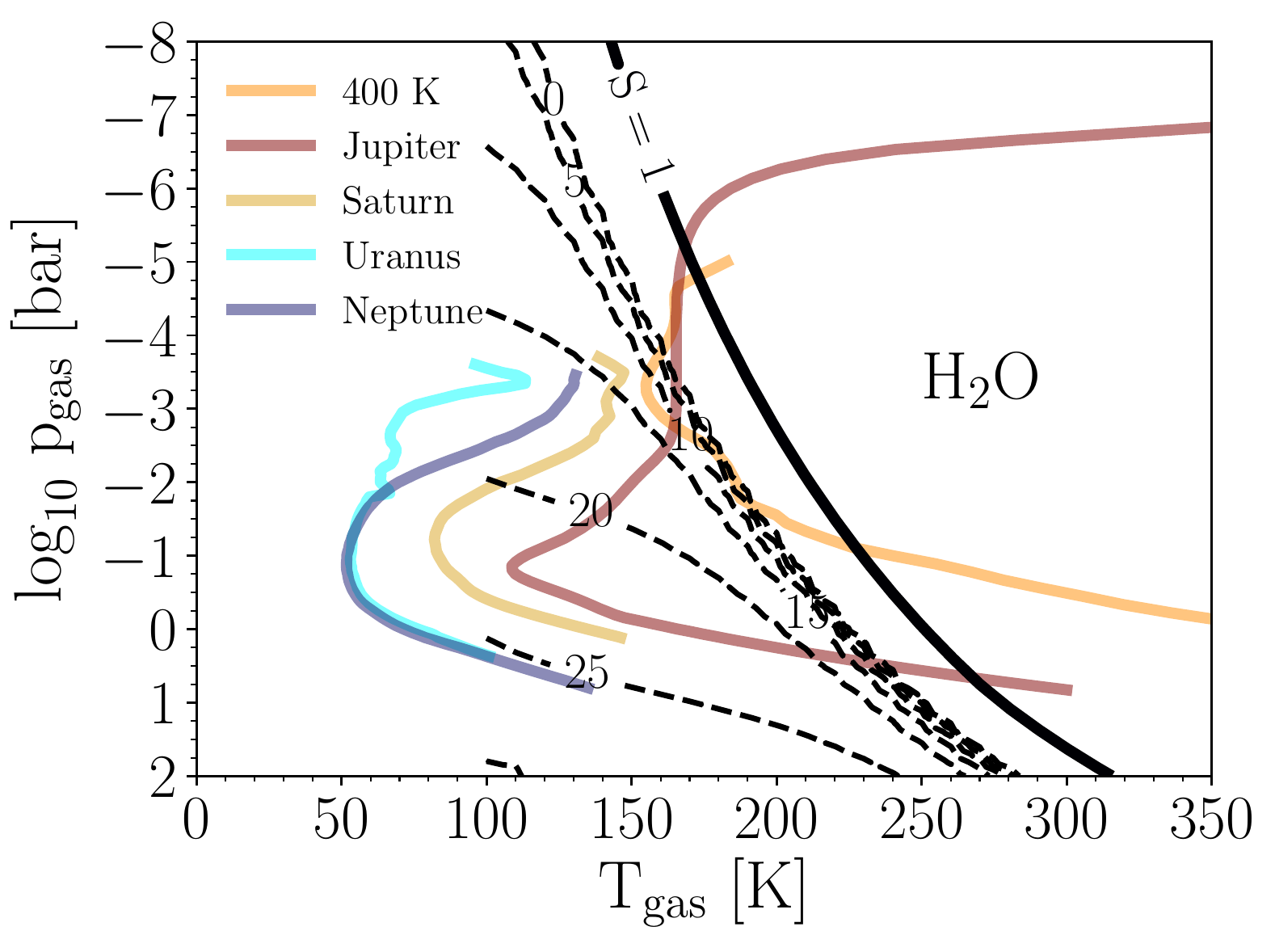}
   \caption{Nucleation rates of H$_{2}$O[s/l] using MCNT (Eq. \ref{eq:J*final}.).
   \textbf{Left:} Contours of $\log_{10}$ J$_{*}$ [cm$^{-3}$ s$^{-1}$] of H$_{2}$O[s/l] (colour bar and thin black lines) across the temperature-pressure grid, and the gas-solid phase equilibrium S(T$_{\rm gas}$,p$_{\rm gas}$) = 1 zone (thick black, solid).
   \textbf{Right:} S = 1 phase equilibrium result (black, solid) and selected $\log_{10}$ J$_{*}$ contour lines (black, dashed) compared the (T$_{\rm gas}$, p$_{\rm gas}$) profiles of a T$_{\rm eff}$ = 400 K, log g = 5 (dark orange, solid), \textsc{Exo-REM} \citep{Baudino2015,Baudino2017} brown dwarf, Jupiter (brown, solid) \citep{Rimmer2016}, Saturn (gold, solid),  Uranus (cyan, solid) and Neptune (dark blue, solid), \citep{Lindal1992}.}
   \label{fig:H2O_res}
\end{figure*}

Figure \ref{fig:H2O_res} (left) suggests that H2O[s/l] requires a supercooling of 10-30 K before nucleation starts becoming efficient.
Figure \ref{fig:H2O_res} (right) suggests that if H$_{2}$O[s/l] nucleation is present, it would occur deep in the atmospheres ($\sim$ 10-100 bar) of the Solar System-like gas/ice giants, assuming solar metallicity.

\subsection{NH$_{3}$[s]}

\begin{figure*}[ht] 
   \centering
   \includegraphics[width=0.49\textwidth]{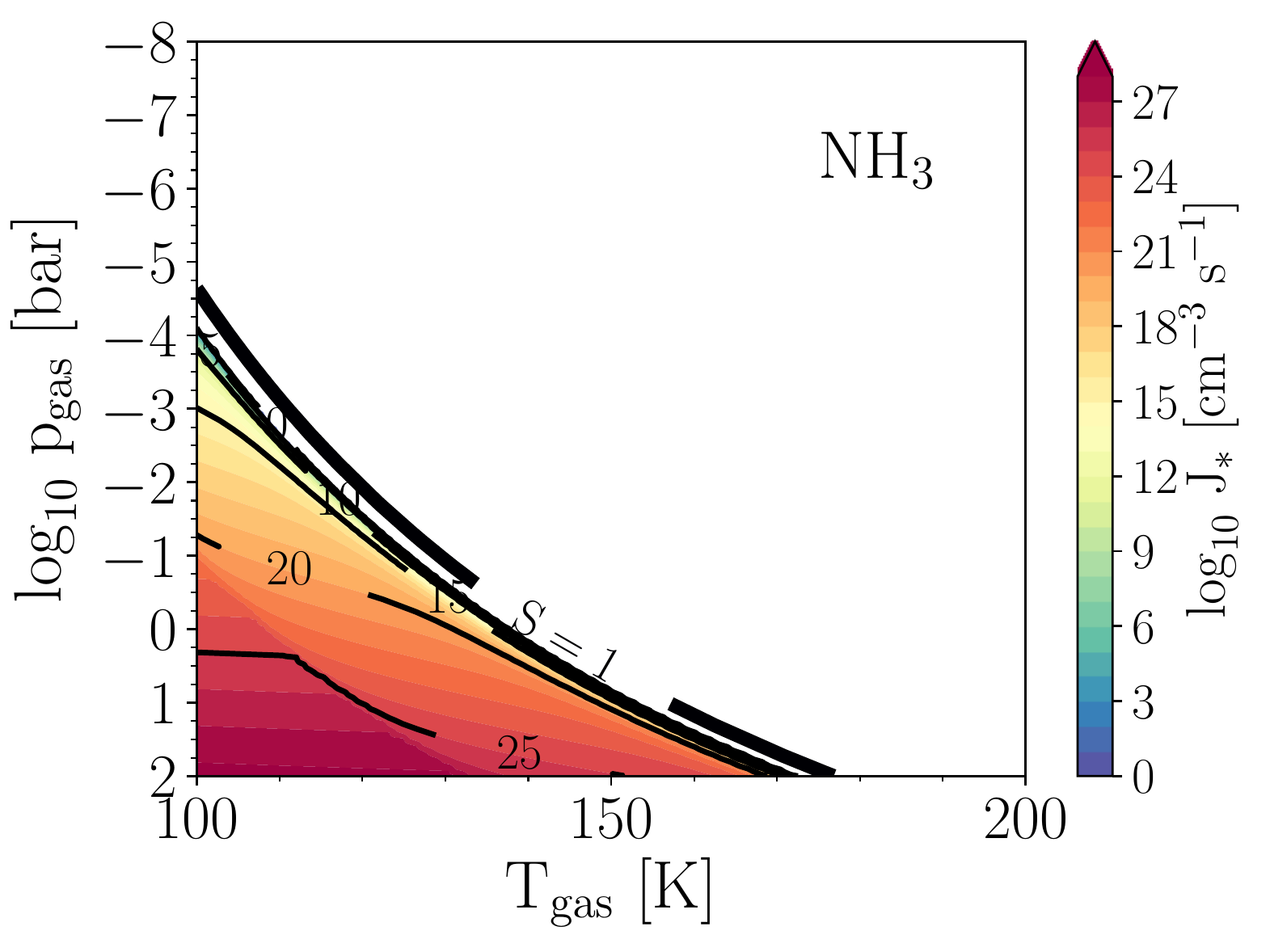}
   \includegraphics[width=0.49\textwidth]{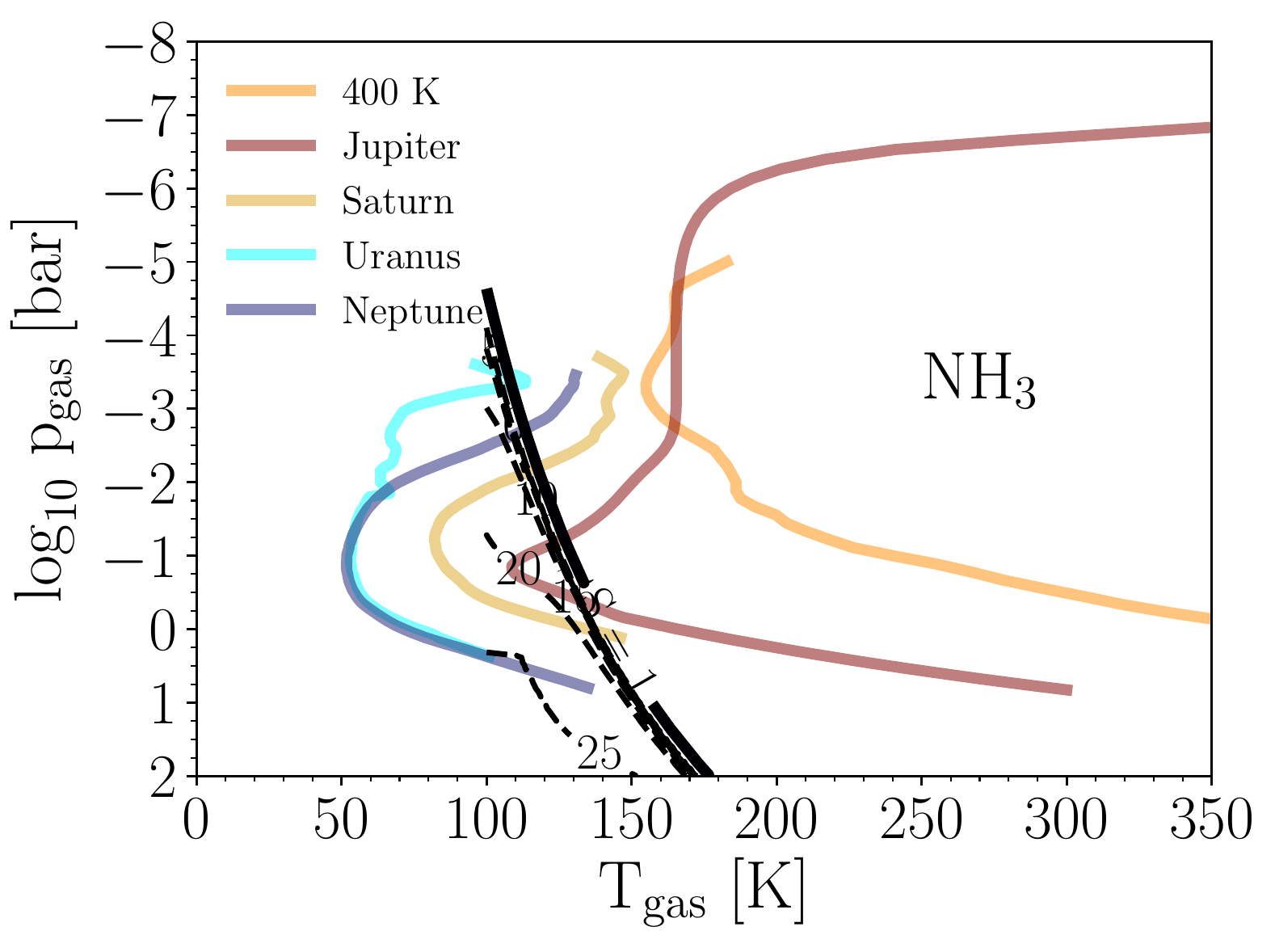}
   \caption{Nucleation rates of NH$_{3}$[s] using MCNT (Eq. \ref{eq:J*final}.).
   \textbf{Left:} Contours of $\log_{10}$ J$_{*}$ [cm$^{-3}$ s$^{-1}$] of NH$_{3}$[s] (colour bar and thin black lines) across the temperature-pressure grid, and the gas-solid phase equilibrium S(T$_{\rm gas}$,p$_{\rm gas}$) = 1 zone (thick black, solid).
   \textbf{Right:} S = 1 phase equilibrium result (black, solid) and selected $\log_{10}$ J$_{*}$ contour lines (black, dashed) compared the (T$_{\rm gas}$, p$_{\rm gas}$) profiles of a T$_{\rm eff}$ = 400 K, log g = 5 (dark orange, solid), \textsc{Exo-REM} \citep{Baudino2015,Baudino2017}, Jupiter (brown, solid) \citep{Rimmer2016}, Saturn (gold, solid),  Uranus (cyan, solid) and Neptune (dark blue, solid), \citep{Lindal1992}.}
   \label{fig:NH3_res}
\end{figure*}

Figure \ref{fig:NH3_res} (left) suggests that NH$_{3}$[s] requires a supercooling of 10-20 K before nucleation starts becoming efficient.
Figure \ref{fig:NH3_res} (right) suggests that if NH$_{3}$[s] nucleation is present, it would occur at pressures of $\sim$ 1-10 bar in the Solar System-like gas/ice giants, assuming solar metallicity.

\subsection{H$_{2}$S[s/l]}

\begin{figure*}[ht] 
   \centering
   \includegraphics[width=0.49\textwidth]{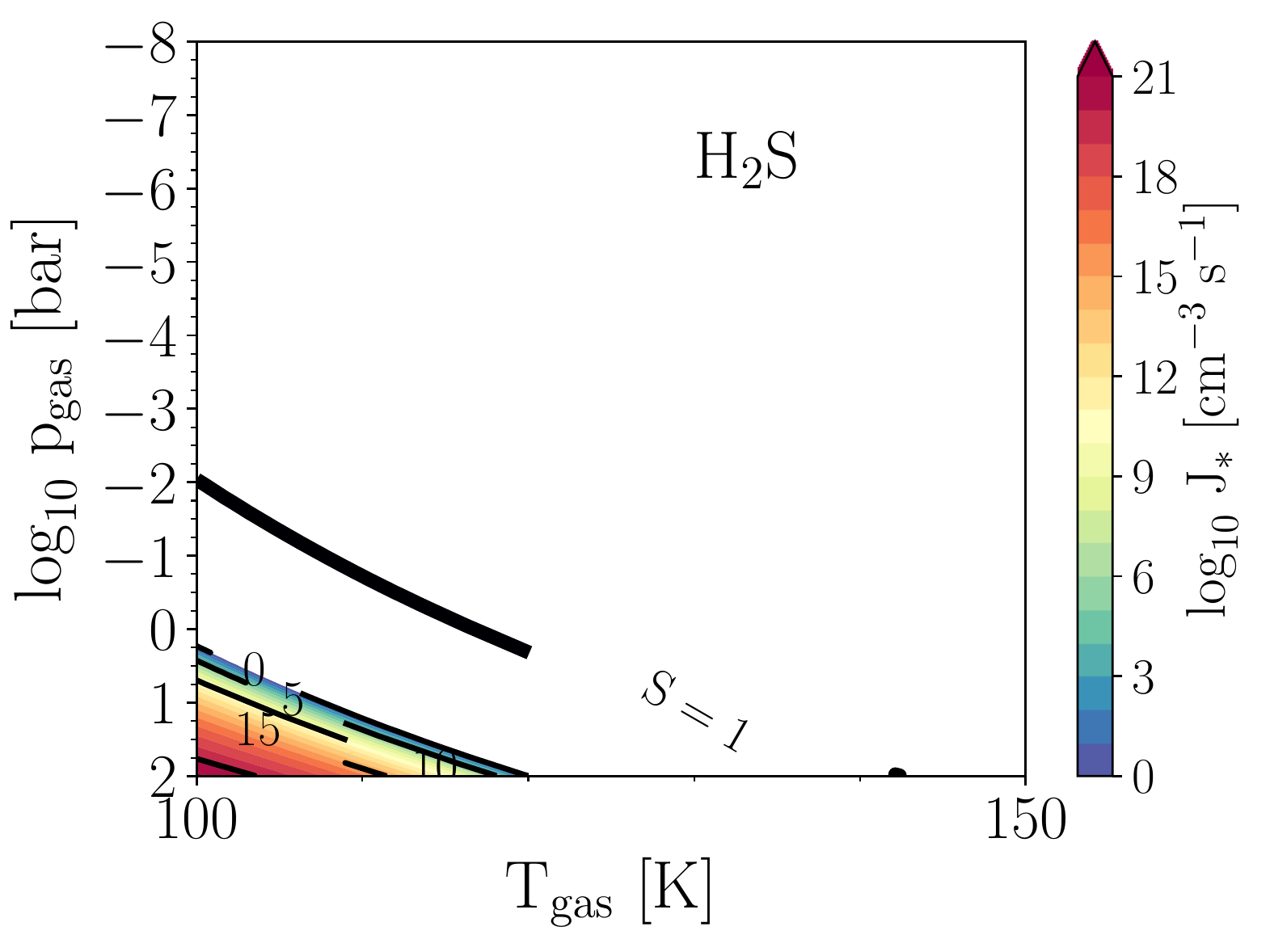}
   \includegraphics[width=0.49\textwidth]{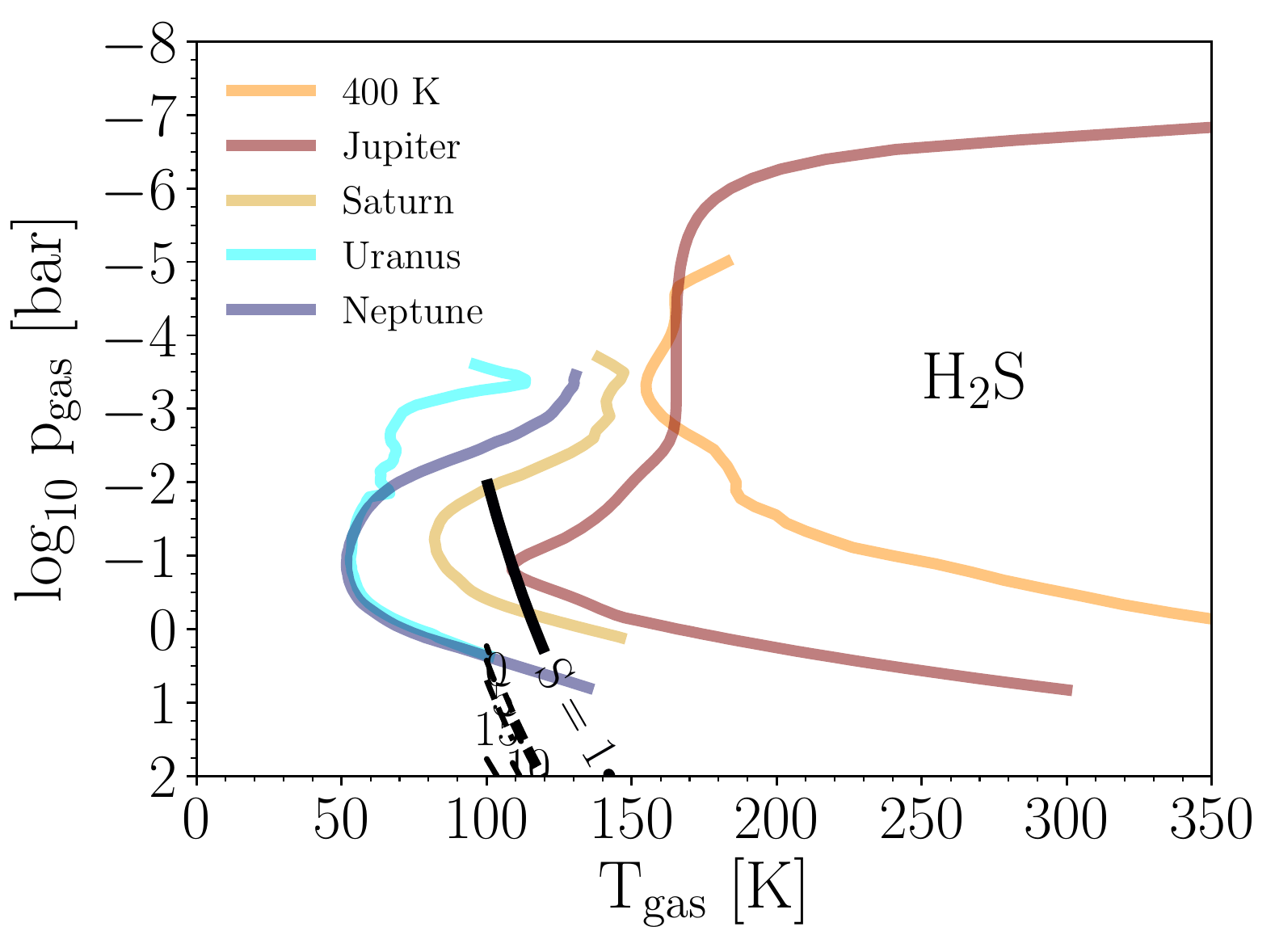}
   \caption{Nucleation rates of H$_{2}$S[s/l] using MCNT (Eq. \ref{eq:J*final}.).
   \textbf{Left:} Contours of $\log_{10}$ J$_{*}$ [cm$^{-3}$ s$^{-1}$] of H$_{2}$S[s/l] (colour bar and thin black lines) across the temperature-pressure grid, and the gas-solid phase equilibrium S(T$_{\rm gas}$,p$_{\rm gas}$) = 1 zone (thick black, solid).
   \textbf{Right:} S = 1 phase equilibrium result (black, solid) and selected $\log_{10}$ J$_{*}$ contour lines (black, dashed) compared the (T$_{\rm gas}$, p$_{\rm gas}$) profiles of a T$_{\rm eff}$ = 400 K, log g = 5 (dark orange, solid), \textsc{Exo-REM} \citep{Baudino2015,Baudino2017}, Jupiter (brown, solid) \citep{Rimmer2016}, Saturn (gold, solid),  Uranus (cyan, solid) and Neptune (dark blue, solid), \citep{Lindal1992}.}
   \label{fig:H2S_res}
\end{figure*}

Figure \ref{fig:H2S_res} (right) suggests that if H$_{2}$S[s/l] nucleation is present, it would occur at pressures of $\sim$ 1-10 bar in Uranus and Neptune-like planets, assuming solar metallicity.
The homogenous H$_{2}$S[s/l] seed particle nucleation is unlikely to occur in the atmospheres of Jupiter and Saturn-like objects, although H$_{2}$S[s/l] cloud formation may occur in Saturn like atmospheres, assuming solar metallicity.

\subsection{CH$_{4}$[s]}

We find that methane does not nucleate at the lowest temperature of 100 K considered in this study.
However, the S = 1 phase equilibrium limit of CH$_{4}$[s] occurs near 100 K.

\section{Multiple seed species and high-altitude clouds}
\label{sec:hierarchy}

Once seed particles have formed, other materials are also thermally stable, S $>$ 1, since the seed particle species requires supercooling before efficient nucleation occurs.
These materials can now form a considerable mantle (the bulk) through chemical surface reactions on top of the seed's surface.
The material composition of local cloud particles is therefore dependent on the availability of a seed particle species and condensation species that are supersaturated (S $>$ 1).

By comparing the zones of efficient nucleation (J$_{*}$ $>$ 1 cm$^{-3}$ s$^{-1}$) for each of our chosen species to the S $>$ 1 zone of other thermally stable materials, the dominating bulk material species can be identified at each (T$_{\rm gas}$, p$_{\rm gas}$) in the atmosphere.
This comparison allows the identification of possible seed forming materials able to trigger cloud formation across the large temperature and pressure ranges found in brown dwarf and exoplanet atmosphere.
To investigate the triggering of cloud formation by formation of seed particles at each (T$_{\rm gas}$, p$_{\rm gas}$), we compare our nucleation rate contour results to the S = 1 phase equilibrium limit of several cloud species presented in previous literature \citep{Visscher2006, Visscher2010, Morley2012, Wakeford2017a}.

\subsection{Titanium/Silicon oxides as possible seed particles}

\begin{figure}[ht] 
   \centering
   \includegraphics[width=0.49\textwidth]{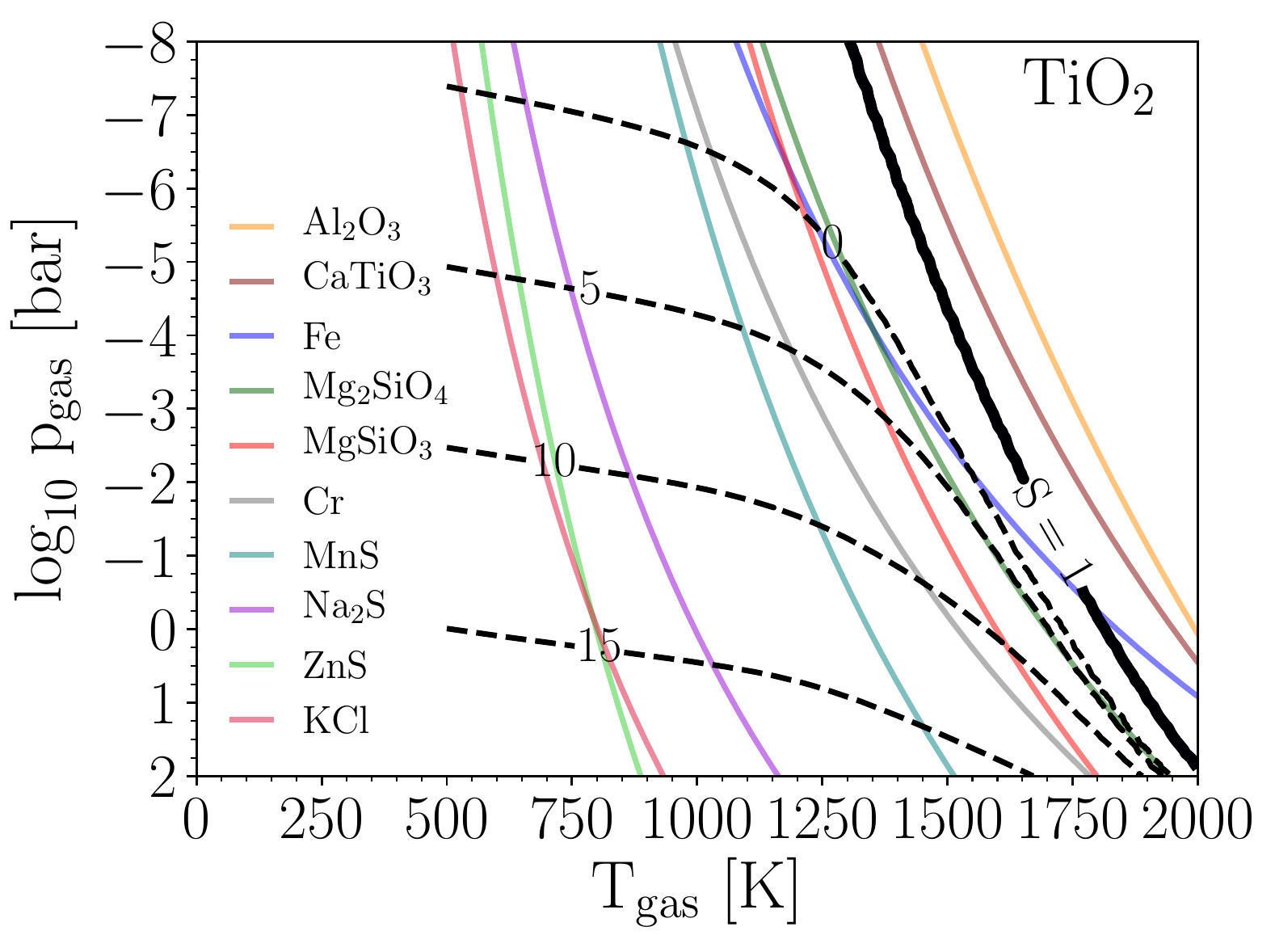}
   \includegraphics[width=0.49\textwidth]{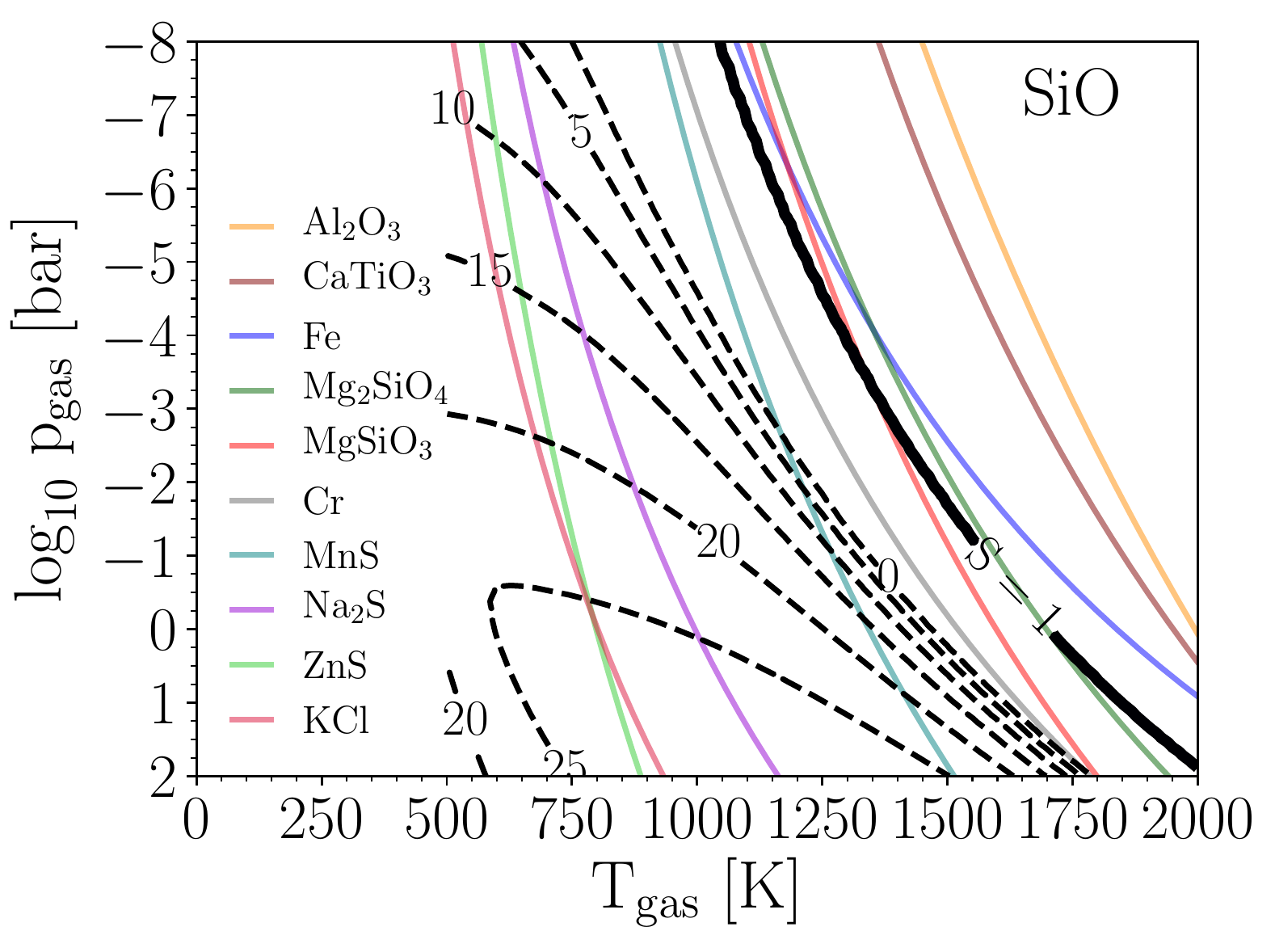}
   \caption{Nucleation rate $\log_{10}$ J$_{*}$ [cm$^{-3}$ s$^{-1}$] contour lines (black, dashed) of TiO$_{2}$[s] (upper) and SiO[s] (lower), and the gas-solid S = 1 phase equilibrium results (thick black, solid) .
   This is compared to the gas-solid S = 1 equilibrium limit equations (coloured, solid) found in \citet{Visscher2006, Visscher2010, Morley2012, Wakeford2017a} for a variety of cloud condensate species.}
   \label{fig:TiSi_seeds}
\end{figure}

Figure \ref{fig:TiSi_seeds} shows our TiO$_{2}$[s] and SiO[s] S = 1 phase equilibrium and nucleation rate J$_{*}$ contour lines against the S = 1 phase equilibrium analytical expressions of \citet{Visscher2006, Visscher2010, Morley2012, Wakeford2017a} for a variety of condensate species.
We find that TiO$_{2}$[s] nucleates efficiently in the same regions where Al$_{2}$O$_{3}$[s], CaTiO$_{3}$[s], Fe[s], MgSiO$_{3}$[s] and Mg$_{2}$SiO$_{4}$[s] species S = 1 phase equilibrium regions are located.
This suggests that TiO$_{2}$[s] is a candidate for providing the condensation surfaces for these clouds species to grow.
SiO[s] nucleation occurs at lower temperatures than TiO$_{2}$[s], with the SiO[s] S = 1 phase equilibrium region occurring at a similar temperature and pressure to the MgSiO$_{3}$[s] and Mg$_{2}$SiO$_{4}$[s] species.

This suggests, in the case of sub-stellar atmospheres, that TiO$_{2}$[s] is the preferred condensation surface provider for the refractory cloud formation species since SiO[s] will prefer to condense onto available TiO$_{2}$[s] seed particles lofted from deeper in the atmosphere.
\citet{Lee2015a} also conclude that TiO$_{2}$[s] is the primary seed particle for refractory cloud species, rather than SiO[s], due to the competition for gas phase Si atoms from condensing MgSiO$_{3}$[s] and Mg$_{2}$SiO$_{4}$[s] species, which quickly reduces the number density of gas phase SiO to where it can no longer nucleate efficiently..

\subsection{Chromium as a possible seed particle}

\begin{figure}[ht] 
   \centering
   \includegraphics[width=0.49\textwidth]{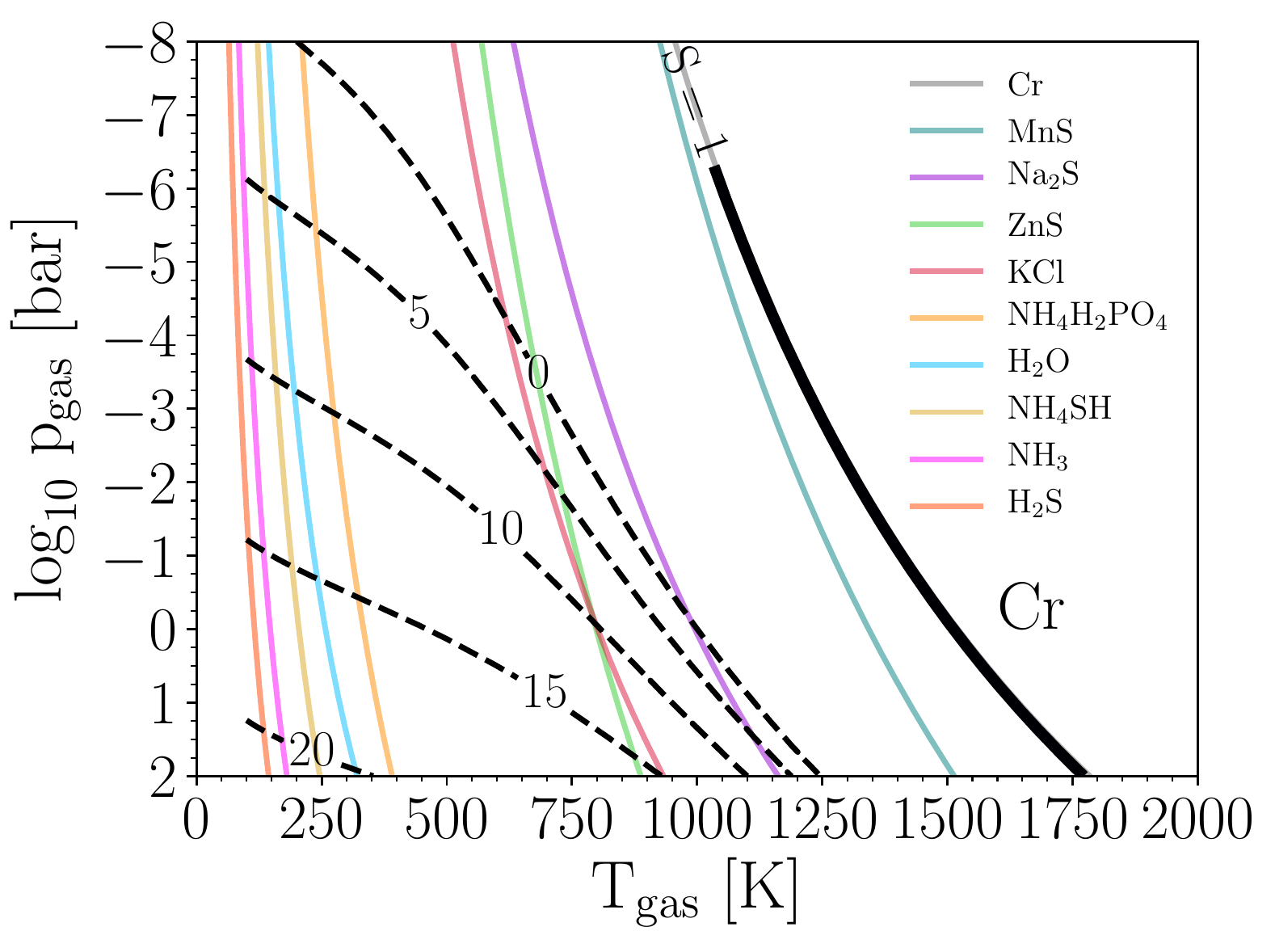}
   \caption{Cr[s] S = 1 phase equilibrium result (thick black, solid) and nucleation rate $\log_{10}$ J$_{*}$ contour lines (black, dashed),
   compared to the S = 1 equilibrium limit expressions (coloured, solid) found in \citet{Visscher2006, Visscher2010, Morley2012, Wakeford2017a} for a variety of cloud condensate species.}
   \label{fig:Cr_seeds}
\end{figure}

We examine the role of Cr[s] as a condensation surface provider to the sulfide cloud species MnS[s], Na$_{2}$S[s] and ZnS[s].
Figure \ref{fig:Cr_seeds} presents our Cr[s] nucleation rate contours compared to the S = 1 phase equilibrium calculations of the more volatile solid species found in \citet{Visscher2006, Visscher2010, Morley2012, Wakeford2017a}.
The regions of efficient Cr[s] seed particle nucleation overlaps with the regions of thermal stability for these species.
This suggests that Cr[s] seed particles are a candidate for providing condensation surfaces to the sulfide (and also possibly the KCl[s]) cloud layer \citep[e.g.][]{Morley2012}.

\subsection{Chlorides as possible seed particles}

\begin{figure}[ht] 
   \centering
   \includegraphics[width=0.49\textwidth]{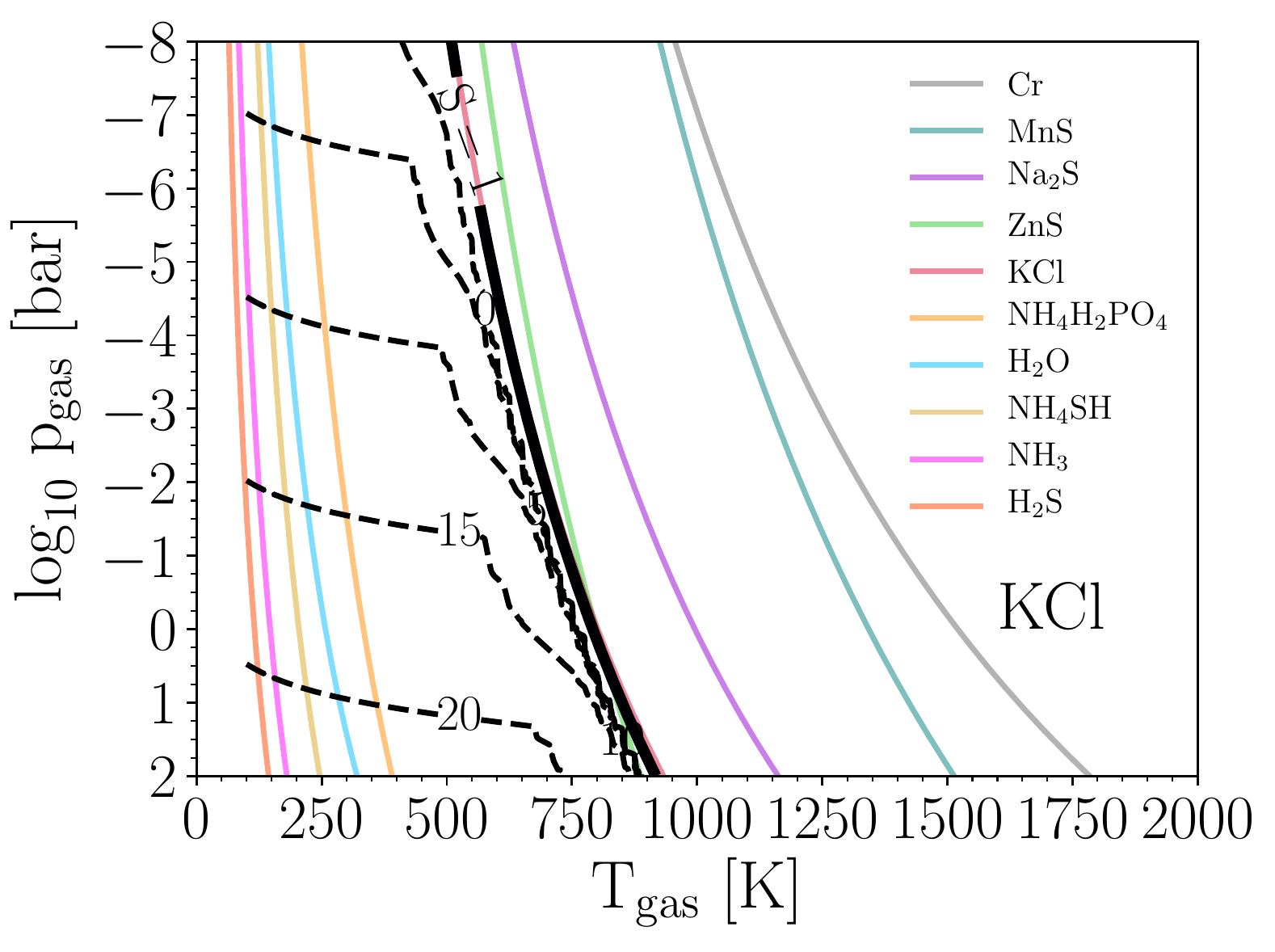}
   \includegraphics[width=0.49\textwidth]{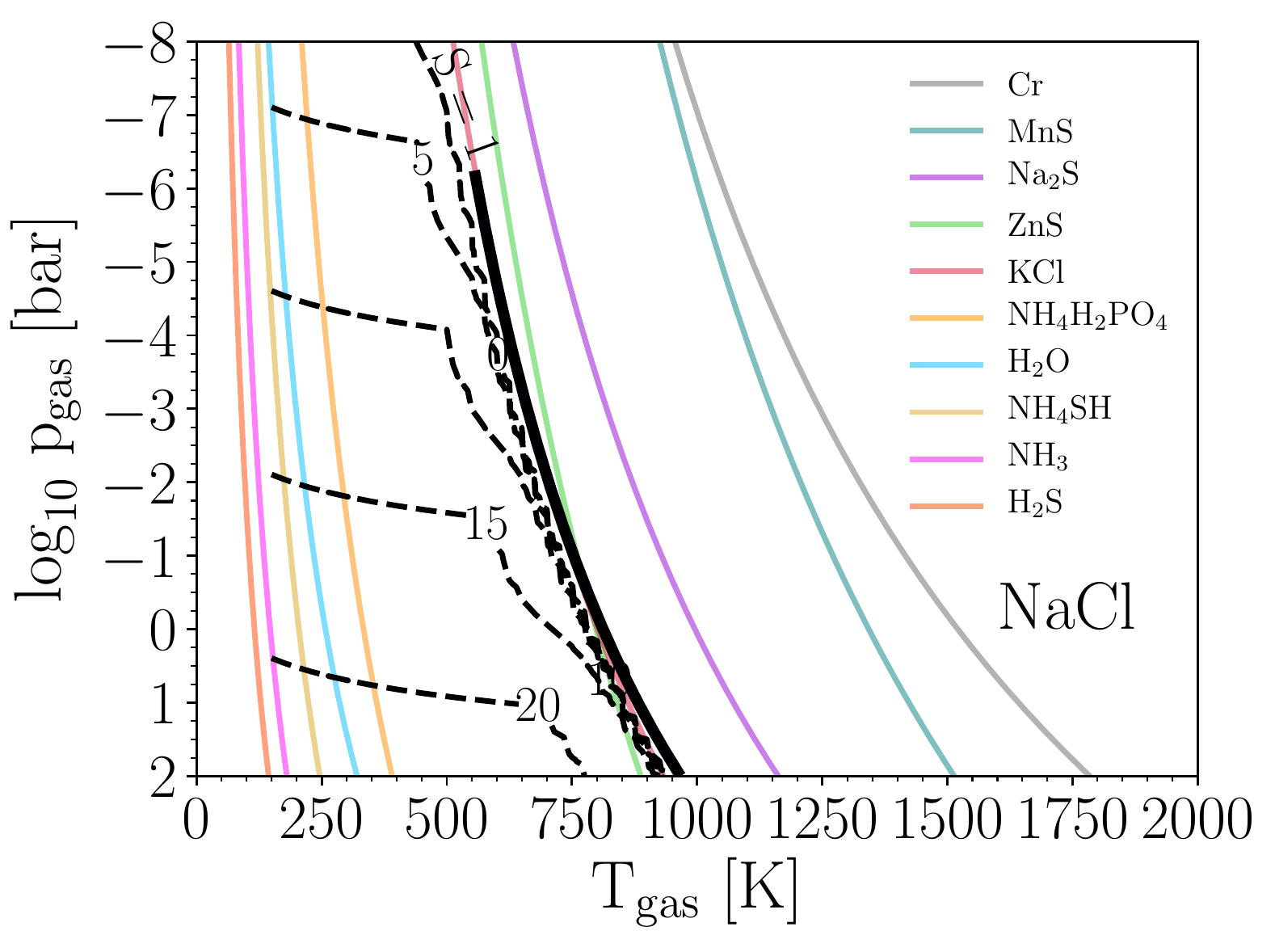}
   \caption{KCl[s] (upper) and NaCl[s] (lower) S = 1 phase equilibrium result (thick black, solid) and nucleation rate $\log_{10}$ J$_{*}$ contour lines (black, dashed),
   compared to the S = 1 equilibrium limit equations (coloured, solid) found in \citet{Visscher2006, Visscher2010, Morley2012, Wakeford2017a} for a variety of cloud condensate species.}
   \label{fig:KClNaCl_seeds}
\end{figure}

Figure \ref{fig:KClNaCl_seeds} shows the KCl[s] and NaCl[s] nucleation rate contour line compared to the S = 1 phase equilibrium analytical equations of the molecules found in \citet{Visscher2006, Visscher2010, Morley2012, Wakeford2017a}.
KCl[s] and NaCl[s] are possible candidate seed species providers for the NH$_{4}$H$_{2}$PO$_{4}$[s] and H$_{2}$O[s/l] cloud layers in certain sub-stellar atmospheres due to their overlap of efficient nucleation rate regions and the S = 1 phase equilibrium of NH$_{4}$H$_{2}$PO$_{4}$[s] and H$_{2}$O[s].
However, there is a large temperature difference ($\sim$ 500 K) between the onset of efficient nucleation and the S = 1 phase equilibrium for water.
This suggests that should KCl[s] or NaCl[s] be the seed particle surfaces for water clouds, they would have to be mixed upward in the atmosphere by several scale heights.

\subsubsection{Other potential chloride seed particles}

\begin{figure*}[ht] 
   \centering
   \includegraphics[width=0.49\textwidth]{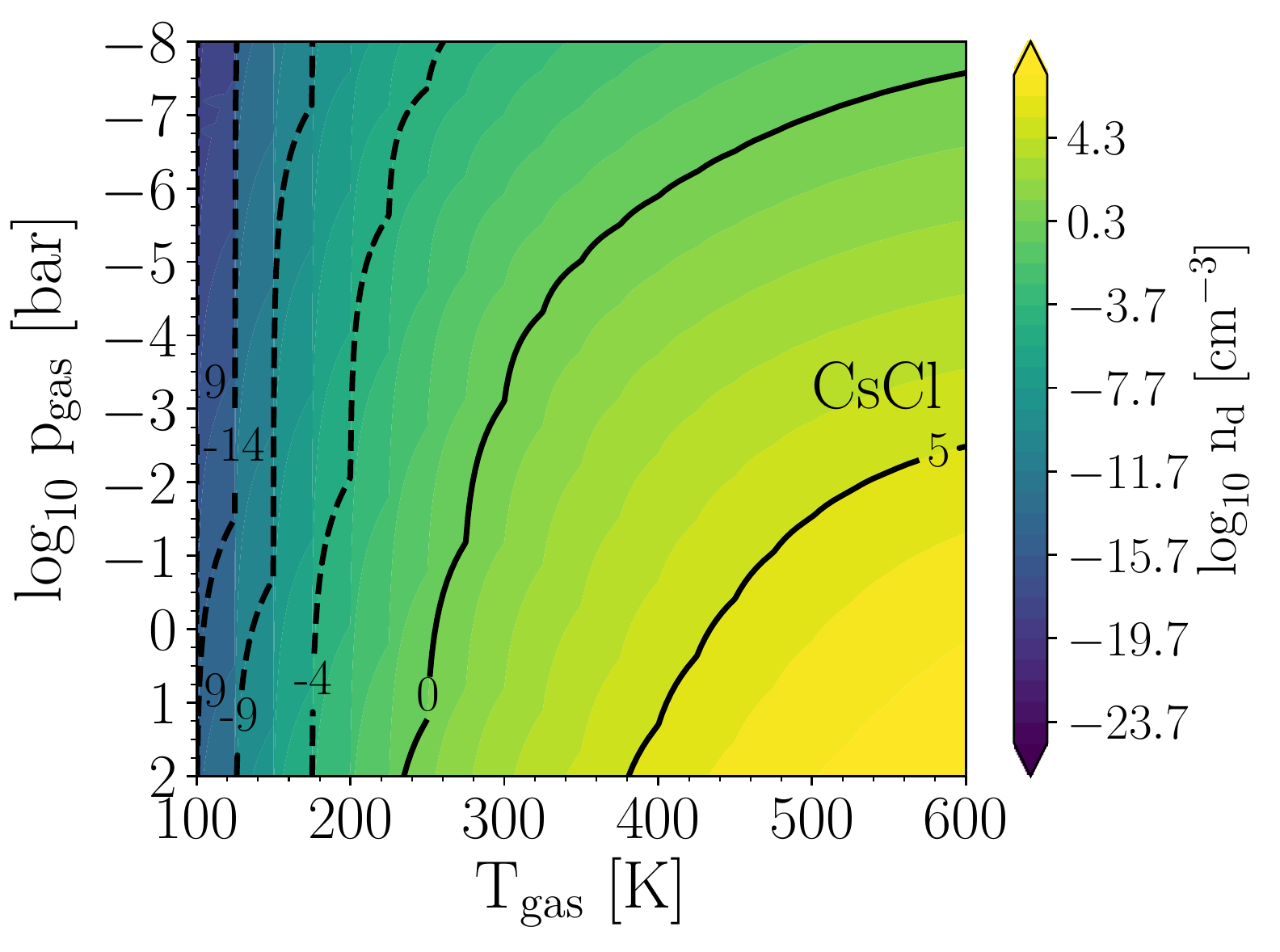}
   \includegraphics[width=0.49\textwidth]{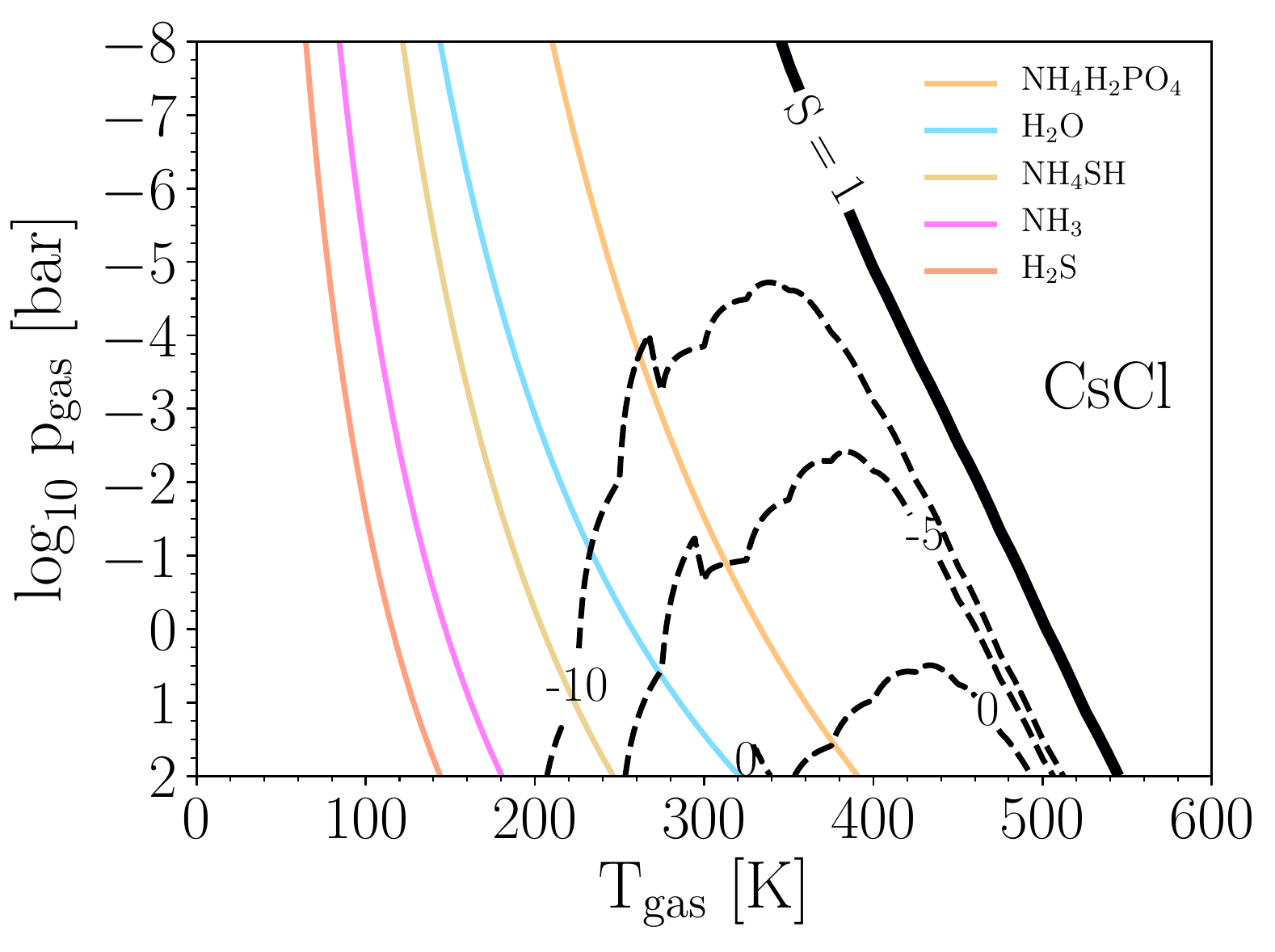}
   \caption{\textbf{Left:} Number density contours of CsCl gas phase monomers, $\log_{10}$ n(CsCl) [cm$^{-3}$].
   \textbf{Right:} CsCl[s] S = 1 phase equilibrium result (thick black, solid) and nucleation rate $\log_{10}$ J$_{*}$ contour lines (black, dashed),
   compared to the S = 1 equilibrium limit equations (coloured, solid) found in \citet{Visscher2006, Visscher2010, Morley2012, Wakeford2017a} for a variety of cloud condensate species.}
   \label{fig:Cs_seeds}
\end{figure*}

CsCl[s] and RbCl[s] cloud species have been suggested to be present in sub-stellar objects at lower temperatures and higher altitudes to KCl[s] and NaCl[s] \citep{Fegley1994,Lodders1999}, closer to the temperature range of the gas-solid S = 1 phase equilibrium limit of H$_{2}$O[s/l].
Should CsCl[s] and/or RbCl[s] nucleate in a similar fashion to KCl[s] and NaCl[s], it is possible that these species (or a similar Cl-bearing molecule) provides the condensation surfaces to the H$_{2}$O[s/l] cloud decks in the Solar System-like gas giant planets and cool (T$_{\rm eff}$ $<$ 450 K) Y brown dwarfs \citep{Burrows2003,Morley2014}.
The solar composition mole fraction of Cs and Rb (1.2 $\cdot$ 10$^{-11}$ and 3.3 $\cdot$ 10$^{-10}$ respectively) are magnitudes lower than Na and K (1.74 $\cdot$ 10$^{-6}$  and 1.07 $\cdot$ 10$^{-7}$ respectively) \citep{Asplund2009}.
This suggests that CsCl[s] and RbCl[s] are unfavourable seed particle candidates due to their very low atomic abundances.
However, the calculations in \citet{Fegley1994, Lodders1999} and \citet{Burrows1999} show that gas phase CsCl is the most dominant Cs-bearing molecule by several orders of magnitude at T$_{\rm gas}$ $<$ 1000 K in sub-stellar atmospheres.
This suggests that despite the low atomic abundance of Cs, the majority of the Cs-bearing gas molecules will be CsCl and potentially be in large enough in number density to nucleate seed particles in regions where S(CsCl[s]) $>$ 1.

To examine this, we calculate the vapour pressure of CsCl[s], $p^{\rm vap}_{\rm CsCl}$ [dyn cm$^{-2}$], between a temperature range of 100-800 K from the relation \citep[e.g.][]{Woitke1999}
\begin{equation}
p^{\rm vap}(T) = p^{\st} \exp\left(\frac{\dG(s) - \dG(1)}{RT}\right) ,
\end{equation}
where the Gibbs formation energies of gas phase CsCl ($\dG(1)$) and solid CsCl[s] ($\dG(s)$) were taken from the JANAF thermochemical tables \citep{Chase1986}, with p$^{\st}$ = 1 bar.
Using the \textsc{SciPy} \textsc{curvefit} optimal parameter fitting routine yields
\begin{equation}
\label{eq:CsCl_vap}
p^{\rm vap}_{\rm CsCl}(T) = \exp\left(29.967 - \frac{23055}{T}\right) .
\end{equation}
Gas phase CsCl, CsO and CsOH molecules were added to the $<$ 500 K TEA species list (Section \ref{sec:TEA}) and an additional temperature grid from 100-600 K in steps of 25 K was re-run, as this covers the region where S(CsCl[s]) $\approx$ 1.
A surface tension of $\sigma$ = 100 [erg cm$^{-2}$] was assumed for CsCl[s], assuming similarity to the KCl[s] and NaCl[s] values.
Figure \ref{fig:Cs_seeds} (left) presents the contour plot of the number density of CsCl gas phase monomers.
We find that, as in \citet{Fegley1994, Lodders1999} and \citet{Burrows1999}, that CsCl is a dominant Cs-bearing molecule, however, CsOH molecules become more abundant as the temperature lowers towards 200 K.
This results in a decreasing trend in the number of CsCl molecules at temperatures $<$ 500 K.

Figure \ref{fig:Cs_seeds} (right) shows our results for CsCl[s] nucleation rates compared to the S = 1 phase equilibrium threshold of species from Figure \ref{fig:KClNaCl_seeds}.
We find that, although CsCl[s] shows similar supercooling and nucleation contour behaviours to KCl[s] and NaCl[s], the magnitude of the nucleation rate is $\approx$ 15 orders of magnitudes lower compared to KCl[s] and NaCl[s] for similar $\Delta T_{\rm cool}$.
This is a direct result of the reduced number of CsCl gas phase mononomers as well as dominant Cs-bearing species moving from CsCl to CsOH.
The temperature difference required between the onset of efficient nucleation and the S = 1 phase equilibrium for H$_{2}$O[s/l] is now $\sim$ 250 K for CsCl[s], compared to KCl[s] and NaCl[s] which is $\sim$ 500 K.
However, the regions of efficient nucleation occurs only for gas pressures p$_{\rm gas}$ $>$ 1 bar.

We find that CsCl[s] is able to theoretically produce seed particles due to three effects
\begin{enumerate}
\item The vapour pressure curve of CsCl[s] (Eq. \ref{eq:CsCl_vap}) is very similar to KCl[s] and NaCl[s], therefore the vapour pressure is exponentially decreasing at temperatures T$_{\rm gas}$ = 100-700 K, with a value p$^{\rm vap}_{\rm CsCl}$ $\approx$ 10$^{-6}$ dyn cm$^{-2}$ at 600 K.
\item This exponentially decreasing vapour pressure allows an S $>$ 1 zone to occur at temperatures T$_{\rm gas}$ $<$ 600 K and subsequent nucleation, despite the low number density of the CsCl gas phase monomers in the gas phase (Fig. \ref{fig:Cs_seeds}, left).
\item Since gas phase CsCl is present with CsOH, combined with the naturally low number densities of CsCl, the efficient nucleation regions of CsCl[s] are confined to atmospheric pressures of p$_{\rm gas}$ $>$ 1 bar.
\end{enumerate}

\section{Discussion}
\label{sec:discussion}

In this section, we discuss several possible implications of our seed particle results.

\subsection{Potential seed particle hierarchy}

\begin{figure*}[ht] 
   \centering
   \includegraphics[width=0.65\textwidth]{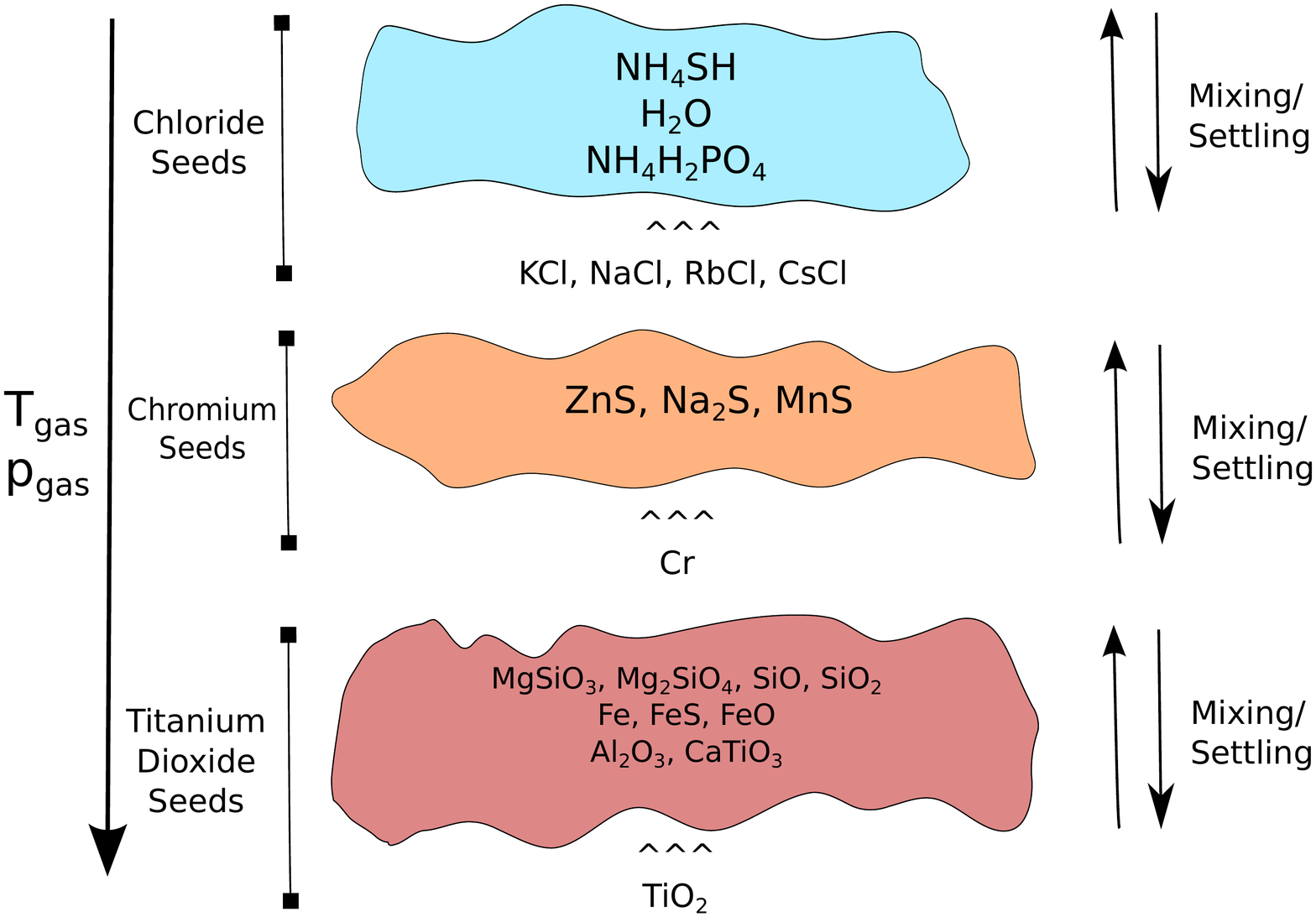}
   \caption{Schematic diagram of a potential cloud formation and seed particle hierarchy scenario.
   We depict a three cloud layer scenario where TiO$_{2}$[s] provides seed particles for the deepest cloud layers, Cr[s] for the `middle' cloud layer and a chloride species (e.g. KCl[s], NaCl[s], RbCl[s], CsCl[s]) for the colder H$_{2}$O[s/l] cloud layers.}
   \label{fig:hierarchy}
\end{figure*}

The results of our nucleation rates suggest the presence of a possible (T$_{\rm gas}$, p$_{\rm gas}$) hierarchy trio of cloud seed particles leading to the formation of different cloud layers.
In summary, TiO$_{2}$[s] and/or SiO[s] nucleation supply seed particle surfaces to the Al, Ti, Fe, Mg and Si-bearing refractory clouds, which then can condense on these surfaces and grow the cloud particles.
At lower temperatures, Cr[s] may provide the condensation surfaces to the S-bearing cloud species (MnS[s], Na$_{2}$S[s], ZnS[s]) and possibly KCl[s].
In the coolest layers, the nucleation of Cl-bearing seed particles are candidates for providing surfaces for the NH$_{4}$H$_{2}$PO$_{4}$[s], NH$_{4}$SH[s] and H$_{2}$O[s/l] cloud layers.
Figure \ref{fig:hierarchy} shows a schematic representation of our results.

We suggest that it is possible a Cl-bearing species may be the seed particle provider to H$_{2}$O[s/l] cloud layers in sub-stellar atmospheres.
However, despite our theoretical calculations showing that efficient nucleation of Cl-bearing species can occur in the H$_{2}$O[s/l] cloud zones, more detailed modelling of each chloride species individually will be required to answer which chloride (if any) can act as the seed particle to the H$_{2}$O[s/l] cloud layers.
A more volatile Cl-bearing candidate for seed particle formation is NH$_{4}$Cl[s], which has been included as a cloud species in the \citet{Fegley1994} models, and was shown to saturate closer to the water S = 1 line than CsCl[s] and RbCl[s].
\citet{Showman2001} also investigated the possible nucleation of NH$_{4}$Cl[s] in Jupiter and Saturn.
However, the nucleation of NH$_{4}$Cl[s] and NH$_{4}$SH[s] may occur in a heteromolecular reaction \citep{Banic1980}, beyond the scope of the current study.

Should the condensation of NH$_{4}$H$_{2}$PO$_{4}$[s] also be possible due to favourable atmospheric mixing timescales of phosphorus bearing gas molecules \citep[see discussion in][]{Visscher2006}, chloride seed particles may act as the surface provider for its condensation.
The formation of complex condensates like NH$_{4}$H$_{2}$PO$_{4}$[s] requires three body collisions \citep{Visscher2006} which occur with low probabilities, despite the phosphorus atomic solar molar ratio ($\epsilon$(P) = 2.57 $\cdot$ 10$^{-7}$) being of comparable abundances to potassium \citep{Asplund2009}.
It is likely that should NH$_{4}$H$_{2}$PO$_{4}$[s] form in an atmosphere, a seed surface is required to facilitate these three body reactions.

\subsection{Consistent modelling with nucleation}

The nucleation rate calculations presented here are not fully consistent with respect to the element conservation compared to the full kinetic cloud formation model \citep[e.g.][]{Woitke2004,Helling2016,Lee2016,Helling2017}.
We only consider the seed formation process as described in Section \ref{sec:nuctheory}, but we do not take into account the feedback mechanisms of element consumption by seed formation or the mixing and settling of gas/solid phase material in the atmosphere.
We further do not consider the effect of surface growth and evaporation on the seed formation.
Therefore, our results are only valid at the very onset of cloud formation with an initially pure gas phase composition.
The element depletion through seed formation is small compared to the element depletion by condensation which will eventually cause the termination of the nucleation process as demonstrated in \citet[][Figure 8]{Lee2015b}.
Neglecting element depletion will also influence the location of the S = 1 gas-solid phase equilibrium threshold.

Our nucleation results depend on the metallicity (assumed here to be solar), because it determines the gas-phase composition and number density of gas phase monomers available to the nucleation process.
The supersaturation ratio, S, is dependent on the partial pressure of a species, which is determined by the temperature and number density of that species in the local gas phase.
This in turn affects the efficiency of the nucleation process at a specific temperature and pressure.
\citet{Witte2009} study the dependence of cloud formation in brown dwarf atmospheres on the metallicity by applying their \textsc{Drift-Phoenix} model atmosphere code.
They found that the cloud structures do not scale linearly with [M/H], but with a more complex feedback on the sub-stellar atmospheric structures due to the changes in the cloud structure and seed particle formation rates from metal poor to metal rich environments.

\citet{Mahapatra2017} summarised in their Figures. 2-5 show the atomic and molecular number densities change if the element abundances change from solar to Earth-crust abundances and also hydrogen poor environments.
For example, they show that the number density of gas phase SiO monomers is increased by several orders of magnitude for their bulk silicate earth (BSE) results compared to solar metallicity.
This will have consequences on the seed particle nucleation regions when considering SiO nucleation in BSE metallicity atmospheric environments.

The validity of our proposed seed particle scheme will be verified in future, more complete modelling efforts.

\subsection{Potential observational consequences}

Our results suggest that Chromium seed particle formation may be an important factor for the cloud formation in L and T brown dwarfs.
Should the (T$_{\rm gas}$, p$_{\rm gas}$) profile of the L/T brown dwarf straddle the efficient Cr[s] nucleation zone, Cr[s] seed particles may or may not be able to form and provide surfaces for the sulfide species at higher altitudes, therefore changing the cloud structure composition present in the upper atmosphere.
If the temperature and velocity profiles of brown dwarfs are inhomogeneous in 3D, suggested from the hydrodynamical studies by \citet{Freytag2010,Showman2013a,Zhang2014}, the observational differences between global hotter/less cloudy and colder/more cloudy regions (patchiness) may also depend on the availability of seed particles.

Our results show that detached, high-altitude aerosol layers may form in brown dwarf and exoplanet atmospheres, dependent on the nucleation rate of materials in these low pressure, low temperatures regions of the atmosphere.
This may result in potential forward scattering of starlight in the atmosphere, impacting the observed brightness of the planet dependent on the phase, similar to the greater twilight albedo compared to the dayside for the \textit{Cassini} observations of Titan \citep{Munoz2017}.
The scattering of photons by these aerosol layers may have an impact on the observed transmission spectra, secondary eclipse spectra and phase curves of transiting brown dwarfs and exoplanets.
\citet{Vahidinia2014} show that small particle, high altitude haze/cloud layers above a main cloud base can result in a steepening of the Rayleigh slope feature observed in hot Jupiter transmission spectroscopy.

\citet{Parmentier2016} investigated the potential cloud compositions by post-processing 3D GCM simulations of hot Jupiter atmospheres across the \textit{Kepler} equilibrium temperature range (T$_{\rm eq}$ = 1000-2200 K).
They found that by including cold trapping of cloud species that formed deeper in the atmospheres of their simulations, the geometric albedos of the \textit{Kepler} hot Jupiter sample could be consistently reproduced.
Due to the long mixing timescales ($\tau_{\rm mix}$ = 10$^{6}$-10$^{9}$ s) in the deep atmosphere of hot Jupiters \citep[e.g.][]{Parmentier2013,Agundez2014,Mayne2014,Lee2015b}, both the already formed seed particles and their gas phase constituents cannot reach to higher altitudes in large enough number density.
Should a key seed particle species be cold trapped (i.e. unable to mix upward or form in the higher atmosphere), it may lead to inefficient cloud formation at higher altitudes due to a lack of available condensation surfaces.
These dynamical processes that affect the seed particle formation and global locations may lead to an insight into the physical and chemical differences between cloudy and non-cloudy signatures in hot Jupiter transmission spectra \citep[e.g.][]{Sing2016, Bruno2017}.
Future 3D cloud modelling efforts will attempt to capture these processes by including a large and concurrently modelled set of seed particle species and growth species.

This process also has consequences for the observed geometric albedos of hot Jupiter exoplanets.
\citet{Parmentier2016} (Figure 12) predict that should MnS[s] clouds form on the dayside of hot Jupiters in the T$_{\rm eq}$ = 1100-1500 K range, then the geometric albedo of the hot Jupiter will be A$_{\rm g}$ $>$ 0.1 in the \textit{Kepler} bandpass.
However, should Cr[s] seed particles be unable to form, or reach the saturation layer of MnS, then MnS[s] cannot condense and we expect the dayside of the T$_{\rm eq}$ = 1100-1500 K hot Jupiter population to remain cloud free and with low A$_{\rm g}$ $\sim$ 0.1 geometric albedo.
The upcoming CHEOPS \citep{Broeg2013} and TESS \citep{Ricker2014} missions will be able to discover and measure the geometric albedos of several hot Jupiters in this T$_{\rm eq}$ range, providing an observational test of these cloud property predictions.

\subsection{Mn/Zn/Na sulfides and heteromolecular nucleation}

We do not consider the homogenous nucleation of MnS[s], ZnS[s] and Na$_{2}$S[s] since our CE schemes lack the available input data for gas phase MnS, ZnS and Na$_{2}$S number density calculations so far.
Gas phase Manganese Sulfide (MnS) number densities have been calculated in \citet{Sharp1990} and \citet{Burrows1999} which includes data derived from \citet{Tsuji1973} for gas phase MnS thermochemical data.
\citet{Visscher2006} also include MnS as a gas and solid phase species.
MnS[s] is listed in \citet{Burrows1999} and \citet{Visscher2006} as a key species for T $<$ 1300 K.
However, gas phase MnS is not discussed and it is listed as a minor species in \citet{Burrows1999} and \citet{Visscher2006}, suggesting that it is not a very abundant species in the gas phase compared to other sulfides.
ZnS[s] and Na$_{2}$S[s] are included as a solid phase species in \citet{Visscher2006}.
However, the gas phase ZnS and Na$_{2}$S number densities are not presented in \citet{Visscher2006}.

The most abundant S-bearing gas phase molecules are H$_{2}$S and HS \citep{Visscher2006}.
The net cloud formation surface chemical reaction for these species involve H$_{2}$S \citep{Visscher2006,Morley2012}
\begin{equation}
  Mn + H_{2}S \rightarrow MnS[s] + H_{2},
\end{equation}
where Mn and H$_{2}$S are generally found with greater number densities in CE than gas phase MnS \citep{Visscher2006}.
This suggests that the sulfide materials may heteromolecularly nucleate seed particles, rather than homogeneously nucleate, which is beyond the scope of this study.

\subsection{The need for small cluster thermochemical data}
\label{sec:smallcluster}

\begin{figure*}[ht] 
   \centering
   \includegraphics[width=0.49\textwidth]{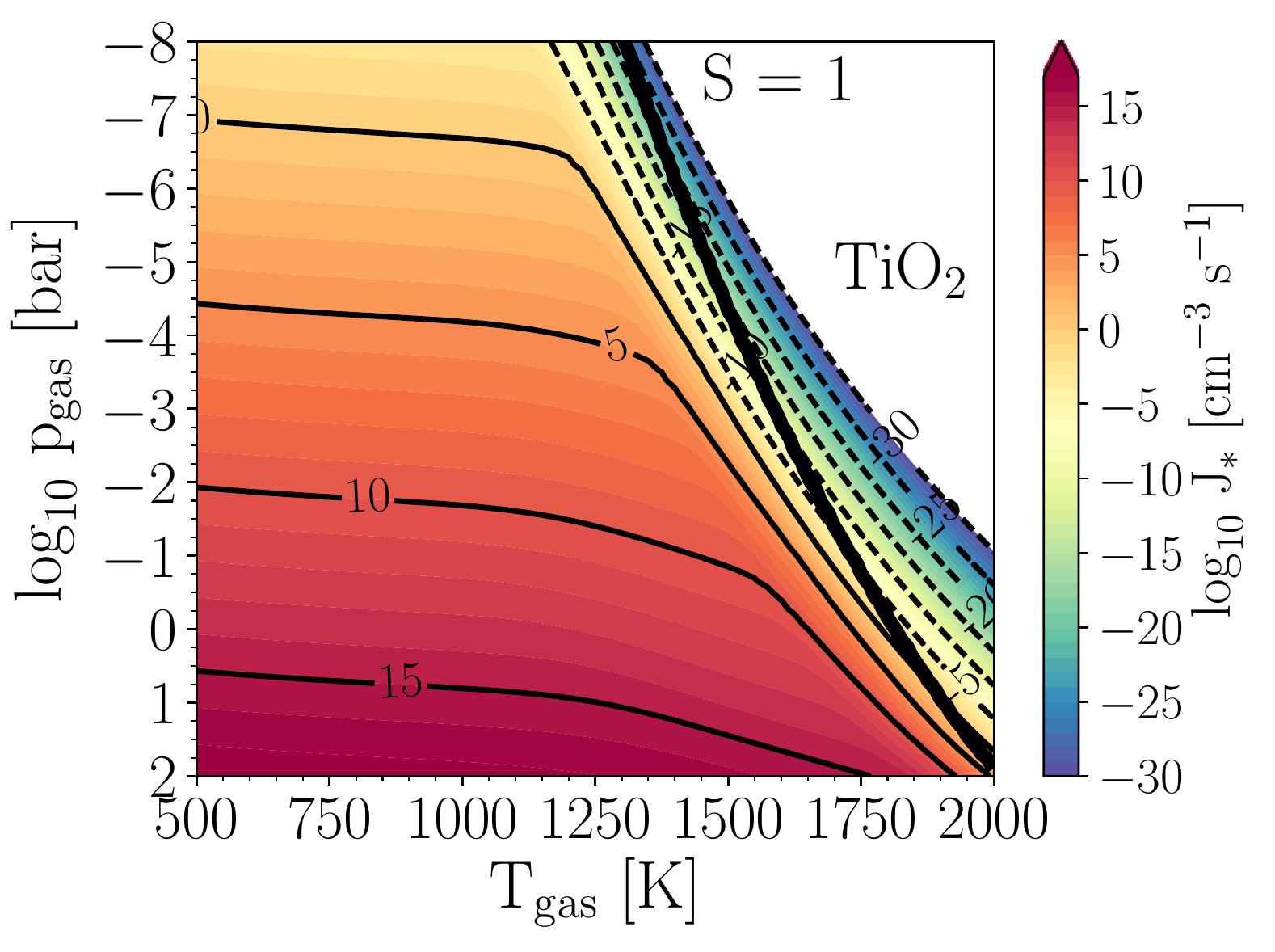}
   \includegraphics[width=0.49\textwidth]{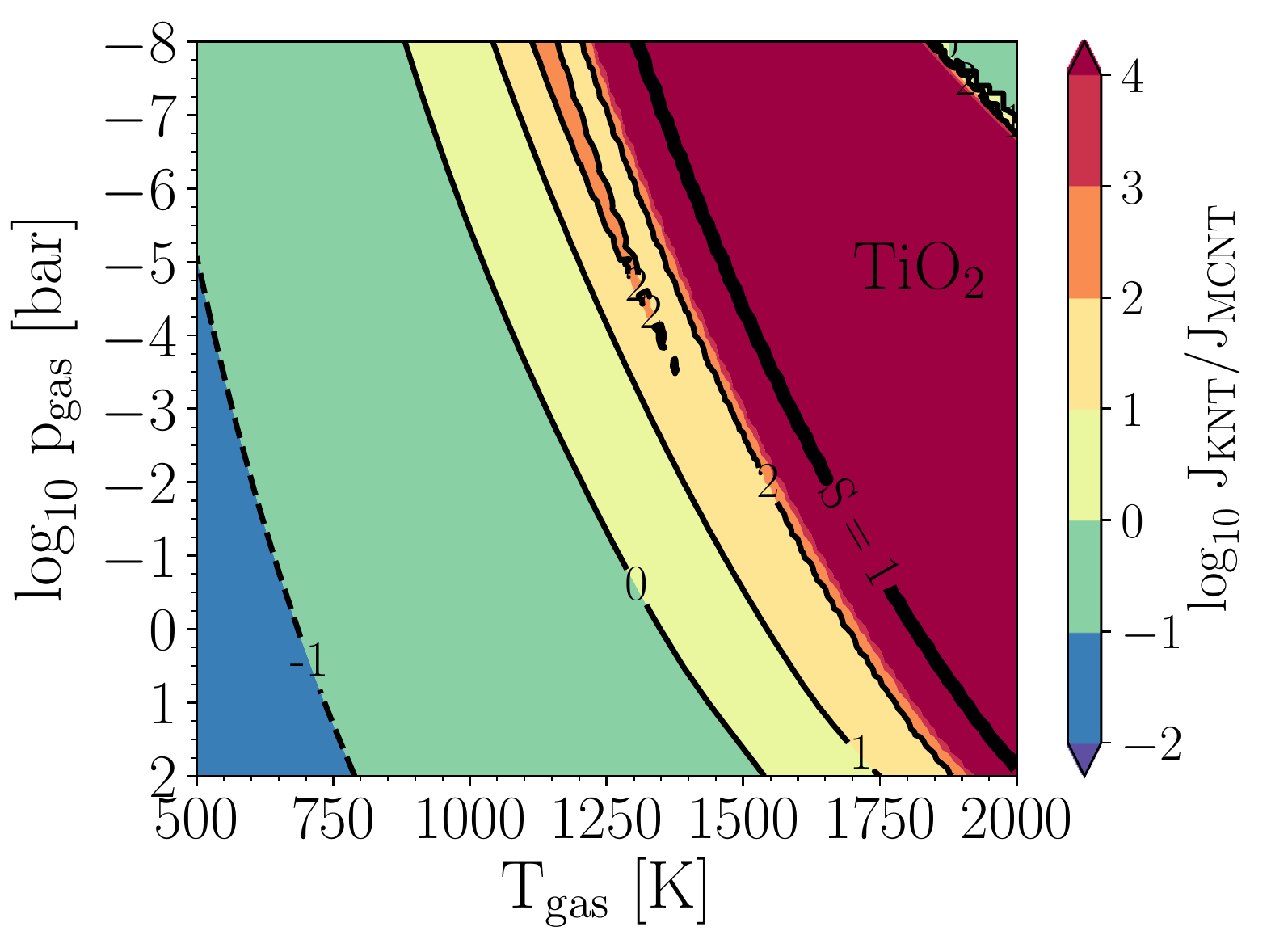}
   \caption{
   \textbf{Left:} Contours of $\log_{10}$ J$_{*}$ [cm$^{-3}$ s$^{-1}$] of TiO$_{2}$[s] (colour bar and thin black lines) applying KNT across the temperature-pressure grid, and the gas-solid phase equilibrium S(T$_{\rm gas}$,p$_{\rm gas}$) = 1 zone (thick black, solid).
   \textbf{Right:} The $\log_{10}$ ratio of J$_{\rm KNT}$/J$_{\rm MCNT}$ for TiO$_{2}$[s] nucleation.}
   \label{fig:TiO2_KNT}
\end{figure*}

One key assumption of classical nucleation theory is that the surface tension of the bulk material is representative of the bond energies through small cluster space, which is generally not the case \citep[e.g.][]{Gail2014}.
A better estimate of nucleation rates are usually acquired by performing laboratory experiments or computational chemistry calculations \citep[e.g.][]{Bhatt2007,Mauney2015,Bromley2016} in order to catalogue and evaluate small cluster thermochemical data.

To illustrate this point, we apply kinetic nucleation theory to TiO$_{2}$[s] seed particle nucleation (Section \ref{sec:KNT}), using the small cluster (TiO$_{2})_{N}$, N = 1-10 thermochemical data from \citet{Lee2015a}, and compare to the results of modified classical nucleation theory presented in Section \ref{sec:TiO2_res}.
To generalise the temperature dependence of the Gibbs energy of formation data, $\dG(N)$, the tables in \citet{Lee2015a} were fit using the polynomial expression of \citet{Sharp1990, Burrows1999}
\begin{equation}
\dG(N) = aT^{-1} + b + cT + dT^{2} + eT^{3} ,
\label{eq:dfG_polyfit}
\end{equation}
the coefficients of which can be found in Appendix \ref{app:dfG_polyfit}.
Figure \ref{fig:TiO2_KNT} (Left) shows the nucleation contour results of the KNT expression Eq. \eqref{eq:J*KNT}.
It is clear that an unphysical nucleation rate is produced where S $<$ 1, where the small clusters are more thermochemicaly favourable than the bulk material, a situation that does not occur.
This is due to the critical cluster size N$_{*}$ increasing rapidly as S $\rightarrow$ 1 \citep{Gail2014}, to beyond N = 10 cluster sizes.
Thus, by truncating Eq. \eqref{eq:J*KNT} at N = 10, an unphysical nucleation rate is produced in regions at S $<$ 1, where N$_{*}$ $>$ 10.
The KNT and MCNT results are within 1 order of magnitude for regions where S $\gg$ 1 (Figure \ref{fig:TiO2_res}, Right).
However, as (T$_{\rm gas}$,p$_{\rm gas}$) approaches the S = 1 equilibrium zone, larger differences between the two methods occur.

Care must therefore be taken when considering a temperature and pressure range where the use of KNT data becomes inaccurate.
\citet{Bromley2016} found that KNT and a full kinetic model agree well in range T = 300-620 K for their SiO nucleation data, but show large divergences for higher temperatures.
\citet{Mauney2015} show excellent agreement between KNT and CNT that made use of computational chemistry derived variables for carbon clusters of N = 2-99.
By considering larger cluster geometries, they were able to capture N$_{*}$ across a wide temperature and pressure space for carbon nucleation.

Computational chemistry and experimental based investigations into the small cluster properties of each of the cloud species discussed here is warranted for future sub-stellar cloud formation modelling.
Most of our seed particle nucleation data and modelling stems from the works of \textit{ab initio} DFT computational chemistry efforts by the \citet{John1997, Chang1998, Patzer1999, Jeong2000} studies for AGB outflows.
A similar effort in the investigation of sub-stellar seed particles is required in order that a more complete, microphysical based cloud formation theory be available to the brown dwarf and exoplanet community.
Until the properties of these small clusters are more complete, classical nucleation theory remains a reasonable, physically based starting point for sub-stellar seed particle formation studies \citep[e.g. see discussion in][]{Gail2014}.

\section{Summary and conclusion}
\label{sec:conclusion}

Seed particle formation is important to the cloud formation process, as without a supply of active condensation surfaces, large, global scale atmospheric cloud structures cannot grow.
Inspired by past chemical equilibrium studies of cloud species in sub-stellar objects, we investigated the nucleation properties of previously identified cloud species (TiO$_{2}$[s], SiO[s], Cr[s], KCl[s], NaCl[s], CsCl[s], H$_{2}$O[s/l], NH$_{3}$[s], H$_{2}$S[s/l], CH$_{4}$[s]) across an atmospheric gas temperature and gas pressure range of 100-2000 K and 10$^{-8}$-100 bar respectively, assuming solar metallicity.

Our results suggest a possible hierarchy of condensation surface providers in brown dwarf and exoplanet atmospheres.
TiO$_{2}$[s] provides the seed particles for the most refractory species at the deepest, hottest cloud layers of the atmosphere (e.g. Fe[s], Al$_{2}$O$_{3}$[s]).
TiO$_{2}$[s] seed particles also provide surfaces for the silicate cloud layer composed of SiO[s], SiO$_{2}$[s], MgSiO$_{3}$[s] and Mg$_{2}$SiO$_{4}$[s] species.
Cr[s] may provide seed particles for the sulfide species MnS[s], Na$_{2}$S[s] and ZnS[s].
A selection of chloride species (KCl[s], NaCl[s], RbCl[s], CsCl[s], NH$_{4}$Cl[s]) may act as a potential seed particle for the NH$_{4}$H$_{2}$PO$_{4}$[s] and H$_{2}$O[s/l] cloud layers in the coldest sub-stellar objects.

The consequences of such a hierarchy on the atmospheric properties of sub-stellar objects are potentially highly significant, since the global cloud cover depends on the availability of solid seed particles, but also the gas phase chemical and temperature-pressure conditions for other species to condense on the seed particle surfaces.
Sub-stellar objects which fall into the three suggested seed particle hierarchies, or have a combination of two or more, can be expected to exhibit very different observational signatures due to the large effect clouds have on the atmospheric profiles.
Should efficient nucleation of a species occur at high altitudes in the atmosphere, a detached aerosol zone may form.
This would affect the observed transit spectra and phase curves, by wavelength dependent absorption and scattering of photons.
Additionally, should a key seed particle species not be present in the photosphere (for example, cold trapped and unable to mix upward to higher altitudes), then the photosphere of the sub-stellar object can be expected to be cloud free, even if other solid species are supersaturated.
Such a process has consequences on the observed geometric albedo and transmission spectra of an exoplanet.

The calculation of seed particle nucleation rates for each species can be further improved with a thorough investigation of the small cluster properties of each seed particle formation species using computational chemistry techniques or lab based nucleation experimentation.
Detailed small cluster geometry and thermochemical properties for sub-stellar atmosphere seed particles is warranted for future, holistic kinetic modelling of sub-stellar atmospheric clouds.

\begin{acknowledgements}
We thank the anonymous referee for their careful and thoughtful review and suggestions on the manuscript content.
GKHL and ChH highlight the financial support of the European community under the FP7 ERC starting grant 257431.
GKHL acknowledges support from the Universities of Oxford and Bern through the Bernoulli fellowship.
JB is supported by NASA through the NASA ROSES-2016/Exoplanets Research Program, grant NNX1:7ACO3G.
We thank P. Rimmer for providing the extended Jupiter (T$_{\rm gas}$, p$_{\rm gas}$) profile.
We thank P. Woitke for discussions on CE chemistry and discussions on the manuscript content.
We thank V. Parmentier for additional discussion topic suggestions.
We thank J.-L. Baudino for the Exo-REM Y Dwarf T-p profile.
Our local HPC computational support at Abu Dhabi, Oxford and St Andrews is highly acknowledged.
Most plots were produced using the community open-source Python packages Matplotlib, SciPy and AstroPy.
A large portion of the vapour pressure and thermochemical data were found using the NIST-JANAF thermochemical databases.
\end{acknowledgements}

\clearpage
\bibliographystyle{aa}
\bibliography{bib2}{}

\appendix

\clearpage
\section{Bulk properties of seed particle candidates}
\label{app:bulk_properties}

\begin{table*}
\caption{Table of species vapour pressure expressions used in this study.
T$_{C}$ denotes a temperature in degrees Celsius.
References: \textbf{a} - \citet{Woitke2004}, \textbf{b} - \citet{Wetzel2013}, \textbf{c} - \citet{Morley2012}, \textbf{d} - \citet{Stull1947}, \textbf{e} - \citet{Ackerman2001}, \textbf{f} - \citet{Prydz1972}, \textbf{g} - this work.}
\begin{center}
\begin{tabular}{c l c} \hline \hline
Solid s & Vapour pressure & Reference \\ \hline
TiO$_{2}$[s] & $\ln$P(dyn cm$^{-2}$) = 35.8027 - 74734.7/T & \textbf{a} \\
SiO[s] &  $\ln$P(dyn cm$^{-2}$) = 32.52 - 49520/T & \textbf{b} \\
Cr[s]  & $\log_{10}$P(bar) = 7.490 - 20592/T & \textbf{c} \\
KCl[s]  & $\log_{10}$P(bar) = 7.611 - 11382/T & \textbf{c} \\
NaCl[s] & $\log_{10}$P(bar) = 5.07184 - 8388.497 / (T - 82.638)   & \textbf{d} \\
CsCl[s] & $\ln$P(dyn cm$^{-2}$) = 29.9665 - 23055.3/T & \textbf{g} \\
H$_{2}$O[s/l] & P(dyn cm$^{-2}$) =  6111.5 $\exp$[(23.036 T$_{C}$ -  T$_{C}^{2}$/333.7)/(T$_{C}$ + 279.82)]  & (T $<$ 273.16 K)$^{\textbf{e}}$ \\
 & P(dyn cm$^{-2}$) =  6112.1 $\exp$[(18.729 T$_{C}$ -  T$_{C}^{2}$/227.3)/(T$_{C}$ + 257.87)]  & (T = 273.16-1048 K)$^{\textbf{e}}$ \\
NH$_{3}$[s] & P(bar) = $\exp$[10.53 - 2161/T - 86596/T$^{2}$]  & \textbf{e} \\
H$_{2}$S[s/l] & $\log_{10}$P(bar) = 4.43681 - 829.439 / (T - 25.412)  & (T = 138.8-212.8 K)$^{\textbf{d}}$  \\
 & $\log_{10}$P(bar) = 4.52887 - 958.587 / (T - 0.539)  & (T = 212.8-349.5 K)$^{\textbf{d}}$ \\
CH$_{4}$[s] & $\log_{10}$P(bar) = 3.9895 - 443.028 / (T - 0.49) & \textbf{f} \\ \hline
\end{tabular}
\end{center}
\label{tab:vapp}
\end{table*}%

\begin{table*}
\caption{Surface tensions, $\sigma$, used in the modified classical nucleation theory scheme.}
\begin{center}
\begin{tabular}{c c c}  \hline \hline
Solid s & $\sigma$ [erg cm$^{-2}$] &  Reference  \\ \hline
TiO$_{2}$[s] & 480.6  &  \citet{Lee2015a} \\
Cr[s] & 3330  & \citet{Tran2016}\\
KCl[s] & 100.3 &  \citet{Janz2013} \\
NaCl[s] & 113.3 & \citet{Janz2013}  \\
CsCl[s] & 100.0 & Est. \\ \hline
H$_{2}$O[s] & 109 &  \citet{Ketcham1969}  \\
NH$_{3}$[s] & 23.4 & \citet{Weast1988} \\
H$_{2}$S[s] & 58.1 & \citet{Nehb2006}  \\
CH$_{4}$[s] & 14.0 &  \citet{UScoast1984} \\  \hline
\end{tabular}
\end{center}
\label{tab:s_tensions}
\end{table*}%

\begin{table*}
\caption{Molar mass, bulk density ($\rho_{d}$) and hypothetical monomer radii ($a_{0}$) for each seed particle candidate species.
The bulk density of H$_{2}$O[s] (water ice) is taken from \citet{Ackerman2001} and references therein.
The bulk density of NH$_{3}$[s] (ammonia ice) and CH$_{4}$[s] (methane ice) are taken from \citet{Satorre2013} and \citet{Satorre2008} respectively.}
\begin{center}
\begin{tabular}{c c c c} \hline \hline
Solid s &  Molar mass [amu] & $\rho_{d}$ [g cm$^{-3}$] & $a_{0}$ [\AA] \\ \hline
TiO$_{2}$[s] & 79.866 & 4.23 & 1.956  \\
SiO[s] & 44.085 & 2.18 & 2.001 \\
Cr[s] & 51.996 & 7.19 & 1.421 \\
KCl[s] & 74.551 & 1.98 & 2.462 \\
NaCl[s] & 58.443 & 2.16 & 2.205  \\
CsCl[s] & 168.359 & 3.99 & 2.557 \\ \hline
H$_{2}$O[s] & 18.015 & 0.93 & 1.973  \\
NH$_{3}$[s] & 17.031 & 0.87 & 1.980 \\
H$_{2}$S[s] & 34.081 & 1.12 & 2.293  \\
CH$_{4}$[s] & 16.043 & 0.47 & 2.383 \\ \hline
\end{tabular}
\end{center}
\label{tab:radii}
\end{table*}%

\clearpage
\section{Nucleation parameter sensitivity}
\label{app:paramaters}

\begin{figure*}[ht] 
   \centering
   \includegraphics[width=0.42\textwidth]{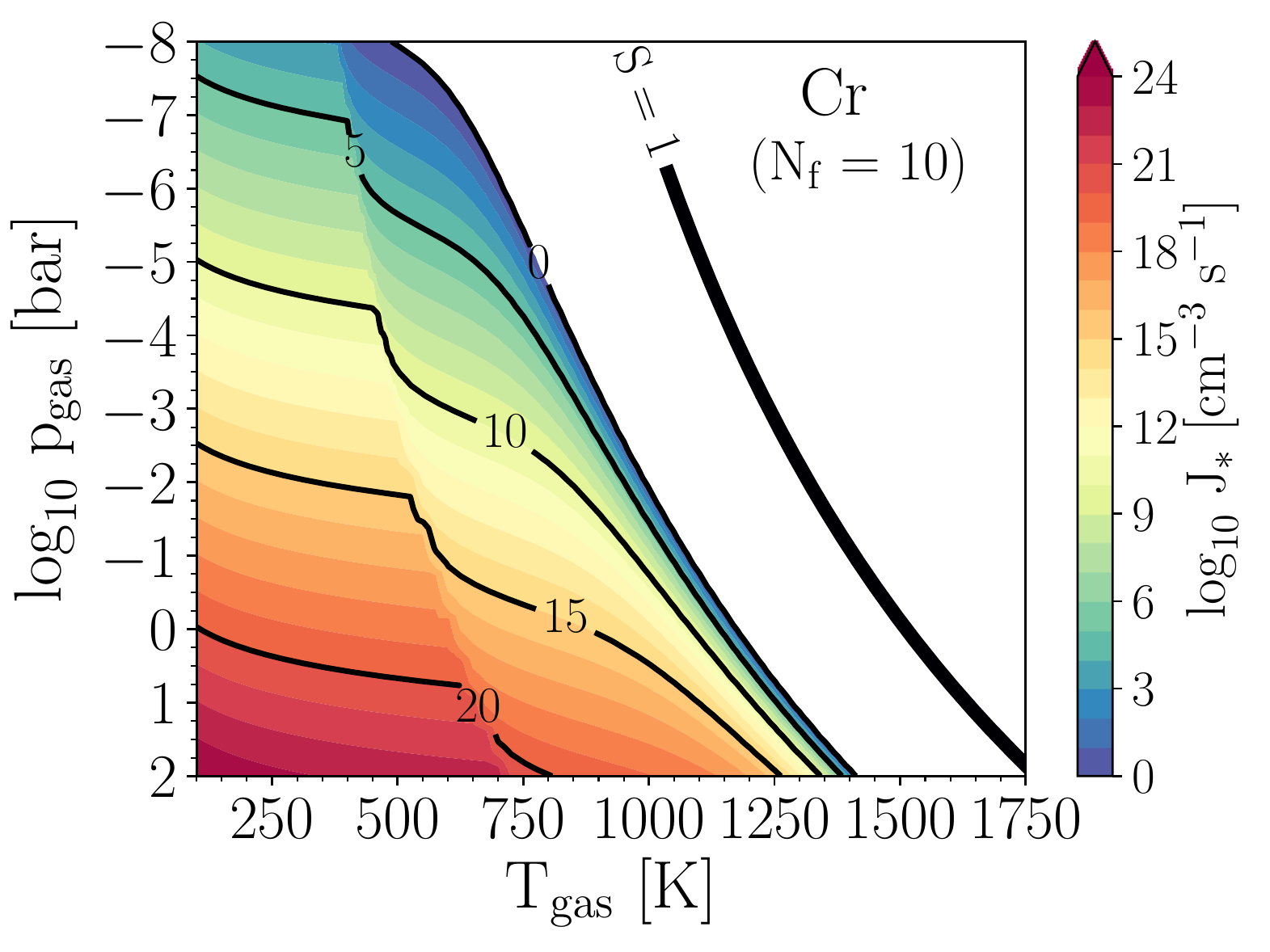}
   \includegraphics[width=0.42\textwidth]{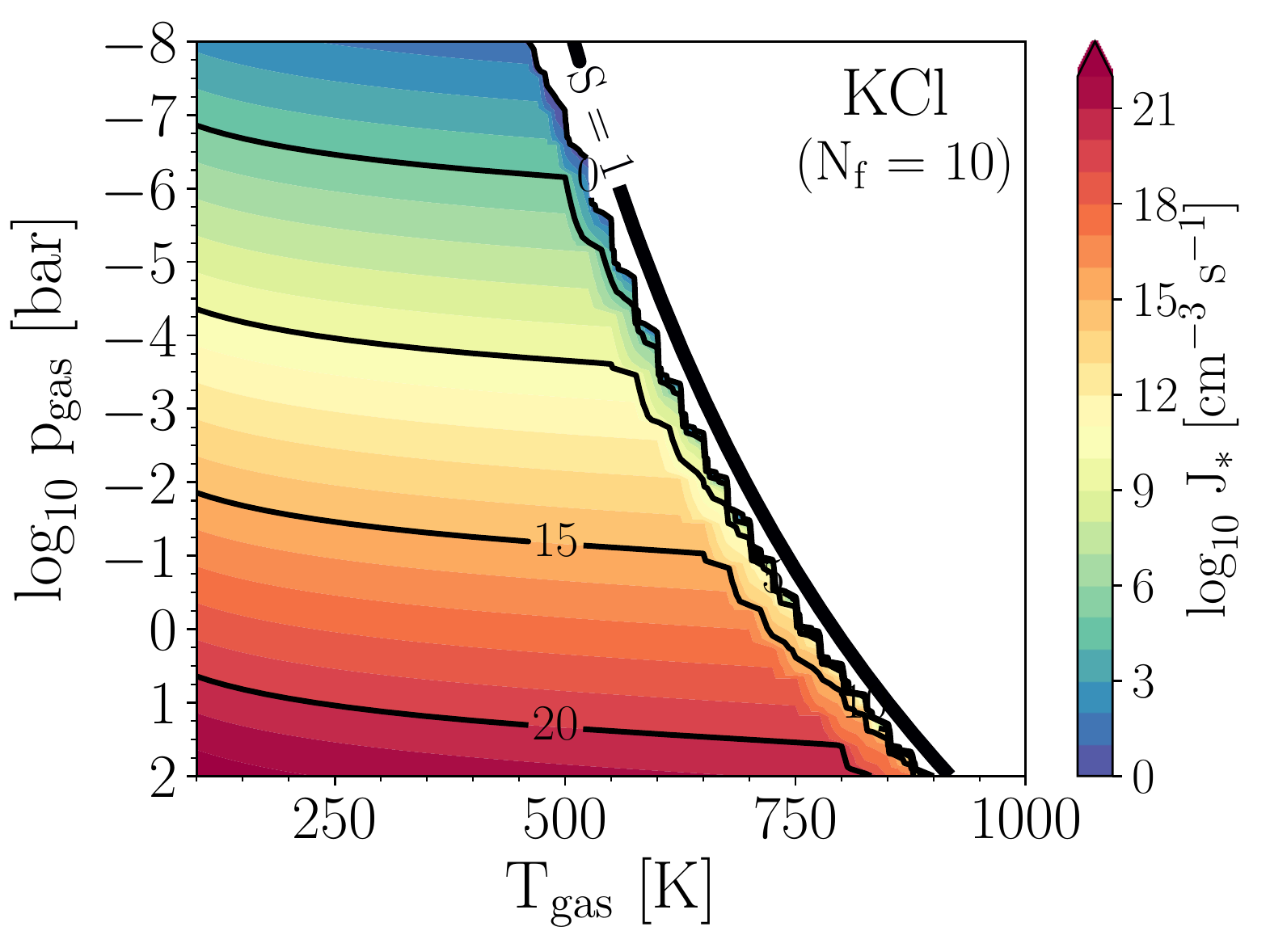}
   \includegraphics[width=0.42\textwidth]{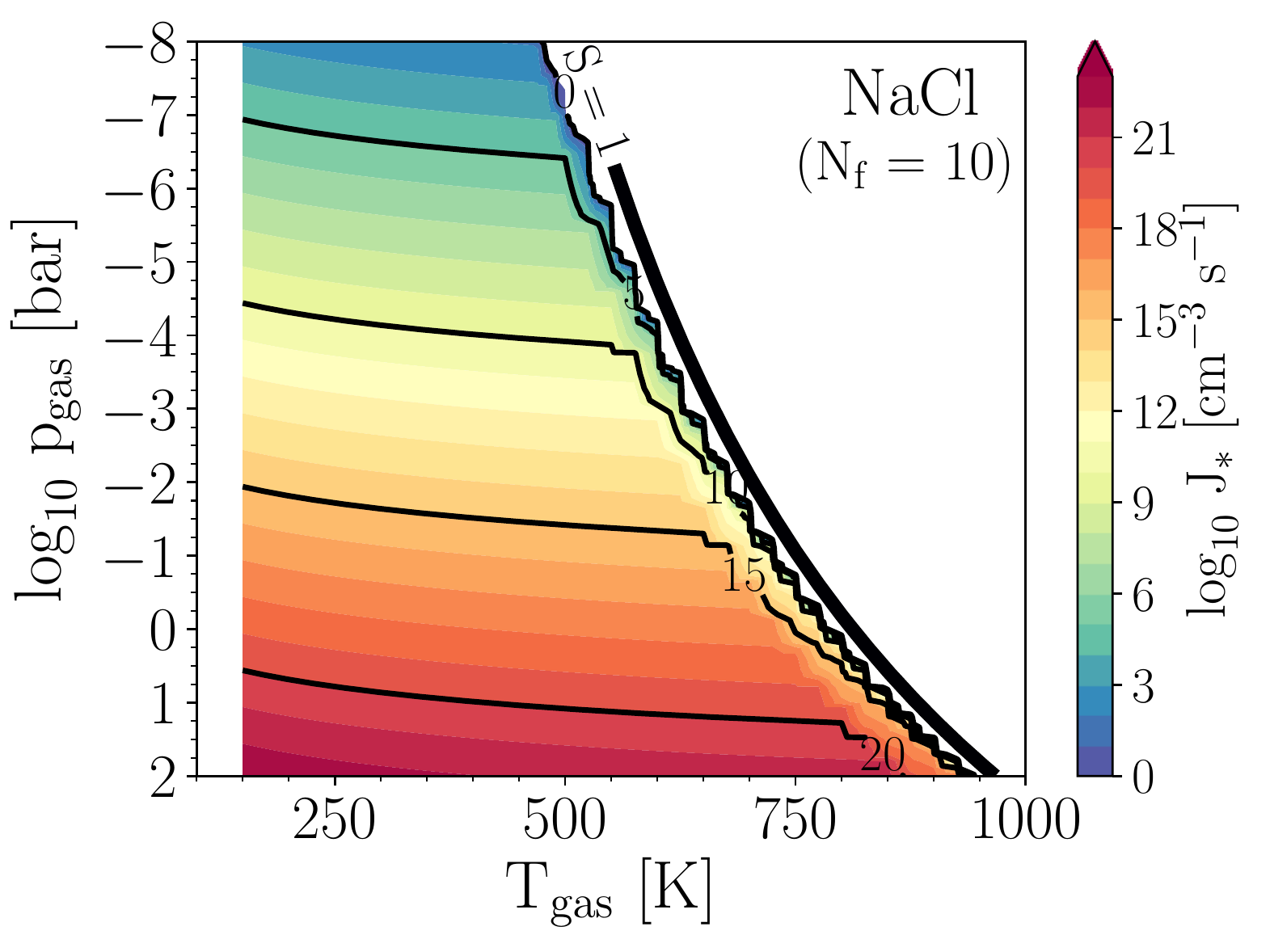}
   \includegraphics[width=0.42\textwidth]{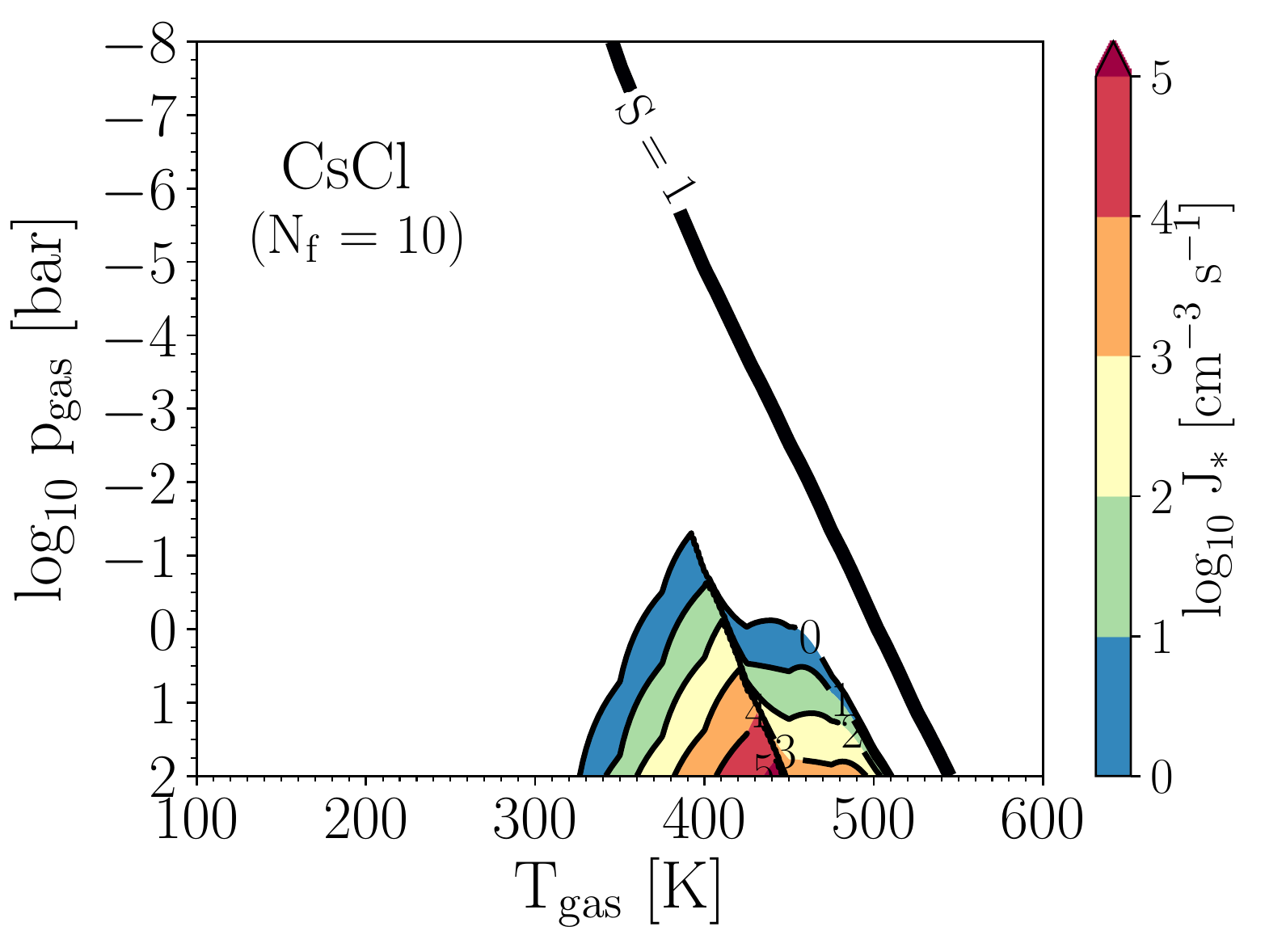}
   \includegraphics[width=0.42\textwidth]{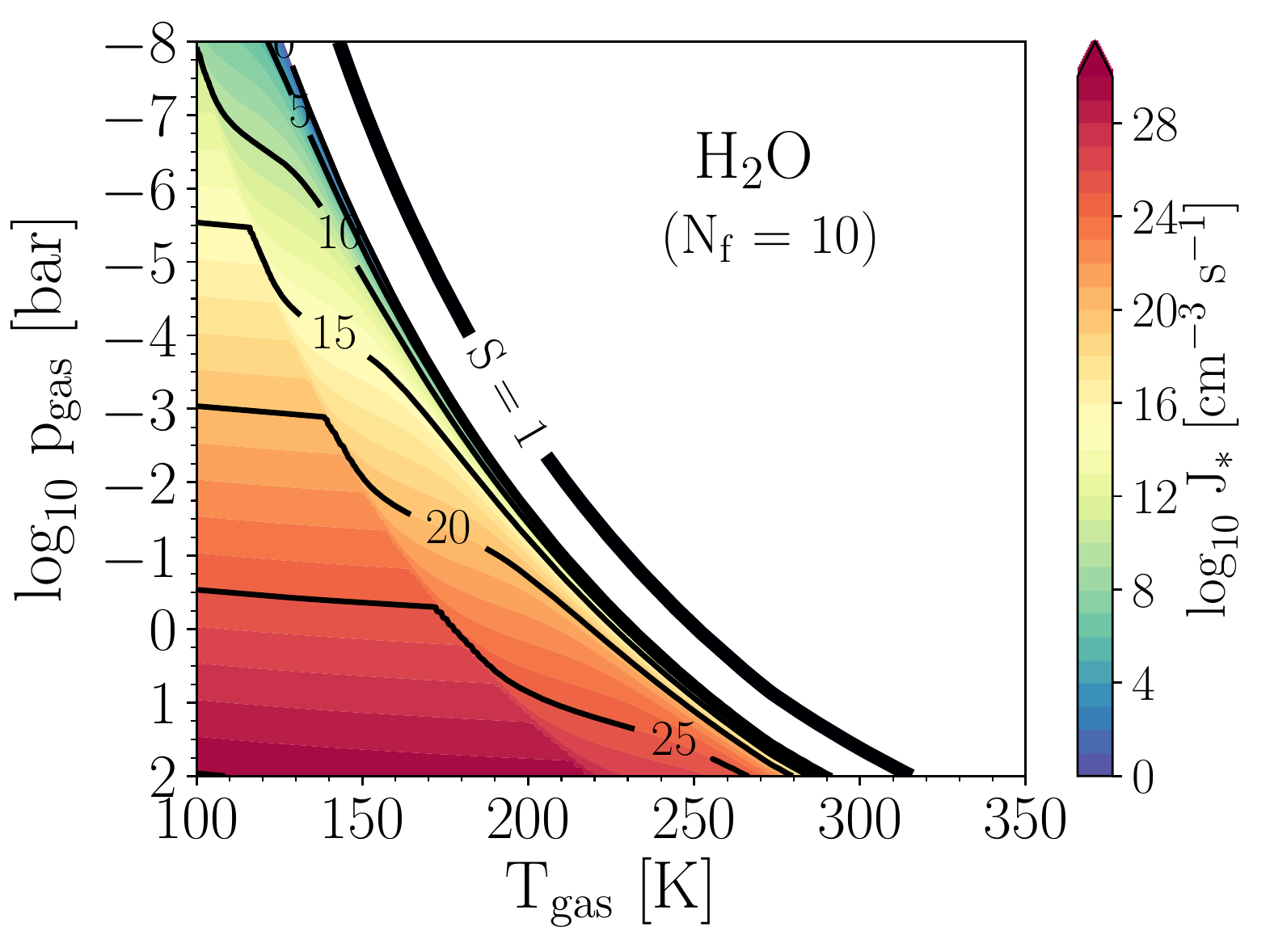}
   \includegraphics[width=0.42\textwidth]{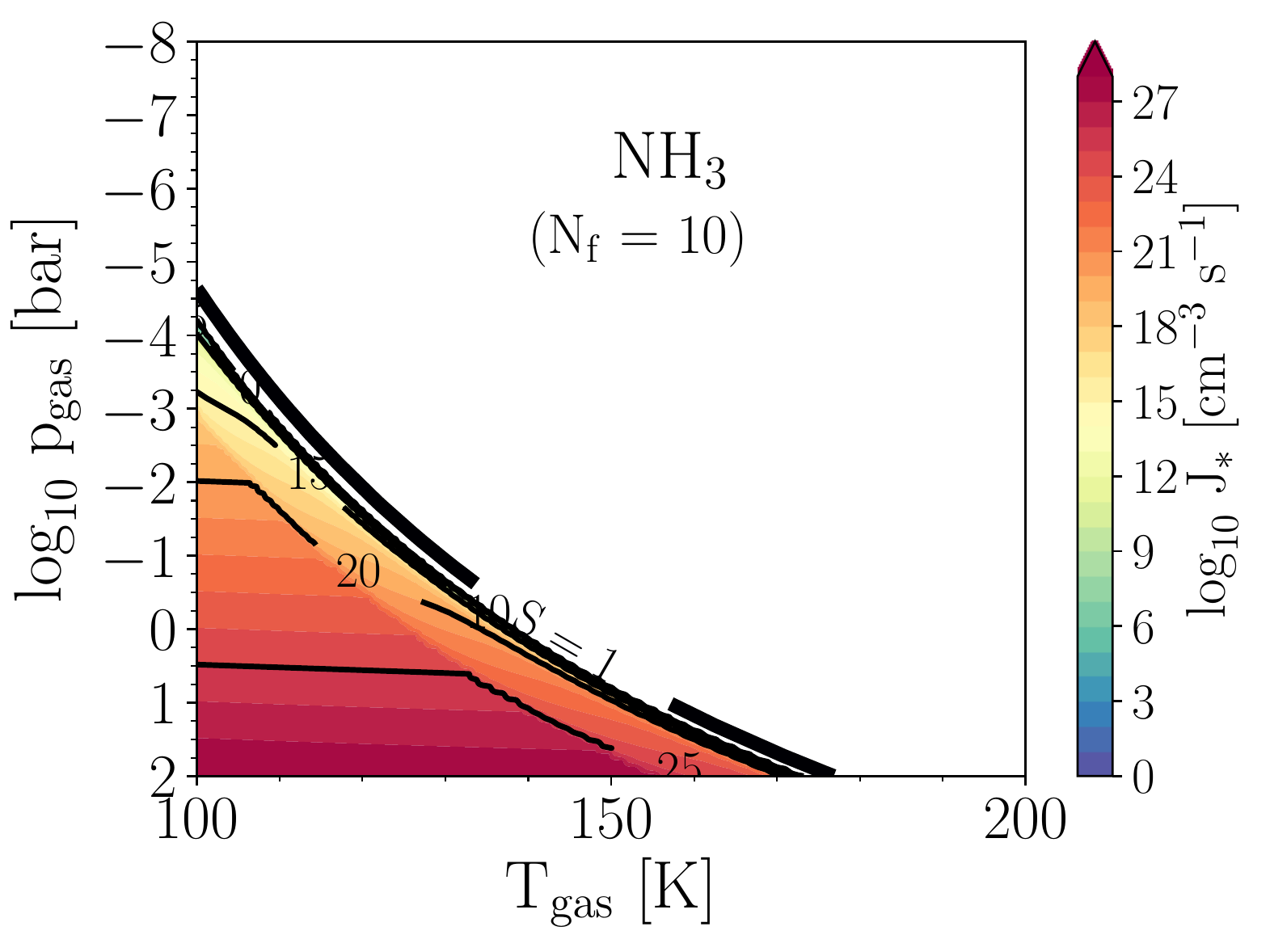}
   \includegraphics[width=0.42\textwidth]{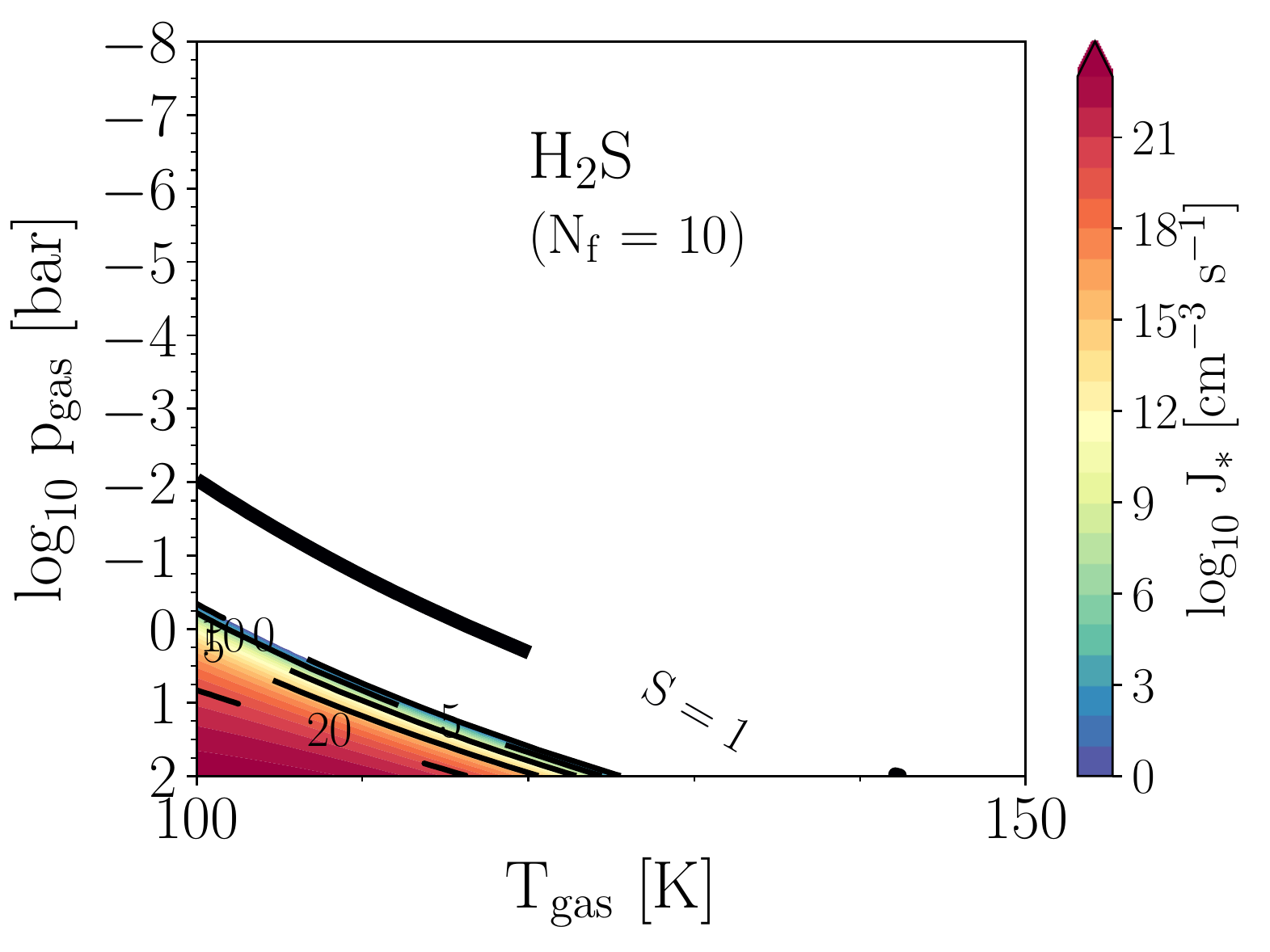}
   \caption{Contour plots of the nucleation species assuming a fitting factor of N$_{\rm f}$ = 10, except TiO$_{2}$[s] and SiO[s], compared to the main text where N$_{\rm f}$ = 1.
   The change of the fitting factor does not alter the conclusions of the main text.}
   \label{fig:paramaters}
\end{figure*}

\clearpage
\section{$\dG(N)$ polynomial coefficients}
\label{app:dfG_polyfit}

\begin{table*}
\caption{$\dG(N)$ [kJ mol$^{-1}$] temperature dependent polynomial fit coefficients for Eq. \ref{eq:dfG_polyfit}.
The coefficients were calculated using the SciPy (\textsc{curve\_fit}) non-linear least squares fitting routine.
Valid over the temperature range T = 500-2000 K.}
\begin{center}
\begin{tabular}{c c c c c c}  \hline \hline
N & a & b & c & d & e \\ \hline
1 & -1.63472903E+03 &  -2.29197239E+02 &  -3.60996766E-02 &  1.60056318E-05  & -2.02075337E-09 \\
2 & -4.39367806E+03 &  -9.77431160E+02 &   1.01656231E-01 &   2.16685151E-05 &  -2.90960794E-09 \\
3 & -7.27464297E+03 &  -1.72789122E+03 &   2.40409836E-01 &   2.74002833E-05 &  -3.81294573E-09 \\
4 & -1.02808569E+04 &  -2.51074121E+03 &   4.15061961E-01 &   3.30076021E-05 &  -4.69138304E-09 \\
5 & -1.37139638E+04 &  -3.27506794E+03 &   5.73212328E-01 &   4.12461166E-05 &  -6.14829810E-09 \\
6 & -1.60124756E+04 &  -4.13772573E+03 &   7.32672450E-01 &   4.44131101E-05 &  -6.48290229E-09 \\
7 & -1.89334054E+04 &  -4.91964308E+03 &   8.93689186E-01 &   4.99942488E-05 &  -7.35905348E-09 \\
8 & -2.17672541E+04 &  -5.72392348E+03 &   1.05703014E+00 &   5.57819924E-05 &  -8.27043313E-09 \\
9 & -2.48377680E+04 &  -6.51357184E+03 &   1.22288686E+00 &   6.10116309E-05 &  -9.08225913E-09 \\
10 & -2.76078426E+04 &  -7.34516329E+03 &   1.37500651E+00 &   6.70631142E-05 &  -1.00410219E-08 \\ \hline
\end{tabular}
\end{center}
\label{tab:app:dfG_polyfit}
\end{table*}%

\end{document}